%% file: main.tex
\documentclass[12pt]{article}
\usepackage[margin=1in]{geometry}

\usepackage{color,
setspace,sectsty,comment,footmisc,caption,
pdflscape,subcaption,array,hyperref,adjustbox, longtable
}
\usepackage{minitoc}
\usepackage{times}
\usepackage[toc,page,header]{appendix}
\usepackage{subfiles}
\usepackage{url}            % simple URL typesetting
\usepackage{booktabs}       % professional-quality tables
\usepackage{nicefrac}       % compact symbols for 1/2, etc.
\usepackage{microtype}      % microtypography
\usepackage{amsmath,amssymb,amsfonts,amsthm}
\usepackage[round]{natbib} % cannot use with AAAI! A.S.

\usepackage{float}
\usepackage{sgame, tikz} % Game theory packages
\usepackage{algorithm,algpseudocode}
\usepackage{makecell}
 \usepackage{multirow}
 \usepackage{graphicx}

\usepackage{booktabs} % For formal tables

\newtheorem{finding}{Finding}

\usepackage{slivkins-setup,slivkins-theorems}
\renewcommand{\xhdr}[1]{\vspace{1mm} \noindent{\bf #1}}

\newcommand{\TheoryModel}{Bayesian-choice model\xspace}
\newcommand{\ExptsModel}{reputation-choice model\xspace}

\newcommand{\term}[1]{\ensuremath{\mathtt{#1}}\xspace}

\newcommand{\TS}{\term{TS}}    % Thompson Sampling
\newcommand{\DEG}{\term{BEG}}  % "Dynamic eps-greedy"
\newcommand{\DG}{\term{BG}}    % "Dynamic Greedy"

\newcommand{\HMR}{\term{HMR}} % HardMaxRandom
\newcommand{\HM}{\term{HM}}    % HardMax

\newcommand{\Beta}{\term{Beta}} % for Beta distribution
\newcommand{\Eeog}{\term{EoG}} % shorthand for "effective end of game"

\newcommand{\MRV}{mean reward vector\xspace} % mean reward vector
\newcommand{\MRVs}{mean reward vectors\xspace} % mean reward vector

\newcommand{\gr}{\texttt{gr}}
\newcommand{\BIRgr}{\BIR^{\gr}}
\newcommand{\alggr}{\alg[\gr]}
\newcommand{\rewgr}{\rew_{\gr}}

\DeclareMathOperator*{\Expectation}{\mathbb{E}}
\newcommand{\Ex}[2]{\Expectation_{#1}\left[#2\right]}
\newcommand{\FiniteGame}{finite competition game\xspace}

\newcommand{\rew}{\term{rew}}  % Bayesian-expected reward after n local rounds
\newcommand{\PMR}{\term{PMR}} % posterior mean reward
\newcommand{\support}{\term{support}}

\newcommand{\BIR}{\term{BIR}} % Bayesian Instantaneous Regret

\newcommand{\respF}{f_{\term{resp}}}

\newcommand{\BReg}{\term{BReg}}
\newcommand{\HardMax}{\term{HardMax}}
\newcommand{\HardMaxRandom}{\term{HardMax\&Random}}
\newcommand{\SoftMaxRandom}{\term{SoftMax}}
\newcommand{\Uniform}{\term{Uniform}}

\newcommand{\StaticGreedy}{\term{StaticGreedy}}
\newcommand{\DynGreedy}{\term{BayesGreedy}}
\newcommand{\DynamicGreedy}{\DynGreedy}
\newcommand{\DynamicEpsGreedy}{\term{BayesEpsilonGreedy}}
\newcommand{\Thompson}{\term{ThompsonSampling}}

\newcommand{\bmonotone}{Bayesian-monotone\xspace}
\newcommand{\bmonotonicity}{Bayesian-monotonicity\xspace}

\newcommand{\termSub}[2]{\ensuremath{\mathtt{#1}_{#2}}\xspace}
\newcommand{\alg}[1][]{\termSub{alg}{#1}}
\newcommand{\prior}{\ensuremath{\mP}\xspace}
\newcommand{\priorMu}{\ensuremath{\prior_\mathtt{mean}}\xspace}
\newcommand{\posteriorN}[2]{\mN_{#1,#2}}  % \posteriorN{principal}{round}

\newcommand{\termTXT}[1]{{\em {#1}}\xspace}

\newcommand{\rationality}{\termTXT{rationality}}

\newcommand{\innovation}{\termTXT{innovation}}

\newcommand{\competition}{\termTXT{competition}}
\newcommand{\Competition}{\termTXT{Competition}}
\newcommand{\competitiveness}{\termTXT{competitiveness}}
\newcommand{\exploration}{\termTXT{exploration}}

\usepackage[suppress]{color-edits}  % use this to suppress the package
\addauthor{as}{red}    % as for Alex
\addauthor{ga}{blue}  % ga for Guy
\addauthor{sw}{orange} % sw for Steven
\addauthor{ym}{magenta} % ym for Yishay
\renewcommand{\arraystretch}{1.5} % Default value: 1

\begin{document}

%\begin{titlepage}

\title{\vspace{-12mm}Competing Bandits:\\
The Perils of Exploration under Competition%
\thanks{
All theoretical results are from \citet{CompetingBandits-itcs18}, and all numerical simulations are from \citet{CompetingBandits-ec19} (which was published as a 2-page abstract in ACM EC 2019). This manuscript features a unified and streamlined presentation, expanded related work, and revised background materials (\eg Appendices~\ref{app:bg},\ref{app:examples} are new). \vspace{1mm}

We thank Ian Ball, Yeon-Koo Che, Bar Light, Sven Rady, Sara Shahanaghi and Glen Weyl for helpful comments and conversations. We also thank the audience of the conference and seminar talks on this work, and the anonymous conference and journal referees. All errors are our own.\vspace{1mm}}}
%seminar participants at Columbia and conference participants at Innovations in Theoretical Computer Science 2018, ACM Economics and Computation 2019 and the MIT Conference on Digital Experimentation 2020, for helpful comments and conversations.

% \textit{Competing Bandits: Learning under Competition}, at Innovations in Theoretical Computer Science 2018, and \textit{The Perils of Exploration under Competition: A Computational Modeling Approach}, at ACM Economics and Computation 2019.}}

\author{Guy Aridor
\footnote{Northwestern University Kellogg, Department of Marketing. Email: guy.aridor@kellogg.northwestern.edu.}
\and \hspace{-0.75cm}
\rule{0.0in}{0pt}
Yishay Mansour
\footnote{Tel Aviv University, Department of Computer Science, and Google Tel Aviv. Email: mansour.yishay@gmail.com. \newline
Supported in part by Israel Science Foundation grant number 993/17.}
\and \hspace{-0.75cm}
\rule{0.0in}{0pt}
Aleksandrs Slivkins
\footnote{Microsoft Research New York City. Email: slivkins@microsoft.com.}
\and \hspace{-0.75cm}
\rule{0.0in}{0pt}
Zhiwei Steven Wu%
\footnote{Carnegie Mellon University, Pittsburgh, PA
Email: zstevenwu@cmu.edu.\newline
Research done when Z.S. Wu was an intern and a postdoc at Microsoft Research NYC.}
}

\date{First version: July 2020\\ This version: October 2024}

\maketitle

\vspace{-5mm}
\begin{abstract}
\input{abstract-short}
\end{abstract}

%\noindent\textbf{Keywords:}  Competition vs. innovation, exploration vs. exploitation, multi-armed bandits, regret. \\
%\noindent\textbf{JEL Codes}: D83, L15, O31

%D83 (Microeconomics -> Information, Knowledge, and Uncertainty -> Search • Learning • Information and Knowledge • Communication • Belief • Unawareness), L15 (Industrial Organization -> Market Structure, Firm Strategy, and Market Performance -> Information and Product Quality • Standardization and Compatibility), O31 (Economic Development, Innovation, Technological Change, and Growth ->Innovation • Research and Development • Technological Change • Intellectual Property Rights ->Innovation and Invention: Processes and Incentives)

%\setcounter{page}{0}
\thispagestyle{empty}
%\end{titlepage}

\newpage
%\addtocontents{toc}{\protect\setcounter{tocdepth}{0}}
\begin{small}
\setcounter{tocdepth}{2}
\tableofcontents
\end{small}
\thispagestyle{empty}
\newpage

\section{Introduction}
\label{sec:intro}

\subfile{content/sec-intro}

\section{Related work}
\label{sec:related-work}
\subfile{content/related_work}

\section{Our model in detail}
\label{sec:model}
\subfile{content/sec-model}

\section{Theoretical results: the \TheoryModel}
\label{sec:theory}
\subfile{content/sec-theory}

%\section{Full rationality (HardMax)}
%\label{sec:rational}
%\subfile{itcs18paper/sec-rational}

%\section{Relaxed rationality: HardMax \& Random}
%\label{sec:random}
%\subfile{itcs18paper/sec-random}

%\section{SoftMax response function}
%\label{sec:soft}
%\subfile{itcs18paper/sec-soft}

%\section{Economic implications}
%\label{sec:welfare}
%\subfile{itcs18paper/sec-welfare}

\section{Numerical simulations: the \ExptsModel}
\label{sec:sim}
%\label{sec:sim_details}

\subfile{content/sim_details}

\subfile{ec19paper/content/perf_in_iso}

\subfile{ec19paper/content/inverted_u}

\subfile{ec19paper/content/barriers}

\subfile{ec19paper/content/non_greedy_choice}

\subfile{ec19paper/content/revisited}

\subfile{content/conclusion}

%\newpage
\addcontentsline{toc}{section}{References}

\bibliographystyle{ecta.bst}
\begin{small}
\bibliography{bib-abbrv,bib-ML,refs,bib-bandits,bib-AGT,bib-slivkins, bib-random}
\end{small}

\clearpage

\renewcommand{\appendixname}{Appendix}
\renewcommand{\appendixtocname}{List of appendices}

      % body goes here
\appendix
\renewcommand{\contentsname}{Appendix}

%\tableofcontents
%\addtocontents{toc}{\protect\setcounter{tocdepth}{2}}
      % appendices go here
%\newpage

%In the appendix, we provide background on multi-armed bandits as well as several discussions and proofs omitted from the main text. Furthermore, we provide plots and tables for our experiments, which were omitted from the main text. In all cases, the plots and tables here are in line with those in the main text, and lead to similar qualitative conclusions.

\section{Notation and Definition Tables}
\label{app:notation}
\subfile{content/app-tables}

\section[Background for non-specialists: multi-armed bandits]{Background for non-specialists: \\multi-armed bandits}
\label{app:bg}
%\label{sec:related-classes}
\subfile{content/sec-bg}

\section{Monotone MAB algorithms}
\label{app:examples}

\subfile{itcs18paper/app-examples}

\section{Non-degeneracy via a random perturbation}
\label{app:perturb}
\subfile{itcs18paper/app-perturb}

\section{Full proofs for Section~\ref{sec:theory}}
\label{sec:theory-proofs}
\subfile{content/sec-theory-proofs}

\newpage
\section{Full experimental results}
\label{app:expts}

\subfile{ec19paper/content/appendix_for_one_version}

%\subfile{content/old-intro}

\end{document}

%% file: abstract-short.tex
Most online platforms learn from interactions with users, and engage in \emph{exploration}: making potentially suboptimal choices in order to acquire new information. We study the interplay between \emph{exploration} and \emph{competition}: how such platforms balance the exploration for learning and competition for users.
%Here users play three distinct roles: they are customers that generate revenue, they are sources of data for learning, and they are self-interested agents which choose among the competing platforms.

We consider a stylized duopoly in which two firms face the same multi-armed bandit problem. Users arrive one by one and choose between the two firms, so that each firm makes progress on its bandit problem only if it is chosen. We study whether competition incentivizes the adoption of better algorithms. We find that stark competition disincentivizes exploration, leading to low welfare.
%induces firms to commit to a ``greedy" algorithm that leads to low welfare.
However, weaker competition
%by providing firms with some ``free" users
incentivizes better exploration algorithms and increases welfare.  We investigate two channels for weakening the competition: stochastic user choice models and a first-mover advantage.
Our findings speak to the competition-innovation relationship and the first-mover advantage in the digital economy.
%Our findings are closely related to the ``competition vs. innovation" relationship, and elucidate the first-mover advantage in the digital economy.

%% file: content/sec-intro.tex
Learning from interactions with users is ubiquitous in modern customer-facing platforms, from product recommendations to web search to content selection to fine-tuning user interfaces. Many platforms purposefully implement \emph{exploration}: making potentially suboptimal choices for the sake of acquiring new information. Online platforms routinely deploy A/B tests, and are increasingly adopting  more sophisticated exploration methodologies based on \emph{multi-armed bandits}, a standard and well-studied framework for exploration and making decisions under uncertainty. This trend has been stimulated by two factors: almost-zero cost of deploying iterations of a product (provided an initial infrastructure investment), and the fact that many online platforms primarily compete on product quality, rather than price
(\eg because they are supported by ads or cheap subscriptions).
%\citep{Gittins-book11,Bubeck-survey12,slivkins-MABbook,LS19bandit-book}.

%~\cite{KohaviAB-2015,KohaviLSH09}

In this paper, we study the interplay between \exploration and \competition.%
\footnote{\Ie we add \competition to the standard exploration-exploitation tradeoff studied in multi-armed bandits.}
Platforms that engage in exploration typically need to compete against one another. Most importantly, platforms compete for users, who benefit them in two ways:
generating revenue and providing data for learning. This creates a tension:
%between \exploration and \competition.
while exploration may be essential for improving the service tomorrow, it may degrade the service quality \emph{today}, in which case some of the users can leave and there will be fewer users to learn from. This may create a ``data feedback loop" when the platform's performance further degrades relative to competitors who keep learning and improving from \emph{their} users, and so forth. Taken to the extreme, such dynamics may cause a ``death spiral" effect when the vast majority of customers eventually switch to competitors.

%\asdelete{Users therefore serve three distinct roles: they are customers that generate revenue, they are sources of data for learning, and they are self-interested agents who choose among the competing systems.}

The main high-level question we ask is:
%that we focus on in this paper is:
%\begin{align}\label{eq:main-Q}
%\asedit{We ask:}
\textbf{Whether and how does competition between platforms incentivize the adoption of better exploration algorithms?}
%\end{align}
This translates into a number of more concrete questions. While it is commonly assumed that better technology always helps, is this so under competition? Does increased competition lead to higher consumer welfare? How significant are the data feedback loops
%--- when more data leads to more users, which leads to even more data, etc. ---
and how they relate to the anti-trust considerations?
%Finding formalizations that admit meaningful answers is a major part of the overall challenge.
We offer a mix of theoretical results and numerical simulations, in which we study complex interactions between platforms' learning dynamics and users' self-interested behavior. Prior work on exploration vs. competition targets technically very different models of competition which are not amenable to our high-level question (as we discuss in Section~\ref{sec:related-work}).

%The choice of a particular technology (exploration algorithm) not an abstract, static choice with a predetermined outcome for the platform. Instead, we model the algorithms explicitly, and investigate how they play out in competition over an extended period of time.

%\footnote{This is a fundamental question which is part of a larger policy discussion around whether data can serve as an indirect network effect and lead to similar ``market tipping" results as is standard in the literature on competition in markets with network effects (see \cite{jullien2019economics} for a policy oriented discussion of this).}

% the extent to which the game between the two principals is competitive
% degree of innovation that these models incentivize.
% the extent to which agents make rational decisions

%\subsection{Our model}
%\label{sec:intro-model}

%\subsection{Our model: competing bandits}
%\label{sec:intro-model}

%We investigate these questions with
\xhdr{Our model: competition game.} We consider a stylized duopoly model in which two firms (\emph{principals}) compete for users (\emph{agents}). Principals compete on quality rather than on prices, and engage in exploration in order to learn which actions lead to high quality products. Agents arrive sequentially. A new agent arrives and chooses a principal (more on this below). The principal selects an action which affects the quality of service provided to this agent, \eg a list of web search results. The agent experiences this action and the resulting reward from this action is observed by the principal. Each principal only observes its own users. Principals commit to their strategies in advance, so as to maximize their market share.

%\asedit{For tractability, our theoretical results and our numerical simulations adopt closely related but technically different model variants (respectively, Bayesian and frequentist), with similar findings.}

%to one of the principals.
%In all variants, agents have little or no information about other agents' choices and rewards.
In more detail, the principal-side model is as follows. Each principal faces a basic and well-studied version of the multi-armed bandit problem, where each reward is drawn independently from a fixed, action-specific distribution. Each principal's pure strategy is a multi-armed bandit algorithm,
which dynamically adjusts to the observed rewards.
%\footnote{\label{fn:mixed}
%There is no distinction between pure and mixed strategies, because bandit algorithms can be randomized, and a distribution over bandit algorithms is also a bandit algorithm.}
However, it is oblivious to all signals on competition (such as the market share or the competitor's choices or rewards), even when such signals are available.
%\footnote{\gaedit{In particular, the algorithm does not respond to neither the competitors' rewards (which it does not observe) nor its current market share.}}
This modeling choice reflects the reality of industrial applications, which follow a huge body of knowledge in machine learning; more on this in Section~\ref{sec:discussion}.
Due to similar practical considerations, we expect the actual strategy choice to reflect the competition game only in a crude, qualitative way. Hence,  basic outcomes under competition are worth studying \emph{per se}, not only as a stepping stone to equilibrium characterization.
%Responding to the competition is another layer of complexity which has not been previously studied in this context, let alone made practical.

To flesh out the meaning of \emph{better} exploration algorithms, as per the main question, we draw on the literature from machine learning.  We consider algorithm's performance \emph{in isolation}: in a standalone exploration problem without competition. Rewards are not discounted with time, and we focus on big, qualitative differences in asymptotic regret rates.
%We draw on the machine learning literature to compare bandit algorithms to one another in isolation, so that we can talk about \emph{better} bandit algorithms in a principled way.\footnote{\gaedit{This literature typically compares the performance of different algorithms in a stand-alone exploration problem according to their asymptotic regret, which can be interpreted as maximizing consumer welfare in our context (see Appendix~\ref{app:bg} for self-contained background). Thus, we can utilize this comparison measure to assess the quality of the algorithms adopted under competition.}}
%\footnote{Such comparisons are somewhat subtle, as some algorithms may be better for some problem instances and/or time intervals, and worse for some others. More on this in Appendix~\ref{app:bg}. Generally,  ``better" algorithms are better in the long run, but could be worse initially. }
One baseline is algorithms that do not purposefully explore, and instead make myopically optimal decisions; we call them \emph{greedy algorithms}. In isolation, they are known to perform poorly for a wide variety of problem instances.

The agent-side model is as follows. When an agent arrives, she forms a reward estimate for each principal, and then chooses a principal using these reward estimates according to some fixed decision rule. Modeling the reward estimates is subtle, as one needs to specify how the  agents know and interpret the principals’ algorithms and the algorithms' past performance. We consider two extremes for this issue: Bayesian agents that know the algorithms but do not observe the past performance, and frequentist agents that observe each algorithm’s recent average performance (“reputation score”) but have no prior knowledge or beliefs on the algorithms. The Bayesian model follows a standard rational “template” in theoretical economics, whereas the frequentist model is more realistic. We find that the former is amenable to theoretical analysis, and the latter to simulations. Our main findings are similar for both.

Throughout our results, we investigate two channels for weakening the competition: relaxing the rationality of users (via their decision rule) and giving one firm a first-mover advantage.

\xhdr{Theoretical results in a Bayesian model.}
%We endow agents with Bayesian rationality, a common modeling approach for a theoretical investigation.
We consider a Bayesian model (called the \emph{\TheoryModel}), where agents have a common Bayesian prior on reward distributions, know the principals' algorithms and their own arrival times, but do not observe the previous agents' choices or rewards. Each agent computes  Bayesian-expected rewards for both principals, and uses them as reward estimates to decide which principal to choose.
%We focus on qualitative, asymptotic differences in the algorithms' performance.
Our results depend crucially on agents' decision rule:
%\begin{itemize}

\textbf{(i)} The most obvious decision rule maximizes the reward estimate; we call it \HardMax. We find that it is not conducive to adopting better algorithms: each principal's dominant strategy is to choose the greedy algorithm. Further,
%\HardMax is very sensitive to tie-breaking:
if the tie-breaking is probabilistically biased in favor of one principal, the latter
%then this principal
can always prevail in competition.
% has a simple ``winning strategy" no matter what the other principal does.

\textbf{(ii)} We dilute the \HardMax agents with a small fraction of ``random agents" who choose a principal uniformly at random.
(They can be interpreted as consumers that are oblivious to the principals' reputation.) We call this model \HardMaxRandom. Then better algorithms help in a big way: when two algorithms compete against one another, a sufficiently better algorithm is guaranteed to win all non-random agents after an initial learning phase. There is a caveat, however: any algorithm can be defeated by interleaving it with the greedy algorithm. Consequently,
%This caveat has two undesirable consequences:
a better algorithm may sometimes lose in competition, and a pure Nash equilibrium typically does not exist.

\textbf{(iii)} We further soften the decision rule so that the selection probabilities vary smoothly in terms of the reward estimates.
%as a function of the difference between  principals' Bayesian-expected rewards;
We call it \SoftMaxRandom, a more realistic middle ground between \HardMax and random agents.%
%\footnote{Alternatively, one can obtain a \SoftMaxRandom decision rule using a mixture of more ``basic" agent types that follow \HardMax unless the principal's Bayesian-expected rewards are too close to each other.}
In the most technical result of the paper, we find that a sufficiently better algorithm prevails under much weaker assumptions.

These findings are consistent with Schumpeter's inverted-U relationship between competition and innovation, whereby too little or too much competition is bad for innovation, but intermediate levels of competition tend to be better
\citep{Schumpeter-42,aghion2005competition,Vives-08}.
We interpret innovation as the adoption of better exploration algorithms,%
\footnote{Adoption of exploration algorithms tends to require substantial R\&D effort in practice, even if the algorithms are well-known and/or similar technologies already exist elsewhere \citep[\eg see][]{DS-arxiv}.}
and control the severity of the competition by varying the agents' decision rule from \HardMax (cut-throat competition) to \HardMaxRandom to \SoftMaxRandom and all the way to the uniform selection. See Figure~\ref{fig:inverted-U2} for a stylized representation. 
%\footnote{Agents' decision rule also controls the agents' rationality. While agents' rationality and severity of competition are often modeled separately, it is not unusual to have them modeled with the same ``knob" \cite[\eg][]{Gabaix-16}.}
Another, technically different inverted-U relationship (detailed in Section~\ref{sec:theory-welfare}) zeroes in on the \HardMaxRandom model.

\begin{figure}[t]
\begin{center}
\begin{tikzpicture}[scale=1]
      \draw[->] (-.5,0) -- (9.5,0) node[above]
        {\qquad\qquad Competitiveness};
      \draw[->] (0,-.5) -- (0,3) node[above] {Better algorithm in equilibrium};
      \draw[scale=0.8,domain=0.5:9.5,smooth,variable=\x,blue, line width=0.3mm] plot ({\x},{3.5 - 0.15*(\x - 5)^2});
     \node[below] at (1, 0) {\footnotesize \Uniform};
     \node[below] at (3.9, 0) {\footnotesize \SoftMaxRandom};
     \node[below] at (6, -.5) {\footnotesize \HardMaxRandom};
     \node[below] at (8, 0) {\footnotesize \HardMax};
      % \draw[scale=0.5,domain=-3:3,smooth,variable=\y,red]  plot ({\y*\y},{\y});
 \end{tikzpicture}

\caption{The stylized inverted-U relationship.}
\label{fig:inverted-U2}
\end{center}
\end{figure}

\xhdr{Numerical simulations in a frequentist model.}
We then consider a frequentist model (called the \emph{\ExptsModel}), where agents observe signals about the principals' past performance and make their decisions naively, without invoking any prior knowledge or beliefs. The performance signals are aggregated as a scalar \emph{reputation score} for each principal, modeled as a sliding window average of its rewards. Thus, agents' decision rule depends only on the two reputation scores. While this model uses more realistic assumptions about agent choices and allows us to characterize outcomes in finite samples, we provide numerical simulations to characterize the outcomes of interest. This allows us to refine and expand the results from the Bayesian model in several ways:

%\begin{itemize}
\textbf{(i)}
%We compare \HardMax and \HardMaxRandom decision rules.
We find that the greedy algorithm often wins under the \HardMax decision rule, with a strong evidence of the ``death spiral" effect mentioned earlier. As predicted by the theory, better algorithms prevail under \HardMaxRandom with enough ``random" users.
%if the expected number of ``random" users is sufficiently large.
 %\asdelete{However, this effect is negligible for smaller parameter values.}

%\footnote{\asedit{Reputation scores already introduce some noise into users' choices. However, the amount of noise due to this channel is typically small, both in our simulations and in practice, because reputation signals average over many datapoints.}}

\textbf{(ii)} Focusing on \HardMax, we investigate the first-mover advantage as a different channel to vary the intensity of competition: from the first-mover to simultaneous entry to late-arriver. We find that the first-mover is incentivized to choose a more advanced exploration algorithm, whereas the late-arriver is often incentivized to choose the ``greedy algorithm" (more so than under simultaneous entry). Consumer welfare is higher under early/late arrival than under simultaneous entry. We frame these results in terms of an inverted-U relationship.

%\footnote{\asedit{We consider the ``permanent monopoly" scenario for comparison only, without presenting any findings. We just assume that a monopolist chooses the greedy algorithm, because it is easier to deploy in practice. Implicitly, users have no ``outside option": the service provided is an improvement over not having it (and therefore the monopolist is not incentivized to deploy better learning algorithms). This is plausible with free ad-supported platforms such as Yelp or Google.}}

\textbf{(iii)}
However, the greedy algorithm is sometimes \emph{not} the best strategy under high levels of competition.\footnote{In our theoretical results on \HardMax, the greedy algorithm is always the best strategy, mainly because it is aware of the Bayesian prior (whereas in the simulations  the prior is not available).}
We revisit algorithms' performance in a standalone bandit problem, \ie without competition. We find that the most natural performance measure does not explain this phenomenon, and suggest a new, more nuanced one that does.
%performance measure that does.

%\textbf{(iii)} We investigate the algorithms' performance without competition. We suggest a new performance measure to explain why the greedy algorithm is sometimes not the best strategy under high levels of competition.\footnote{In our theoretical results on \HardMax, the greedy algorithm is always the best strategy, mainly because it is aware of the Bayesian prior (whereas in the simulations  the prior is not available).} We find that mean reputation -- arguably, the most natural performance measure -- is sometimes \emph{not} a good predictor for the outcomes under competition.

\textbf{(iv)} We decompose the first-mover advantage into two distinct effects: free data to learn from (\emph{data advantage}), and a more definite, and possibly better reputation compared to an entrant (\emph{reputation advantage}), and run additional experiments to separate and compare them. We find that either effect alone leads to a significant advantage under competition. The data advantage is larger than reputation advantage when the incumbent commits to a more advanced bandit algorithm. Finally, we find an ``amplification effect" of the data advantage: even a small amount thereof gets amplified under competition, causing a large difference in eventual market shares.

% AS: moved the below to "related work"
%While traditional models of innovation study lab-based R\&D, we consider data-driven innovation, which crucially depends on data generated by the firm's customers. We focus on innovation in \emph{exploration technology} which systematically improves the firm's products, whereas prior work would define innovation as improvement in the products themselves. We recover the inverted-U relationship purely through the reputational consequences of exploration. By contrast, the inverted-U relationships from prior work rely on the monetary aspects: investments into  R\&D and profits from innovation.

\xhdr{Further economic interpretations.}
Our model speaks to policy discussions on regulating data-intensive digital platforms \citep{furman2019unlocking, scott2019committee}, and particularly to the ongoing debate on the role of data in the digital economy. One fundamental question in this debate is whether data can serve a similar role as traditional ``network effects", creating scenarios when only one firm can function in the market \citep{Rysman09, jullien2019economics}.
%whereby, when these effects are present, in many cases only one firm can function in the market, leading to competition \emph{for} the market being more important than competition \emph{in} the market \citep{Rysman09, jullien2019economics}.
The death spiral/amplification effects mentioned above have a similar flavor: a relatively small performance loss due to exploration (resp., data advantage)  gets amplified under competition and causes the firm to be starved of users (resp., take over most of the market).
%\gaedit{We further find that a small data advantage for one firm gets amplified under competition and leads to that firm taking the entire market, showing that data can provide a similar incumbency advantage as those provided by traditional network effects and can serve as a barrier to entry in online markets.}
However, a distinctive feature of our approach is that the network effects arise endogenously.
%\gadelete{However, a distinctive feature of our approach is that we explicitly model the learning problem of the firms and consider them deploying algorithms for solving this problem.  Thus, we do not explicitly model the network effects, but they arise endogenously from our setup.}

Our results highlight that understanding the performance of learning algorithms in isolation does not necessarily translate to understanding their impact in competition, precisely due to the fact that competition leads to the endogenous generation of observable data. Approaches such as \citet{lambrecht2015can, bajari2018impact, varian2018artificial} argue that the diminishing returns to scale and scope of data in isolation mitigate such data feedback loops,
%as non-existent
but ignore the differences induced by learning in isolation versus under competition. Explicitly modeling the interaction between learning technology and data creation allows us to speak on how data advantages are characterized and amplified by the increased \emph{quality} of data gathered by better learning algorithms, not just the quantity thereof. In particular, we find that incumbency is good for innovation and welfare, \emph{and} % but also
creates a barrier to entry,
all % precisely
due to data feedback loops.
%\gadelete{Interestingly, data feedback loops can be bad for innovation (as in ``death spiral"), or good for innovation (as in ``amplification effect"), depending on the level of competition.}

%Interestingly, we find that data feedback loops can be bad for innovation (as in ``death spiral"), or good for innovation (as in ``amplification effect"), depending on the level of competition.

\xhdr{Significance.}
Our results have a dual purpose: shed light on real-world implications of some typical scenarios, and investigate the space of models for describing the real world. As an example for the latter: while the \HardMax model with simultaneous entry is arguably the most natural model to study \emph{a priori}, our results elucidate the need for more refined models with ``free exploration" (\eg via random agents or early entry). On a technical level, we connect a literature on regret-minimizing bandits in machine learning and that on competition in economics.

The two technical parts of the paper, Bayesian/theoretical and frequentist/experimental, are on equal footing. While one does not provide direct experimental (resp., theoretical) justification for the other, they yield consistent conclusions, and present two complementary but different approaches to attack the same problem. Our theory takes a Bayesian perspective, standard in economic theory, and discovers several strong asymptotic results. Much of the difficulty, both conceptual and technical, is in setting up the model and the theorems. In particular, it was crucial to interpret the results and intuitions from the literature on multi-armed bandits so as to formulate meaningful and productive assumptions on bandit algorithms and Bayesian priors. The numerical simulations for the frequentist model provide a more nuanced and ``non-asymptotic" perspective. In essence, we look for substantial effects within relevant time scales. (In fact, we start our investigation by determining what time scales are relevant in the context of our model.) The central challenge is to capture a huge variety of bandit algorithms and bandit problem instances with only a few representative examples, and arrive at findings that are consistent across the entire space.

The Bayesian model is suitable for analysis and the frequentist model for simulations, \emph{but not vice versa}. A natural implementation of the Bayesian model requires running time quadratic in the number of rounds,%
\footnote{\label{fn:Tsquared}\Eg this is because at each round $t$, one needs to recompute, and integrate over, a discrete distribution with $t$ possible values, namely the number of agents that have chosen principal $1$ so far.}
which precludes numerical simulations at a sufficient scale. The frequentist model features an intricate feedback loop between algorithms' performance, their reputations and agents' choices, which simplifies the simulations but does not appear analytically tractable.

%% file: content/related_work.tex
\xhdr{Exploration.} Multi-armed bandits (\emph{MAB}) is an elegant and tractable abstraction for tradeoff between \emph{exploration} and \emph{exploitation}: essentially, between acquisition and usage of information. MAB problems have been studied for many decades by researchers from computer science, operations research, statistics and economics, generating a vast and multi-threaded literature.  The most relevant thread concerns the basic model of regret-minimizating bandits with stochastic rewards and no auxiliary structure (which is the problem faced by each principal in our model), see Appendix~\ref{app:bg} for background. This basic model has been extended in many different directions, with a considerable amount of work on each: \eg payoffs with a specific structure (\eg combinatorial, linear, convex or Lipschitz), payoff distributions that change over time, and auxiliary payoff-relevant signals.
%\footnote{There is a superficial similarity, in name only, between this paper and the work on ``dueling bandits" \citep[starting from][]{Yue-dueling12,Yue-dueling-icml09}. In this work, there is only one bandit algorithm which chooses two arms in each round, and the only observes which arm has ``won the duel".}
Dedicated monographs \citep{Bubeck-survey12,slivkins-MABbook,LS19bandit-book} cover the work on regret-minimizing formulations (which mainly comes from computer science). The classic book
\citep{Gittins-book11} focuses on the Markovian formulations, which predate regret-minimization. Connections to economics are detailed in
books \citep{CesaBL-book,slivkins-MABbook} and surveys \citep{Bergemann-survey06,Horner-survey16}. Industrial applications are discussed in \citep{DS-arxiv}.

A monopolistic bandit algorithm may interact with self-interested parties, leading to a tension between exploration and incentives. This tension has been studied in several scenarios:
%\asedit{The tension between exploration and incentives}
%The three-way tradeoff between exploration, exploitation and incentives
%has been studied in several scenarios very different from ours:
incentivized exploration in recommendation systems
(starting from \citet{Kremer-JPE14,Che-13}, see \citet[Ch. 11]{slivkins-MABbook}),
    %\citep[\eg][]{Che-13,Frazier-ec14,Kremer-JPE14,ICexploration-ec15,Bimpikis-exploration-ms17,Bahar-ec16,Jieming-unbiased18},
dynamic auctions
    \citep{DynAuctions-survey10},
    %\citep[\eg][]{AtheySegal-econometrica13,DynPivot-econometrica10,Kakade-pivot-or13},
pay-per-click ad auctions
    \citep[\eg][]{MechMAB-ec09,DevanurK09},
coordinating search and matching
    \citep{Bobby-Glen-ec16},
and human computation
    \citep[\eg][]{RepeatedPA-ec14,Ghosh-itcs13}.
Unlike this work, we focus on incentives created in a competition.
%A literature review of this work can be found in
%\citep[Ch. 11.6]{slivkins-MABbook}.

\xhdr{Exploration and competition.}
Several papers consider exploration algorithms in scenarios when the explorer is not a monopolist. The technical models are very different, and not amenable to the high-level question articulated in the Introduction.

\citet{bergemann1997market,bergemann2000experimentation} and \citet{keller2003price} study the interplay of exploration and competition for users when the competing firms experiment with \emph{prices} (whereas in our model the firms experiment with design alternatives). All three papers consider environments with fixed product quality and dynamic strategies that respond to competition, and analyze Markov-perfect equilibria. In contrast, we consider a one-shot game where firms commit to algorithms for their bandit problem and the goal is to learn the best product alternative. This results in the nature of exploration being fundamentally different relative to these papers and, as such, we focus on different outcomes of interest relative to these papers.

In the line of work on \emph{strategic experimentation} (starting from \citet{Bolton-econometrica99,Keller-econometrica05}, see \citet{Horner-survey16} for a survey), agents explore and learn over time in a shared environment. Thus, we have exploration algorithms which interact with each other strategically, \eg each agent prefers to free-ride on someone else's exploration. However, this work is all about cooperation (or lack thereof), rather than competition.
%the agents do not compete with each other in any meaningful sense.

Several papers study competition between two principals who run algorithms but do not interact, directly or indirectly, until the very end of the game. \cite{Ufuk-jeea15} consider a ``research competition" between two firms racing towards a big discovery.
Each firm deploys a bandit algorithm with two arms, corresponding to safe and risky lines of research. The firms do not interact until one of them makes the discovery and wins the game. In the ``dueling algorithms" framework of \citet{DuelingAlgs-stoc11}, each principals runs an algorithm for the same problem. All inputs are observable at once, and principals' payoffs are binary (win/lose). \citet{ben2019regression} study competition between ``offline" machine learning algorithms. In comparison, we study a ``product competition" in which the two firms interact continuously (via the customers' choices), accrue rewards incrementally, and compete for individual customers.

A long line of work from electrical engineering and computer science, starting from
\citet{MultiPlayerMAB-Poor08,MultiPlayerMAB-Liu10}, focuses on competition for resources, not competition for consumers. Specifically, this literature targets an application to \emph{cognitive radios}, where multiple radios transmit simultaneously in a shared medium and compete for bandwidth. Each radio chooses channels over time using a multi-armed bandit algorithm. This work studies a repeated game between bandit algorithms, and focuses on designing algorithms which work well in this game.
%\gadelete{, under various assumptions on communication, synchronization and collisions}.

%and continuing to, \eg \citet{MultiPlayerMAB-Mannor14,MultiPlayerMAB-Shamir-icml16,MultiPlayerMAB-Perchet-18,MultiPlayerMAB-Sellke-19}.

\xhdr{Competition.} The competition vs. innovation relationship and the inverted-U shape thereof have been introduced in a classic book \citep{Schumpeter-42}, and remained an important theme in the literature ever since \cite[\eg][]{aghion2005competition,Vives-08}. This literature treats innovation as R\&D that improves the products and, R\&D costs aside, is a priori beneficial for the firm. In contrast, we focus on innovation in \emph{exploration technology} which systematically improves the firm's products and crucially depends on data generated by the firm's customers. In particular, we find that such innovation may potentially hurt the firm. We recover the inverted-U relationship purely through the reputational consequences of exploration, whereas prior work relies on costs and profits.

% COPIED THIS PARA FROM THE INTRO, EDITED IT IN
%While traditional models of innovation study lab-based R\&D, we consider data-driven innovation, which crucially depends on data generated by the firm's customers. We focus on innovation in \emph{exploration technology} which systematically improves the firm's products, whereas prior work would define innovation as improvement in the products themselves. We recover the inverted-U relationship purely through the reputational consequences of exploration. By contrast, the inverted-U relationships from prior work rely on the monetary aspects: investments into  R\&D and profits from innovation.

The literature on learning-by-doing vs. competition \citep[\eg][]{fudenberg1983learning, dasgupta1988learning, cabral1994learning}
studies firms that learn while competing against each other, so that a firm attracting more consumers reduces its production costs. Our model differs in several important respects. First, firms learn to improve product quality rather than to reduce production costs. Second, the firms' current actions have consequences (via reputation and/or data collected by the algorithm) that directly impact consumer choices in the future. Third, we endogenize the technology behind learning-by-doing by explicitly considering bandit algorithms.

%\gaedit{However, our model differs from these models in two important respects. The first is that in our model firms focus on learning product quality (as opposed to reducing production costs) and the second is that firms' current period actions have reputational consequences that directly impact consumer choices in future periods. These features are absent from traditional learning-by-doing models, but important components of a model that tries to capture data-based innovation.}

A line of work on \emph{platform competition} (starting with \cite{Rysman09}, see \citet{Weyl-White-14} for a survey) concerns competition between firms that improve as they attract more users. This literature is not concerned with \innovation, and typically models network effects exogenously, whereas they are endogenous in our model.
%: they are created by MAB algorithms, an essential part of the model.
A nascent literature studies
%whether and when network effects manifest themselves
network effects
in data-intensive markets \citep{prufer2017competing, hagiu2020data}, but typically models learning as a reduced-form function of past consumer history and focuses on the role of prices.
%as opposed to the reputational consequences of learning.

\cite{schmalensee1982product, bagwell1990informational} investigate how buyer uncertainty about product quality can serve as a barrier to entry for late arrivers; we find a similar effect with ``reputation advantage". \cite{de2020data} note the role of data as a barrier to entry in online markets; we find a similar effect with ``data advantage". \citet{kerin1992first} overview other channels through which first-mover advantage can affect competition.% lead to a competitive advantage.

%\cite{schmalensee1982product, bagwell1990informational} investigate how buyer uncertainty about product quality can serve as a barrier to entry for late arrivers. We observe a similar effect when we investigate the role that reputation can serve as a barrier to entry. In our model the first-mover advantage further provides the incumbent with a ``head start" on data collection relative to the late arrivers. Thus, our model also highlights the role that data can serve as a barrier to entry in online markets which has similarly been noted in \cite{de2020data}. For an extensive overview of the other channels through which first-mover advantages can lead to a competitive advantage, see \cite{kerin1992first}.

%We use first-mover advantage and \asedit{agents' decision rule}
%relaxed versions of rationality
%to model varying competition, instead of

While we use first-mover advantage and agents' decision rule, classic ``market competitiveness" measures, such as the Lerner Index or the Herfindahl-Hirschman Index
\citep{tirole1988theory}, are not applicable to our setting, as they rely on ex-post observable market attributes such as prices or market shares (which are, resp., absent and endogenous for us).

%Neither is applicable to our setting (since there are no prices, and market shares are endogenous).

\xhdr{Choice models.}
Stochastic choice models similar to ours are widely used in economics. ``Random agents" (a.k.a. noise traders) can side-step the ``no-trade theorem'' \citep{Milgrom-Stokey-82}, a famous impossibility result in financial economics. They play a similar role in our model, side-stepping the dominance of the greedy algorithm.
%as we move from \HardMax to \HardMaxRandom.
%\gadelete{Second, there a large literature on non-existence of equilibria due to small deviations, starting with \cite{Rothschild-Stiglitz-76} in the context of health insurance markets. \footnote{\cite{Veiga-Weyl-16,Azevedo-Gottlieb-17} emphasize the distinction between \HardMax and versions of \SoftMaxRandom in this context.} This is superficially similar to how small deviations towards the greedy algorithm rule out equilibria under \HardMaxRandom.}
Moreover, \SoftMaxRandom subsumes the logit choice rule, a standard behavioral model with strong empirical and microeconomic foundations
\citep[\eg][]{mosteller1951experimental, luce1959choice, matvejka2015rational}.
Choice models similar to \SoftMaxRandom are used to explain  horizontal product differentiation \citep[\eg][]{Hotelling-29, Perloff-Salop-85}.
%\asdelete{While agents' rationality and severity of competition are often modeled separately in the literature, it is not unusual to have them modeled with the same ``knob" \cite[\eg][]{Gabaix-16}.}

%%% Local Variables:
%%% TeX-master: "main.tex"
%%% End: 

%% file: content/sec-model.tex
\xhdr{Principals and agents.} There are two principals and $T$ agents. We denote them, resp., principal $i\in \{1,2\}$ and agent $t\in [T]$,
where $[T] := \{1,2\LDOTS T\}$.
%The game proceeds in $T$ rounds.

In each round $t\in [T]$, the following  interaction takes place. Agent $t$ arrives and chooses a principal $i_t\in \{1,2\}$. The principal chooses action $a_t\in A$, where $A$ is a fixed set of actions.%
\footnote{We use `action' and `arm' interchangeably, as common in the literature on multi-armed bandits.}
The agent experiences this action and receives an associated reward $r_t\in \{ 0,1\}$, which is then observed by the principal. We posit \emph{stochastic rewards}: whenever a given action $a\in A$ is chosen, the reward is an independent draw from Bernoulli distribution with mean $\mu_a$. In particular, the mean rewards $\mu_a$, as well as the action set $A$, are the same for both principals and all rounds. The mean rewards are initially not known to anybody. The principals are completely unaware of the rounds when the opponent is chosen. Thus, each principal follows the protocol of \emph{multi-armed bandits} (henceforth, \emph{MAB}). That is: in each round when it is chosen, the principal picks an action from $A$ and observes a reward for this action (and nothing else).

Each principal $i$ commits to an MAB algorithm \alg[i] before round $1$, and uses this algorithm throughout the game. The algorithm proceeds in time-steps:
%\footnote{These time-steps will sometimes be referred to as \emph{local steps/rounds}, so as to distinguish them from ``global rounds" defined before. We will omit the global vs. local distinction when clear from the context.}
each time it is called, it outputs an arm from $A$, and inputs a reward for this action. The algorithm is called only in game rounds when principal $i$ is chosen.
When the distinction between algorithm's time-steps and game rounds is unclear from the context, we will refer to them as, resp., \emph{local steps/rounds} and \emph{global rounds}.

\newcommand{\est}{\mathtt{EST}}

%\gaedit{\footnote{While our model assumes that agents have correct beliefs about the performance of the algorithms, only the beliefs themselves matter and so our results hold as long as all agents have the same initial beliefs, even if they are not aligned with the true prior.}}

\xhdr{Agent response.} Each agent $t$ forms a reward estimate $\est_i(t)\in [0,1]$ for each principal $i$. (What these estimates are, and how much the agents know in order to form them, depends on the Bayesian vs. frequentist model variant.)
The reward estimates determine the choice of the principal. Specifically, agent $t$ chooses principal $1$ with probability
\begin{align}\label{eq:model-f}
p_t = \respF\rbr{ \est_1(t) - \est_2(t) },
\end{align}
where $\respF:[-1,1]\to [0,1]$ is the \emph{response function}, same for all agents. We assume that $\respF$ is monotonically non-decreasing, is larger than $\nicefrac12$ on the interval $(0,1]$, and smaller than $\nicefrac12$ on the interval $[-1,0)$. We consider three variants for $\respF$, depicted in  \reffig{fig:response-functions}:

\begin{figure}[t]
\begin{center}
  \begin{tikzpicture}[scale=2.0]
    \draw[->] (-1.1,0) -- (1.1,0) node[right]
    {$\Delta_t := \est_1(t) - \est_2(t)$};
    \draw[->] (0,-0.1) -- (0,1.1) node[above]
        {$p_t = \text{prob. of choosing principal 1}$};
    \draw[scale=1.0,domain=-1:0,smooth,variable=\q,blue, line width=0.50mm] plot ({\q},{0});
    \draw[scale=1.0,domain=0:1,smooth,variable=\q,blue,line width=0.50mm] plot ({\q},{1});
    \draw[scale=1.0,domain=-1:0,smooth,variable=\y,red]  plot ({\y},{0.1});
    \draw[scale=1.0,domain=0:1,smooth,variable=\y,red]  plot ({\y},{0.9});
    \draw[scale=1.0,domain=-1:1,smooth,variable=\y,cyan, line width=0.45mm, dash pattern=on 3pt off 2pt]  plot ({\y},{1/(1 + 1/(9^\y))});
    % \node[above, blue] at (0.5, 0.5) {\footnotesize $2 q (1 - q)$};
     \node[left] at (0, 0.5) {\footnotesize $1/2$};
     \node[left] at (0, 1) {\footnotesize $1$};
     \node[below left] at (0, 0) {\footnotesize $0$};
     \node[below ] at (1, 0) {\footnotesize $1$};
     \node[below ] at (-1, 0) {\footnotesize $-1$};
  \end{tikzpicture}
\end{center}
\vspace{-5mm}
\caption{The models for $\respF$: \HardMax is thick blue, \HardMaxRandom is red, and \SoftMaxRandom is dashed.}
\label{fig:response-functions}
\end{figure}

\begin{itemize}
\item \HardMax: $\respF$ equals $0$ on the interval $[-1,0)$ and $1$
  on the interval $(0,1]$. In words, a \HardMax agent
  deterministically chooses a principal with a higher reward estimate.

\item \HardMaxRandom:
    % $\respF$ equals $\eps$ on the interval $[-1,0)$ and $1-\eps'$ on the interval $(0,1]$, where $\eps,\eps'\in (0,\tfrac12)$ are some positive constants. In words, each agent is a \HardMax agent with probability $1-\eps-\eps'$, and with the remaining probability she makes a random choice.
    $\respF$ equals $\eps_0$ on the interval $[-1,0)$ and $1-\eps_0$ on the interval $(0,1]$, for some constant $\eps_0\in (0,\tfrac12)$. In words, each agent is a \HardMax agent with probability $1-2\eps_0$, and makes a random choice otherwise.

\item \SoftMaxRandom: $\respF$  lies in $[\eps_0,1-\eps_0]$, breaks ties fairly, and has a bounded derivative around $0$ (see Definition~\ref{def:SoftMax} for a formal definition). Intuitively, $\respF$ is a smoothed version of \HardMaxRandom, without a sharp threshold therein.
\end{itemize}

\HardMaxRandom and \SoftMaxRandom agents can be interpreted in several ways. First, they make mistakes, due to lack of awareness or interest. Second, they give some chance to the non-preferred principal, due to curiosity or a behavioral effect like probability matching.
%\gaedit{make mistakes due to being oblivious to the principals' reputation, due to lack of awareness or interest. \HardMaxRandom models users as either being aware of the reputation or not, whereas \SoftMaxRandom models user mistakes as following a smooth density.}
Third, they can be realized as a distribution over more ``basic" agent types. Indeed, the \HardMaxRandom distribution is a mixture of \HardMax and ``random agents" (which choose a principal uniformly at random). The latter can be interpreted as consumers that are completely oblivious to principals' reputation.
%, due to the lack of awareness or interest.}
One can obtain a \SoftMaxRandom response function using agent types that choose a principal $i$ with a largest reward estimate $\est_i$, unless $|\est_1-\est_2|$ is upper-bounded by some parameter $\theta$, in which case they choose uniformly. Then, we obtain \SoftMaxRandom as a mixture of random agents and these ``$\theta$-\HardMax'' agents, for a suitable distribution over $\theta$.

% \asedit{(We assume $\respF(-1)+\respF(1)=1$ in \HardMaxRandom and \SoftMaxRandom only to simplify notation.)}

%We say that $\respF$ is \emph{symmetric} if $\respF(-x)+\respF(x)=1$ for any $x\in [0,1]$. This implies \emph{fair tie-breaking}: $\respF(0)=\tfrac12$.

\xhdr{Bayesian vs. frequentist variants.}
We consider two model variants, Bayesian and frequentist (we use them, resp., for theoretical results and numerical simulations). The main difference between the two concerns the agents' reward estimates $\est_i(t)$.

%\begin{OneLiners}
%\item
In the \emph{\TheoryModel}, the mean reward vector $\mu = (\mu_a:\; a\in A)$ is drawn from a common Bayesian prior $\priorMu$. Each agent knows its global round $t$, but not the performance signals such as the current market shares. Her reward estimates are defined as posterior mean rewards:
        $\est_i(t) = \PMR_i(t) := \E\sbr{ r_t\mid i_t = i}$
    for each principal $i$, where the agent knows $t$, the principals' algorithms, $\priorMu$, and $\respF$.
%\footnote{Agents do the Bayesian update knowing $t$, the principals' algorithms, the prior $\priorMu$, and $\respF$.}

%Agents know the principals' algorithms.\gaedit{\footnote{While our model assumes that agents have correct beliefs about the performance of the algorithms, only the beliefs themselves matter and so our results hold as long as all agents have the same initial beliefs, even if they are not aligned with the true prior.}}

%\item
In the \emph{\ExptsModel}, agents' reward estimate for a given principal is the average reward of the last $M$ agents that chose this principal. We call it \emph{reputation score}, and interpret it as the current reputation. To make it meaningful initially, each principal enjoys a ``warm start": additional $T_0$ agents arrive before the game starts, and interact with the principal as described above.

%\end{OneLiners}

\xhdr{Competition game.}
Some of our results explicitly study the game between the two principals, termed the \emph{competition game}. We model it as a simultaneous-move game: before the first agent arrives, each principal commits to an MAB algorithm. Principals are risk-neutral; their utility is defined as their market share, \ie the number of agents they attract. Thus, they aim to select the algorithm that maximizes their expected market share.%
\footnote{The immediate goal of principals' MAB algorithms is (still) to maximize agents' rewards, so as to attract agents. Besides, it is unclear how to maximize market share directly within our model. Note that in extensions (Section~\ref{sec:theory-extensions}) the principals' utility can also depend on rewards.}

%\footnote{\gaedit{We are implicitly assuming that the principals and agents have completely aligned utility. This could be violated if, for instance, the principal has a preference that agents consume some type of content over others or if agents have time inconsistent preferences \citep{kleinberg2024challenge}. Our results can be extended to handling some cases of the former type of misalignment (i.e., by having the principals' utility depend on the rewards), but the latter type of misalignment is an interesting direction for future work.}}

The distinction between a pure and mixed strategy in this game is worth clarifying. For each principal, a pure strategy commits to a particular MAB algorithm. A mixed strategy chooses a particular MAB algorithm at random from a mixture, and then commits to this algorithm. While a mixed strategy induces a randomized MAB algorithm, it differs from a pure strategy with the same algorithm in that the realization of the mixture is revealed to the agents.

%Principal's utility is defined as the market share, \ie the number of agents that chose this principal. Principals are risk-neutral and aim to maximize their expected utility.

\xhdr{Extensions.} The \TheoryModel admits several extensions, detailed in Section~\ref{sec:theory-extensions}. First, all/most results extend to arbitrary reward distributions, allow reward-dependent utility, and carry over to a more general version of multi-armed bandits. Second, agents could have beliefs on $(\alg[1],\alg[2],\priorMu,\respF)$ that need not be correct; then, all results carry over with respect to these beliefs.
%Third, all results carry over \asmargincomment{check}  if each agent $t$ has a prior on her arrival time $t$, rather than know it exactly.
Third, we can handle a limited amount of non-stationarity in $\respF$ for the \HardMaxRandom and \SoftMaxRandom decision rules. Finally, the main result on \HardMax extends to time-discounted utilities.

\xhdr{``Non-strategic" exploration strategies.} We focus on a realistic scenario when the exploration strategies available to the principals are ``non-strategic" in nature. Even though the principals play a multi-step game, they do not react to each other's moves or to the agents' strategic choices. This is how industry approaches exploration algorithms, for several reasons.
First, "non-strategic" exploration is well-studied in machine learning, and yet it remains a very complex and actively studied subject in research. Even the seemingly simple algorithms are not straightforward to deploy in practice, and require a substantial investment in infrastructure
(\eg see the discussions in \citet{DS-arxiv} and \citet[Chapter 8.7]{slivkins-MABbook}).
Responding to the competition represents another layer of complexity which has not been previously studied in this context, to the best of our knowledge, let alone made even remotely practical.
Second, while the principals could potentially react to the market share or the reputation scores, baking these signals into one's exploration strategy runs the risk of over-interpreting our competition model, as they may change for exogenous reasons. Alternatively, one could use such signals, as well as the intuitions coming from this paper, to guide the platform's decisions regarding exploration.

%\gaedit{Finally, the assumption that firms commit to the algorithms at the start of the game deserves further justification. We impose this commitment assumption since it is more realistic than considering dynamic strategies and algorithms are, by definition, a commitment to a set of rules. In many contexts where such exploration algorithms are typically deployed, consumers arrive at such a rapid volume that it it is infeasible to consider a firm dynamically adapting its algorithm to each consumer that arrives. Indeed, \cite{DS-arxiv} points out how difficult it is to deploy standard exploration algorithms in practice and such dynamic strategies are substantially more difficult to deploy in practice. While it is possible to consider share-adaptive algorithms that would still keep the commitment assumption but allow for indirect dynamic responses to market conditions, such algorithms do not exist currently in the literature but are a good avenue for future work.}

\xhdr{Bandit algorithms.}
Our treatment of bandit algorithms is very standard in machine learning –- the primary community where these algorithms are designed and studied –- but perhaps less standard in economic theory. The main tenets are as follows.

\begin{itemize}

\item[(i)] Algorithms are designed for vanishing regret without time-discounting, and compared theoretically based on their asymptotic regret rates (rather than Bayesian-optimal time-discounted reward, as in Gittins index, a more standard economic perspective). Indeed, non-discounted, regret-minimizing formulations has been prevalent in the bandits literature over the past two decades \citep{slivkins-MABbook,LS19bandit-book}, and better correspond to practical deployments \citep[\eg][]{DS-arxiv}.

\item[(ii)]
A key distinction is between no exploration (the “greedy” algorithm), fixing the exploration schedule in advance (“exploration-separating” algorithms, \eg the epsilon-greedy algorithm), and adapting exploration to the past observations (\eg Thompson Sampling). Accordingly, there’s a stark 3-way distinction in asymptotic regret rates. In particular, our numerical results use a standard, representative algorithm from each of the three classes.

\end{itemize}

Self-contained background regarding the two tenets can be found in Appendix~\ref{app:bg}.

\subsection{Discussion: stylized model}
\label{sec:discussion}

%\asdelete{Now that we laid out our model, let us discuss several aspects thereof that are unclear from the bare definitions. In particular, we flesh out several points made in the Introduction.}

Our models are stylized in several important respects. Firms compete only on the quality of service, rather than, say, pricing or the range of products.  Agents are myopic: they do not worry about how their actions impact their future utility.\footnote{So, agents do not attempt to learn over time, game future agents, or manipulate the principals' learning algorithms. This is arguably typical in practice, in part because one agent's influence tends to be small.}
On the machine learning side, we focus on qualitative distinctions described above, rather than state-of-art algorithms for realistic applications. %\gaedit{Nonetheless, we additionally show that our results hold even if we relax some of the model assumptions as well as provide various extensions in Section~\ref{sec:theory-extensions}.}

For the \TheoryModel, agents do not observe any signals about the principals' past performance, making agents' behavior independent of a particular realization of the prior. This allows us to summarize each learning algorithm via its Bayesian-expected rewards, not worrying about its detailed performance on particular realizations of the prior.  Such summarization is essential for formulating lucid and general analytic results, let alone proving them.
%It is unclear how to incorporate performance signals in a theoretically tractable model.

For the \ExptsModel, various performance signals available to the users, from personal experience to word-of-mouth to consumer reports, are abstracted as persistent ``reputation scores" reflecting the current reputation, and further simplified to average performance over a sliding time window. The reputation scores directly account for competition, allowing the users to have no direct information on the algorithms deployed or the bandit problem faced by the firms. The latter property makes our model amenable to numerical simulations.
%and further allows us to separate reputation vs. data advantage.
%(Whereas simulating the intricate interplay of learning dynamics and Bayesian priors appears computationally intractable, see Footnote~\ref{fn:Tsquared}.)

%\gaedit{On the machine learning side, we consider a standard model of stochastic bandits. We are less interested in state-of-art algorithms for realistic applications. Rather, we utilize the intuition from the well-established literature on this bandit model that has a clear demarcation between the asymptotic performance of greedy, exploration-separating, and adaptive-exploration algorithms. In Appendix~\ref{app:bg}, we present sufficient background on this bandit model and the demarcation between these classes of algorithms, accessible to non-specialists. We focus on understanding from which of these classes of algorithms get chosen in the equilibrium of our game theoretic model.}

%\gadelete{On the machine learning side, our model captures big, qualitative differences between bandit algorithms, building on the well-established intuition in the literature. Comparisons between bandit algorithms are generally somewhat subtle, as some algorithms may be better for some problem instances and/or time intervals, and worse for some others. In particular, ``better" algorithms are better in the long run, but could be worse initially. We focus on a standard model of stochastic bandits.  However, we are less interested in state-of-art algorithms for realistic applications.}

%% file: content/sec-theory.tex
In this section, we present our theoretical results for the \TheoryModel. While we provide intuition and proof sketches, the detailed proofs are deferred to Appendix~\ref{sec:theory-proofs}. \asedit{A table summarizing the key notation and definitions can be found in Appendix~\ref{app:notation}.}

\subsection{Preliminaries}
\label{sec:theory-prelims}

%\xhdr{Notation.}
Let $\rew_i(n)$ denote the agent's realized reward observed by principal $i$ at local step $n$, \ie the reward collected by algorithm $\alg[i]$ in this local step. For a global round $t$, let $n_i(t)$ denote the number of global rounds before $t$ in which principal $i$ is chosen. We will use the fact that
    $ \PMR_i(t):=  \E\sbr{ r_t\mid i_t = i }
        = \E\sbr{ \rew_i(n_i(t)+1) }. $

\noindent\textbf{Assumptions.} We make two mild assumptions on the prior. First, each arm $a$ can be best:
\begin{align}\label{eq:assn-prob}
\forall a\in A:\;\;\;
\Pr\sbr{\mu_a  > \mu_{a'}
\quad \forall a'\in A\setminus \{a\}}
> 0.
\end{align}

\noindent Second, posterior mean rewards are pairwise distinct given any feasible history $h$:%
\begin{align}\label{eq:assn-distinct}
    \E[\mu_a \mid h] \neq \E[\mu_{a'}\mid h] \quad \forall a,a'\in A.
\end{align}
The \emph{history} of an MAB algorithm at a given step $t$
  comprises actions $a_s$ and rewards $r_s$ in all
  previous steps $s<t$. The history is \emph{feasible} if for each $s$, reward $r_s$ is in the support of the reward distribution for $a_s$. In particular, \emph{prior} mean rewards are pairwise distinct:
$\E[\mu_a] \neq \E[\mu_a']$ for any $a,a'\in A$.

In Appendix~\ref{app:perturb}, we provide two examples for which property \eqref{eq:assn-distinct} is `generic', in the sense that it can be enforced almost surely by a small random perturbation of the prior. The two examples concern, resp., Beta priors and priors with a finite support, and focus on priors $\priorMu$ that are independent across arms.

\xhdr{MAB algorithms.}
We consider two baseline algorithms. The main one, called  \DynGreedy, chooses an arm $a$ with the largest posterior mean reward $\E[\mu_a \mid \cdot ]$ given all information currently available to the algorithm. A more primitive baseline, called \StaticGreedy, chooses an arm $a$ with the largest prior mean reward $\E[\mu_a]$, and uses this arm in all rounds.

%We consider two (Bayesian) greedy algorithms. The first one, called  \DynGreedy, chooses an arm $a$ with the largest posterior mean reward $\E[\mu_a \mid \cdot ]$ given all information currently available to the algorithm. The second algorithm, called \StaticGreedy, chooses an arm $a$ with the largest prior mean reward $\E[\mu_a]$, and uses this arm in all rounds. \asdelete{In both algorithms, ties are broken arbitrarily.}

We characterize the inherent quality of an MAB algorithm in terms of its \emph{Bayesian Instantaneous Regret} (henceforth, \BIR), a standard notion from machine learning:
\begin{align}\label{eq:theory-BIR-defn}
\BIR_i(n) := \textstyle \E[\; \max_{a\in A} \mu_a - \rew_i(n) \;].
\end{align}
We are primarily interested in how fast \BIR decreases with $n$. (We treat the number of arms as a constant.) Intuitively, (much) better MAB algorithms tend to have a (much) smaller BIR, see Appendix~\ref{app:bg} for background.
An algorithm is called \emph{\bmonotone} if it can only get better over time, in the Bayesian sense: namely, if $\E[\rew_i(\cdot)]$ is non-decreasing, and therefore $\BIR(\cdot)$ is non-increasing. This is a mild assumption, see
Appendix~\ref{app:examples}.

%Let $\BIR_i(n) = \BIR(n)$ for the algorithm of principal $i$.

\subsection{\HardMax response function}
\label{sec:theory-HM}

%We posit fully rational agents, in the sense that their response function is \HardMax.

We consider agents with \HardMax response function, and show that principals are not incentivized to \emph{explore}, \ie to deviate from \DynGreedy. The core technical result is that if one principal adopts \DynGreedy, then the other principal loses all agents as soon as he deviates therefrom. We make this formal below.

%To make this formal, let's define what it means for MAB algorithms to deviate.

\begin{definition}\label{def:deviates}
One MAB algorithm \emph{deviates} from another at (local) step $n$ if there is a set $H$ of histories over the previous local steps such that both algorithms lead to $H$ with positive probability, and choose different distributions over arms given any history $h\in H$. If $n=n_0$ is the smallest such step, we say $\alg$ deviates from $\alg'$ \emph{starting from} step $n_0$.
\end{definition}

\begin{theorem}\label{thm:DG-dominance}
Assume \HardMax response function with fair tie-breaking: $\respF(0)=\nicefrac{1}{2}$. Assume that \alg[1] is \DynGreedy, and \alg[2] deviates from \DynGreedy starting from some (local) step $n_0<T$. Then all agents in global rounds $t\geq n_0$ select principal $1$.
\end{theorem}

\DynGreedy is a weakly dominant strategy in the competition game, and a unique Nash equilibrium.  This is because \DynGreedy receives, in expectation, at least half of the agents before global round $n_0$, and all agents after that; both are the best possible against \alg[2]. Moreover, \DynGreedy guarantees at least $T/2$ agents in expectation, and any other strategy can receive strictly less, \eg if the opponent chooses \DynGreedy.

Likewise, consider any mixed Nash equilibrium. On the one hand, each principal can guarantee exactly half of the market share by mimicking the strategy of the opponent. On the other hand, if one principal's mixed strategy is not \DynGreedy, then the opponent can grab more than a half of the market share by switching to \DynGreedy. It follows that (\DynGreedy,\DynGreedy) is a unique Nash equilibrium, whether pure or mixed.

\begin{corollary}\label{cor:DG-dominance}
\DynGreedy is a weakly dominant strategy in the competition game. The game has a unique mixed Nash equilibrium: both principals choose \DynGreedy.
\end{corollary}

The proof of Theorem~\ref{thm:DG-dominance} relies on two key lemmas: that deviating from \DynGreedy implies a strictly smaller Bayesian-expected reward, and that \HardMax implies a ``sudden-death" property: if one agent chooses principal $1$ with certainty, so do all subsequent agents. We re-use both lemmas in later results, so we state them in sufficient generality.

%In particular, Lemma~\ref{lm:DG-rew} works for any response function, as it only considers the stand-alone performance of each algorithm.

\begin{lemma}\label{lm:DG-rew}
Assume that \alg[1] is \DynGreedy, and \alg[2] deviates from \DynGreedy starting from some (local) step $n_0<T$. Then $\E[\rew_1(n_0)]>\E[\rew_2(n_0)]$. The lemma holds for any response function $\respF$ (as it only considers the stand-alone performance of each algorithm).
\end{lemma}

\begin{lemma}\label{lm:DG-sudden}
Consider \HardMax response function with $\respF(0)\geq\tfrac12$.
Suppose \alg[1] is \bmonotone, and $\PMR_1(t_0)>\PMR_2(t_0)$ for some global round $t_0$. Then $\PMR_1(t)>\PMR_2(t)$ for all subsequent rounds $t$.
\end{lemma}

The sudden-death property in Lemma~\ref{lm:DG-sudden} holds because principal $1$ appears at least as good or better to the next agent (by the Bayesian-monotonicity property), whereas principal $2$ appears the same as before. Principal $2$ needs new data in order to improve, and it does not receive new data unless the response function is randomized.

The remainder of the proof of Theorem~\ref{thm:DG-dominance} uses the conclusion of Lemma~\ref{lm:DG-rew} to derive the precondition for Lemma~\ref{lm:DG-sudden}, \ie goes from $\E[\rew_1(n_0)]>\E[\rew_2(n_0)]$ to $\PMR_1(n_0)>\PMR_2(n_0)$. The subtlety one needs to deal with is that the principal's ``local" round corresponding to a given ``global" round is a random quantity due to the random tie-breaking.

%\subsection{\HardMax with biased tie-breaking}
%\label{sec:HardMax-biased}

\xhdr{Biased tie-breaking.}
The \HardMax model is very sensitive to tie-breaking between the principals.
%For starters,
If ties are  broken deterministically in favor of principal $1$, this principal can get all agents no matter what the other principal does, simply by using \StaticGreedy.

\begin{theorem}\label{thm:HardMax-hardTies}
Assume \HardMax response function with $\respF(0)=1$ (ties are always broken in favor of principal $1$). If \alg[1] is \StaticGreedy, then all agents choose principal $1$.
\end{theorem}

\begin{proof}[Proof Sketch]
Agent $1$ chooses principal $1$ because of the tie-breaking rule. Since \StaticGreedy is trivially \bmonotone, all the subsequent agents choose principal $1$ by an induction argument similar to the one in the proof of Lemma~\ref{lm:DG-sudden}.
\end{proof}

A more challenging scenario is when the tie-breaking is biased in favor of principal 1, but not deterministically so: $\respF(0)>\tfrac12$. Then this principal also has a ``winning strategy" no matter what the other principal does. Specifically, principal 1 can get all but the first few agents, under a mild assumption that \DynGreedy deviates from \StaticGreedy.

%We can generalize the theorem top the case of $q >1/2$ if the
%principal $p_1$ can guarantee better than the a priori best action to
%all the agents following the second.

\begin{theorem}\label{thm:HardMax-biased}
Assume \HardMax response function with $\respF(0)>\tfrac12$ (\ie tie-breaking is biased in favor of principal $1$). Assume the prior $\mP$ is such that \DynGreedy deviates from \StaticGreedy starting from some step $n_0$. Suppose that principal $1$ runs a \bmonotone MAB algorithm that coincides with \DynGreedy in the first $n_0$ steps. Then all agents $t\geq n_0$ choose principal $1$.
\end{theorem}

Thus, Principal 1 can use \DynGreedy, or any other \bmonotone MAB algorithm that coincides with \DynGreedy in the first few steps.
The proof re-uses Lemmas~\ref{lm:DG-rew} and~\ref{lm:DG-sudden}, which do not rely on fair tie-breaking.

%\ascomment{Steven: can you come up with a short proof sketch / intuition why this thm works?}

%%%%%%%%%%
\subsection{\HardMax with random agents}
\label{sec:theory-HMR}

Consider the \HardMaxRandom response model, \ie \HardMax mixed with ``random agents".
Informally, we find that
%\begin{align}\label{eq:theory-HMR-informal}
\emph{a much better algorithm wins big}.
%\end{align}
In more detail, a principal with asymptotically better \BIR wins by a large margin: after a ``learning phase" of constant duration, all agents choose this principal with maximal possible probability $\respF(1)$. For example, a principal with $\BIR(n)\leq \tilde{O}(n^{-1/2})$ prevails over one with $\BIR(n)\geq \Omega(n^{-1/3})$.

%However, this positive result comes with a significant caveat detailed in Section~\ref{sec:random-greedy}.

%We formulate and prove a cleaner version of the result, followed by a more general formulation developed in a subsequent Remark~\ref{rem:random-messy}.

To state this result, we need to express a property that \alg[1] eventually catches up and surpasses \alg[2], even if initially it receives only a fraction of traffic. We assume that both algorithms run indefinitely and do not depend on the time horizon $T$; call such algorithms \emph{$T$-oblivious}. In particular, their \BIR at a given step does not depend on $T$.  Then this property can be formalized as follows:
\begin{align}\label{eq:random-better-clean}
(\forall \eps>0)\qquad
%\frac{\BIR_1(\eps n)}{\BIR_2(n)} \to 0.
\BIR_1(\eps n)\,/\,\BIR_2(n) \to 0.
\end{align}

%In fact, a weaker version suffices:
%denoting $\eps_0 = \respF(-1)$, for some constant $n_0$ we have
%\begin{align}\label{eq:random-better-weaker}
%(\forall n\geq n_0) \qquad
%\BIR_1\rbr{\nicefrac{1}{2}\;\eps_0\, n}\,/\, \BIR_2(n)<\nicefrac12.
%\end{align}
%If this holds, we say that \alg[1] \emph{BIR-dominates} \alg[2] starting from (local) step $n_0$.

In fact, a weaker version suffices:
\begin{definition}\label{def:bir_dominate}
\asedit{Let \alg[1],  \alg[2] be $T$-oblivious MAB algorithms.
Say \alg[1] \emph{BIR-dominates} \alg[2] starting from some (local) step $n_0$ if, letting $\eps_0 := \respF(-1)$,}
\begin{align}\label{eq:random-better-weaker}
(\forall n\geq n_0) \qquad
%\frac{\BIR_1(\eps_0\, n/2)}{\BIR_2(n)} <\nicefrac12.
\BIR_1\rbr{\nicefrac{1}{2}\;\eps_0\, n}\,/\, \BIR_2(n)<\nicefrac12.
\end{align}
\end{definition}

\noindent \asedit{Note that two algorithms cannot BIR-dominate one another.}

We also need a mild technical assumption that $\BIR_2(\cdot)$ is not extremely small:

%for some constant $m_0$,
\begin{align}\label{eq:random-assn}
 (\exists m_0\;\forall n\geq m_0) \qquad
  \BIR_2(n) > 4\,e^{-\eps_0\ n/12}.
\end{align}

%Thus, the main result is stated as follows:

\begin{theorem}\label{thm:random-clean}
Fix a \HardMaxRandom response function $\respF$. Suppose algorithms \alg[1], \alg[2] are \bmonotone and $T$-oblivious, and \eqref{eq:random-assn} holds. If \alg[1] \emph{BIR-dominates} \alg[2] starting from step $n_0$, then each agent $t\geq \max(n_0,m_0)$ chooses principal $1$ with probability $\respF(1) = 1- \eps_0$ (which is the largest possible probability for this response function).
\end{theorem}

%Let \alg be a \bmonotone MAB algorithm and $\mA$ be a finite set of \bmonotone MAB algorithms such that each algorithm in $\mA$ satisfies \eqref{eq:random-assn} and is ``dominated" by \alg in the sense of \eqref{eq:random-better-messy}.
%Assume principals can only choose algorithms from $\mA\cup \{\alg\}$.

%\ascomment{Steven: can you come up with a short proof sketch / intuition why this thm works?}

%We'd like to use Theorem~\ref{thm:random-clean}
To conclude that a (much) better algorithm prevails in equilibrium, we consider a version of the competition game in which the principals are restricted to choosing from a given set of MAB algorithms; the algorithms in this set are called \emph{feasible}.

%Indeed, this holds if the principals can only choose from a finite set of MAB algorithms.
%\begin{definition}\label{def:restricted-competition}
%A \emph{\FiniteGame} is the competition game between the two principals in which they can only choose from a finite set of MAB algorithms (called \emph{feasible}). All feasible algorithms are assumed to be \bmonotone and $T$-oblivious, and to satisfy \eqref{eq:random-assn}.
%\end{definition}

\begin{corollary}\label{cor:random}
Fix a \HardMaxRandom response function $\respF$ with fair tie-breaking: $\respF(0)=\nicefrac{1}{2}$. Consider the competition game in which all feasible MAB algorithms are $T$-oblivious, \bmonotone, and satisfy \eqref{eq:random-assn} for some fixed $m_0$. Suppose some feasible algorithm \alg  BIR-dominates all other feasible algorithms, starting from some local step $n_0$. Then, for any sufficiently large time horizon $T$, \alg is a weakly dominant strategy for each principal, and $(\alg,\alg)$ is a unique mixed Nash equilibrium.
\end{corollary}

This corollary is geared towards a fairly realistic scenario when the principals choose among a small number of \emph{types} of MAB algorithms (\eg Epsilon-Greedy vs. Thompson Sampling), rather than small tweaks within each type. We make no positive prediction when a few feasible algorithms are good, but no one dominates the others. Next we show that such positive predictions are essentially impossible.

%\subsection{A little greedy goes a long way}
%\label{sec:random-greedy}

\subsubsection*{Counterpoint: A little greedy goes a long way}

Given any \bmonotone MAB algorithm \alg other than \DynGreedy, we design a modified algorithm which ``mixes in" some greedy choices (and consequently learns at a slower rate), yet prevails over \alg in the competition game. Thus, we have a counterpoint to ``much better algorithms win": even under \HardMaxRandom, a slower-learning algorithm may lose in competition. A similar counterpoint to Corollary~\ref{cor:random} states that non-greedy algorithms cannot be chosen in a pure Nash equilibrium. This is consistent with Theorem~\ref{thm:random-clean}, because the BIR-dominance condition required therein does not hold here.

The modified algorithm, called the \emph{greedy modification} of \alg  with \emph{mixing parameter} $p\in (0,1)$, is defined as follows. Suppose \alg deviates from \DynGreedy starting from some (local) step $n_0$.
%\begin{enumerate}
%\item
The modified algorithm coincides with \DynGreedy
for the first $n_0-1$ steps.
%\item
In each step $n\geq n_0$, \alg is invoked with probability
  $1-p$, and with the remaining probability $p$ does the ``greedy
  choice": chooses an action with the largest posterior mean reward
  given the current information collected by \alg.
%\item
The data from the ``greedy choice" steps are not recorded.%
\footnote{In other words: the algorithm proceeds as if the ``greedy choice" steps have never happened.  While it is usually more efficient to consider all available data, this modification simplifies analysis.}
This completes the specification.

%note that it is not merely
%another pure strategy in the competition game between the two principals, not
%a mixed strategy that randomizes between \alg and the greedy algorithm.

%    \footnote{In other words: in the subsequent rounds, as far as the modified algorithm is concerned, the ``greedy choice" steps have never happened. While it is usually more efficient to consider all available data, this simplification enables a cleaner comparison between the two algorithms. Also, it makes for  a cleaner setup: otherwise, just to make the modified algorithm well-defined, we'd need to assume that \alg[1] is a mapping from step-$n$ histories to actions, for each step $n$. This is usually OK -- most bandit algorithms have this shape anyway -- but it prohibits \alg[1] to correlate its internal randomness across steps.}

%\end{enumerate}
%The modified algorithm is called \emph{greedy deviation}, with probability parameter $p$.
%Parameter $p>0$ is the same for all steps.

We find that the greedy modification prevails in competition if $p$ is small enough. We focus on \emph{symmetric} response functions: ones with
$f(x)+f(-x)=1$ for any $x\in[0,1]$.

\begin{theorem}\label{thm:random-greedy}
Consider a symmetric \HardMaxRandom response function $\respF$.
%\asdelete{with baseline probability $\eps_0 = \respF(-1)$.}
Suppose \alg[1] is \bmonotone, and deviates from \DynGreedy starting from some step $n_0$. Let \alg[2] be the greedy modification of \alg[1] with mixing parameter $p>0$ such that
    $(1-\eps_0)(1-p)>\eps_0$,
where $\eps_0 = \respF(-1)$ is the baseline selection probability.
Then each agent $t\geq n_0$ chooses principal $2$ with probability $1-\eps_0$ (which is the largest possible).
\end{theorem}

%\ascomment{Refactored and rewrote the rest of this subsection.}

Moreover, the greedy modification preserves \bmonotonicity:

\begin{lemma}\label{lm:random-bmonotone}
The greedy modification of any \bmonotone algorithm is \bmonotone, for any mixing parameter.
\end{lemma}

Thus, the greedy modification is a pure strategy in the competition game restricted to \bmonotone MAB algorithms, and it is beneficial in competition. Consider a pure Nash equilibrium of this game. If one principal chooses a non-greedy algorithm, then the opponent could guarantee strictly more than half of the market share via the greedy deviation. This is a contradiction, because the first principal can always guarantee exactly half of the market share by mimicking the opponent. Therefore, both principals must choose \DynGreedy.

\begin{corollary}\label{cor:random-greedy}
Fix a symmetric \HardMaxRandom response function $\respF$. Consider the competition game in which algorithms are feasible if and only if they are \bmonotone. Then:
\begin{OneLiners}
\item[(a)] the only possible pure Nash equilibrium is (\DynGreedy,\,\DynGreedy).

\item[(b)] If \DynGreedy satisfies $\BIR(n)\cdot n^{\gamma}\to \infty$ for some $\gamma>\nicefrac12$, then there are no pure Nash equilibria, for any sufficiently large time horizon $T$.
%The latter is not a Nash equilibrium when some algorithm BIR-dominates \DynGreedy and time horizon $T$ is sufficiently large.
\end{OneLiners}

\end{corollary}

Let us clarify part (b) of the corollary. The stated precondition is a fairly mild form of inefficiency. A more typical scenario for \DynGreedy is a learning failure, with positive-constant \BIR in each round. \Eg this happens when the Bayesian prior $\priorMu$ is independent across arms, and the prior on each $\mu_a$ has a strictly positive density on $[0,1]$ (see Corollary 11.9 in \citet{slivkins-MABbook}). (Even) if the precondition holds, \DynGreedy is dominated by any \bmonotone, $T$-oblivious algorithm with $\BIR(n) = \tildeO(t^{-1/2})$. One such algorithm is Thompson Sampling \citep{Selke-PoIE-ec21}. By Theorem~\ref{thm:random-clean}, it would be a profitable deviation from the (\DynGreedy,\,\DynGreedy) profile if the time horizon $T$ is large enough.

%In the setting of Corollary~\ref{cor:random-greedy}, the typical scenario is that (\DynGreedy,\,\DynGreedy) cannot be a Nash equilibrium, either, so there are no pure Nash equilibria. Indeed, \DynGreedy typically has positive-constant \BIR in each round, \eg when the Bayesian prior $\priorMu$ is independent across arms, and the prior on each $\mu_a$ has a strictly positive density on $[0,1]$ (see Corollary 11.9 in \citet{slivkins-MABbook}). Then \DynGreedy is BIR-dominated by any \bmonotone, $T$-oblivious algorithm whose BIR converges to $0$ over time (see Appendix~\ref{app:examples} for some examples). By Theorem~\ref{thm:random-clean}, any such algorithm would be a profitable deviation from the (\DynGreedy,\,\DynGreedy) profile if the time horizon $T$ is large enough.

\begin{remark}
While Corollary~\ref{cor:random-greedy} does not restrict \emph{mixed} Nash equilibria, this equilibrium concept appears somewhat dubious for the competition game. Indeed, the underlying premise for mixed Nash equilibria is to allow each principal to respond to the competitor's mixed strategy. However, one could argue that the principal could instead respond to the competitor's \emph{pure} strategy directly, because the latter is revealed via commitment.
\end{remark}

Finally, let us argue how the greedy modification can degrade the algorithm in some precise sense. Formulating this claim precisely is somewhat subtle, as per below. We also prove Lemma~\ref{lm:random-bmonotone} as a by-product of this analysis.

%\asmargincomment{I simplified \eqref{eq:cl:HMR-worse-condition}, but I can't rigorously justify it.}

\begin{claim}\label{cl:HMR-worse}
Let \alg[1] be any \bmonotone algorithm, and let \alg[2] be its greedy modification, with an arbitrary mixing parameter $p\in(0,1)$. Let \alggr be a hypothetical algorithm which at each step $n$ outputs the ``Bayesian-greedy choice" based on the data collected by \alg[1] in the first $n-1$ steps. Let $\BIRgr(n)$ be the \BIR of this algorithm. Suppose there exists a convex, decreasing function $f: \R^+\to [0,1]$ and parameter $q\in (1-p,1)$ such that for any sufficiently large step $n$ it holds that
\begin{align}\label{eq:cl:HMR-worse-condition}
\BIRgr(n) \geq f(n) > \BIR_1(\nicefrac{n}{q}).
\end{align}
Then for any sufficiently large step $n$ we have
    $\BIR_1(n) > \BIR_2(n)$.
\end{claim}

\begin{proof}
Let $M_n$ be the number of times \alg[1] is invoked in the first $n$ steps of \alg[2]. Let
    $\mathtt{reg}_2(n) = n\cdot \max_a \mu_a - \rew_2(n)$
be the (frequentist) instantaneous regret of \alg[2]. Then
\begin{align}
\E\sbr{ \mathtt{reg}_2(n) \mid M_n = m}
    &= (1-p) \cdot \BIR_1(m) + p \cdot \BIRgr(m). \nonumber \\
\BIR_2(n)
    &= \E\sbr{ (1-p) \cdot \BIR_1(M_n) + p \cdot \BIRgr(M_n) }.
    \label{eq:BIR-modification}
\end{align}
Using \eqref{eq:cl:HMR-worse-condition},\eqref{eq:BIR-modification} and Jensen's inequality, for any $q\in(1-p,1)$ and any large enough step $n$ we have
\[\BIR_2(n)
    \geq  \E\sbr{ \BIRgr(M_n) }
    \geq \E\sbr{ f(M_n) }
    \geq f\rbr{ \E[M_n] }
    > f(qn)
    > \BIR_1(n).\qedhere
\]
\end{proof}

Lemma~\ref{lm:random-bmonotone} follows from \refeq{eq:BIR-modification}. Indeed, the lemma asserts that $\BIR_2(n)$ is non-decreasing, which follows from \refeq{eq:BIR-modification} because both \alg[1] and \alggr are \bmonotone. The latter follows from the ``informational monotonicity" of the ``greedy step": it can only get better with more information, see Lemma~\ref{lm:MII}.

\subsection{\SoftMaxRandom response function}
\label{sec:theory-SoftMax}

For the \SoftMaxRandom model, we derive a ``better algorithm wins" result under a much weaker version of BIR-dominance. This is the most technical part of the paper.

We start with a formal definition of \SoftMaxRandom:

\begin{definition}\label{def:SoftMax}
A response function $\respF$ is \SoftMaxRandom if the following conditions hold:
\begin{OneLiners}
\item  $\respF(\cdot)$ is bounded away from $0$ and $1$:
    $\respF(\cdot)\in [\eps,1-\eps]$ for some $\eps\in (0,\tfrac12)$,
\item fair tie-breaking: $\respF(0)=\tfrac12$.
\item  the response function
 $\respF(\cdot)$ is ``smooth" around $0$:
 \begin{align}\label{eq:SoftMax-smooth}
 \exists\, \text{constants $\delta_0,c_0,c'_0>0$}
    \qquad \forall x\in [-\delta_0,\delta_0] \qquad
    c_0 \leq \respF'(x) \leq c'_0.
 \end{align}
\end{OneLiners}
\end{definition}

\begin{remark}
This definition is fruitful when $c_0$ and $c_0'$ are close to $\tfrac12$. Throughout, we assume that \alg[1] is better than \alg[2], and obtain results parameterized by $c_0$. By symmetry, one could assume that \alg[2] is better than \alg[1], and obtain similar results in terms of $c_0'$.
\end{remark}

For the sake of intuition, let us derive a version of Theorem~\ref{thm:random-clean}, with the same assumptions and a similar proof. The conclusion is much weaker, though: we can only guarantee that each agent $t\geq n_0$ chooses principal 1 with probability slightly larger than $\tfrac12$. This is essentially unavoidable in a typical case when both algorithms satisfy $\BIR(n)\to 0$.
%by Definition~\ref{def:SoftMax}.

\begin{theorem}\label{thm:SoftMax-weak}
Assume \SoftMaxRandom response function. Suppose algorithms \alg[1], \alg[2] satisfy the assumptions in Theorem~\ref{thm:random-clean}: algorithms \alg[1], \alg[2] are \bmonotone and $T$-oblivious, and \eqref{eq:random-assn} holds. If \alg[1] \emph{BIR-dominates \alg[2] starting from step $n_0$,} then each agent
  $t\geq n_0$ chooses principal $1$ with probability
\begin{align}\label{eq:thm:SoftMax-weak}
     \Pr[i_t = 1]\geq \tfrac12 +  \tfrac{c_0}{4}\; \BIR_2(t).
\end{align}
\end{theorem}

%\begin{proof}[Proof Sketch]
To prove this theorem, we follow the steps in the proof of Theorem~\ref{thm:random-clean} to derive
%\begin{align*}
$\PMR_1(t) - \PMR_2(t)
    \geq \BIR_2(t)/2 -\exp(-\eps_0\, t/12)$.
%\end{align*}
This is at least $\BIR_2(t)/4$ by \eqref{eq:random-assn}. Then \refeq{eq:thm:SoftMax-weak} follows by the smoothness condition \eqref{eq:SoftMax-smooth}.
%\end{proof}

\OMIT {%%%%%%%%
We recover a version of Corollary~\ref{cor:random}, if each principal's utility is the number of users (rather than the more general model in \eqref{eq:general-utility}).

\begin{corollary}\label{cor:SoftMax}
Assume that the response function is \SoftMaxRandom, and each principal's  utility is the number of users.
%
%Suppose principals can only choose algorithms from $\mA\cup \{\alg\}$, where $\mA\cup \{\alg\}$ is a finite set of \bmonotone MAB algorithms such that each algorithm in $\mA$ satisfies \eqref{eq:random-assn} and $\BReg(n)\to \infty$, and is ``dominated" by \alg in the sense of \eqref{eq:random-better-messy}.
%
Consider the \FiniteGame with special algorithm \alg, and assume that all other allowed algorithms satisfy $\BReg(n)\to \infty$. Then, for any sufficiently large time horizon $T$, this game has a unique Nash equilibrium: both principals choose \alg.
\end{corollary}
} %%%% \OMIT{

Let us relax the notion of BIR-dominance so that the constant multiplicative factors in \eqref{eq:random-better-weaker}, namely
 $\eps_0/2$ and $\tfrac12$, are replaced by constants that can be arbitrarily close to $1$.

\begin{definition}\label{def:weak_bir_dominate}
Let \alg[1],  \alg[2] be $T$-oblivious MAB algorithms. Say that
\alg[1] \emph{weakly BIR-dominates} \alg[2] if there are absolute  constants $\beta_0, \alpha_0\in (0, 1/2)$ and $n_0\in\N$ such that
 \begin{align}\label{eq:SoftMax-better}
   (\forall n\geq n_0) \quad
   \frac{\BIR_1((1-\beta_0)\, n)}{\BIR_2(n)} <1-\alpha_0.
 \end{align}
 \end{definition}

\noindent \asedit{Note that two algorithms cannot weakly BIR-dominate one another.}

Now we are ready to state the main result for \SoftMaxRandom:

\begin{theorem}\label{thm:SoftMax-strong}
Assume the \SoftMaxRandom response function. Suppose algorithms \alg[1], \alg[2] are \bmonotone and $T$-oblivious, and \alg[1] weakly-BIR-dominates \alg[2]. Posit mild technical assumptions:
  $\BIR_1(n) \to 0$ and that $\BIR_2$ cannot be extremely small, namely:
\begin{align}\label{eq:SoftMax-assn-strong}
 % (\forall n\geq n(\eps)) \qquad
 %  \BIR_2(n) > e^{-\eps n}. %used to be 4exp( -\eps n/6)
(\exists m_0\; \forall n\geq m_0) \qquad
\BIR_2(n) \geq \tfrac{4}{\alpha_0}\;
\exp \rbr{ - \tfrac{1}{12}\, n\,\min\{\eps_0, \nicefrac{1}{8}\} } .
\end{align}
Then there
  exists some $t_0$ such that each agent $t\geq t_0$ chooses principal
  $1$ with probability
\begin{align}\label{eq:thm:SoftMax-strong}
     \Pr[i_t = 1]\geq \tfrac12 +  \tfrac14 \,c_0\,\alpha_0\; \BIR_2(t).
\end{align}
\end{theorem}

\begin{proof}[Proof Sketch]
The main idea is that even though \alg[1] may have a
slower rate of learning in the beginning, it will gradually catch up
and surpass \alg[2]. We distinguish two phases. In
the first phase, \alg[1] receives a random agent with probability at
least $\respF(-1) = \eps_0$ in each round. Since $\BIR_1$ tends to 0,
the difference in \BIR{s} between the two algorithms is also
diminishing. Due to the \SoftMaxRandom response function, \alg[1]
attracts each agent with probability at least $1/2 - O(\beta_0)$ after
a sufficient number of rounds. Then the game enters the second phase:
both algorithms receive agents at a rate close to $\tfrac12$, and the
fractions of agents received by both algorithms --- $n_1(t)/t$ and
$n_2(t)/t$ --- also converge to $\tfrac12$. At the end of the second
phase and in each global round afterwards, the counts $n_1(t)$ and
$n_2(t)$ satisfy the weak BIR-dominance condition, in the sense that
they both are larger than $n_0$ and $n_1(t)\geq (1-\beta_0)\; n_2(t)$.
At this point, \alg[1] actually has smaller $\BIR$, which reflected in the {\PMR}s eventually. Accordingly, from then on \alg[1]
attracts agents at a rate slightly larger than $\tfrac12$. We prove
that the ``bump'' over $\tfrac12$ is at least on the order of
$\BIR_2(t)$.
\end{proof}

It follows that a weakly-BIR-dominating algorithm prevails in equilibrium.

%It follows that a weakly-BIR-dominating algorithm prevails in equilibrium. We need a mild technical assumption that cumulative Bayesian regret (\BReg) tends to infinity. \BReg is a standard notion from the literature (along with \BIR):
%\begin{align}\label{eq:SoftMax-BReg}
%\BReg(n) := \textstyle n\cdot \E_{\mu\sim\priorMu}
%    \left[ \max_{a\in A} \mu_a\right] -
%    \E\sbr{ \sum_{m=1}^n \rew(m)}
%    = \sum_{m=1}^n \BIR(m).
%\end{align}

\begin{corollary}\label{cor:SoftMax-strong}
%Assume \SoftMaxRandom response function.
Consider the competition game in which all feasible algorithms are \bmonotone, $T$-oblivious, and satisfy
    $\sum_{m=1}^n \BIR(m)\to_n \infty$.%
\footnote{This is a very mild non-degeneracy condition, see Appendix~\ref{app:bg} for background.}
Suppose some feasible algorithm \alg weakly-BIR-dominates all others. Then, for any sufficiently large time horizon $T$, \alg is a weakly dominant strategy for each principal, and $(\alg,\alg)$ is a unique mixed Nash equilibrium.
%there is a unique Nash equilibrium: both principals choose \alg.
\end{corollary}

%If both $\BIR_1()$ and $\BIR_2()$ are of the form $\tilde{\Theta}(n^{-\gamma})$, $\gamma>0$, then the old condition requires $\BIR_1(\cdot)$ to be better by constant multiplicative factor $C$, with $C$ sufficiently large, whereas the new condition allows any $C>1$.

%%% Local Variables:
%%% mode: latex
%%% TeX-master: "main"
%%% End:

%%%%%%%%%%%%%
\subsection{Economic implications}
\label{sec:theory-welfare}

We frame our contributions in terms of the relationship between \competitiveness (as expressed by the ``hardness" of the response function $\respF$),
 %and \rationality on one side,
 and adoption of better algorithms.

\OMIT{ %%%%%%
We frame our contributions in terms of the relationship between \competition and \innovation, \ie between the extent to which the game between the two principals is competitive, and the degree of innovation --- adoption of better that these models incentivize. \Competition is controlled via the response function $\respF$, and \innovation refers to the quality of the technology (MAB algorithms) adopted by the principals. The \competition vs. \innovation relationship is well-studied in the economics literature, and is commonly known to often follow an inverted-U shape, as in \reffig{fig:inverted-U} (see Section~\ref{sec:related-work} for citations). \Competition in our models is closely correlated with \rationality: the extent to which agents make rational decisions, and indeed \rationality is what $\respF$ controls directly.
} %%%%%%%%

\xhdr{Main story.}
Our main story concerns the \FiniteGame between the two principals where one allowed algorithm \alg is ``better" than the others. {We track whether and when \alg is chosen in an equilibrium.} We vary \competitiveness by changing the response function from \HardMax (very competitive environment) to \HardMaxRandom to  \SoftMaxRandom (less competition).%
\footnote{\asedit{Here, we consider \HardMaxRandom and \SoftMaxRandom with the same parameter $\eps_0<\nicefrac{1}{2}$ .}}
%\footnote{\gaedit{Following the economic interpretation of the \HardMaxRandom decision rule, for this comparison we assume that $\epsilon$ is small.}}
Our conclusions are as follows:
\begin{OneLiners}
\item Under \HardMax, no innovation: \DynGreedy is chosen over \alg.
\item Under \HardMaxRandom, some innovation:  \alg is chosen as long as it BIR-dominates.
\item Under \SoftMaxRandom, more innovation: \alg is chosen as long as it weakly-BIR-dominates.%
\end{OneLiners}
These conclusions follow from Corollaries~\ref{cor:DG-dominance}, \ref{cor:random} and \ref{cor:SoftMax-strong}, respectively. Recall that weak-BIR-dominance is a weaker condition, so that a better algorithm is chosen in a broader range of scenarios. We also consider the uniform choice between the principals, which entails the least amount of competition and (when principals optimize market share) provides no incentives to innovate.%
\footnote{However, if principals' utility is aligned with agents' welfare, then a monopolist principal is incentivized to choose the best possible MAB algorithm (namely, to minimize cumulative Bayesian regret $\BReg(T)$). Accordingly, monopoly would result in better social welfare than competition, as the latter is likely to split the market and cause each principal to learn more slowly. This is a well-known effect of economies of scale.}
Thus, we have an inverted-U relationship, see \reffig{fig:inverted-U2}.

%\begin{figure}[t]
%\begin{center}
%\begin{tikzpicture}[scale=1]
%      \draw[->] (-.5,0) -- (9.5,0) node[above]
%        {\qquad\qquad Competitiveness};
%      \draw[->] (0,-.5) -- (0,3) node[above] {Better algorithm in equilibrium};
%      \draw[scale=0.8,domain=0.5:9.5,smooth,variable=\x,blue, line width=0.3mm] plot ({\x},{3.5 - 0.15*(\x - 5)^2});
%     \node[below] at (1, 0) {\footnotesize \Uniform};
%     \node[below] at (3.9, 0) {\footnotesize \SoftMaxRandom};
%     \node[below] at (6, -.5) {\footnotesize \HardMaxRandom};
%     \node[below] at (8, 0) {\footnotesize \HardMax};
%      % \draw[scale=0.5,domain=-3:3,smooth,variable=\y,red]  plot ({\y*\y},{\y});
% \end{tikzpicture}
%
%\caption{The stylized inverted-U relationship in the ``main story".}
%\label{fig:inverted-U2}
%\end{center}
%\end{figure}

\xhdr{Secondary story.}
Let us zoom in on the symmetric  \HardMaxRandom model. Competitiveness within this model are controlled by the baseline probability $\eps_0 = \respF(- 1)$, which varies  smoothly between the two extremes: \HardMax ($\eps_0=0$, tough competition) and the uniform choice ($\eps_0=\tfrac12$, no competition).
Principals' utility is the number of agents.

\newcommand{\deltaU}{\Delta U}

We consider the marginal utility of switching to a better algorithm. Suppose initially both principals use some algorithm \alg, and principal 1 ponders switching to another algorithm \alg' which BIR-dominates \alg. What is the marginal utility $\deltaU$ of this switch?

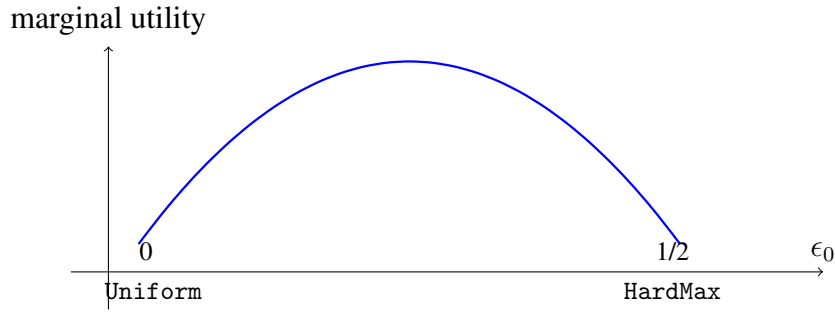
\begin{figure}[t]
\begin{center}
\begin{tikzpicture}[scale=1]
      \draw[->] (-.5,0) -- (9.5,0) node[above]  {$\eps_0$};
      \draw[->] (0,-.5) -- (0,3) node[above] {marginal utility};
      \draw[scale=0.8,domain=0.5:9.5,smooth,variable=\x,blue, line width=0.3mm] plot ({\x},{3.5 - 0.15*(\x - 5)^2});
     \node[below] at (.6, 0) {\footnotesize \Uniform};
     \node[above] at (.5, 0) {\footnotesize 0};
     % \node[below] at (3.9, 0) {\footnotesize \SoftMaxRandom};
     % \node[below] at (6, 0) {\footnotesize \HardMaxRandom};
     \node[below] at (7.5,0) {\footnotesize \HardMax};
     \node[above] at (7.5, 0) {\footnotesize 1/2};
      % \draw[scale=0.5,domain=-3:3,smooth,variable=\y,red]  plot ({\y*\y},{\y});
 \end{tikzpicture}

\caption{The stylized inverted-U relationship from the ``secondary story"}
\label{fig:inverted-U3}
\end{center}
\end{figure}

\begin{OneLiners}
\item if $\eps_0 = 0$ then $\deltaU$ can be negative if \alg is \DynGreedy.

\item if $\eps_0$ is near $0$ then only a small $\deltaU$ can be guaranteed, as it may take a long time for $\alg'$ to ``catch up" with \alg, and hence less time to reap the benefits.

\item if $\eps_0$ is medium-range, then $\deltaU$ is large, as $\alg'$ learns fast and gets most agents.

\item if $\eps_0$ is near $\tfrac12$, the algorithm matters less, so $\deltaU$ is small.
%as principal 1 gets most agents for free no matter what.
\end{OneLiners}
These findings can also be organized as an inverted-U relationship, see Figure~\ref{fig:inverted-U3}.

%%%%%%%%%%%%%
\subsection{Extensions}
\label{sec:theory-extensions}

Our theoretical results can be extended beyond the basic model in Section~\ref{sec:model}.

\xhdr{Reward-dependent utility.}
Except for Corollary~\ref{cor:SoftMax-strong}, our results allow a more general notion of principal's utility that can depend on both the market share and agents' rewards. Namely, each principal $i$ collects $U_i(r_t)$ units of utility in each global round $t$ when she is chosen (and $0$ otherwise), where $U_i(\cdot)$ is some fixed non-decreasing function with $U_i(0)>0$.
%In a formula,
%\begin{align}\label{eq:general-utility}
%\textstyle
%$U_i := \sum_{t=1}^T\; \indicator{i_t=i}\cdot U_i(r_r)$.
%\end{align}

\xhdr{Time-discounted utility.}
Theorem~\ref{thm:DG-dominance} and Corollary~\ref{cor:DG-dominance} holds under a more general model which allows time-discounting: namely, the utility of each principal $i$ in each global round $t$ is $U_{i,t}(r_t)$ if this principal is chosen, and $0$ otherwise, where $U_{i,t}(\cdot)$ is an arbitrary non-decreasing function with $U_{i,t}(0)>0$.

\xhdr{Arbitrary reward distributions.}
Bernoulli rewards can be extended to arbitrary reward distributions. For each arm $a\in A$ there is a parametric family $\psi_a(\cdot)$ of reward distributions, parameterized by the mean reward. Whenever arm $a$ is chosen, the reward is drawn independently from distribution $\psi_a(\mu_a)$. The prior $\priorMu$ and the distributions $(\psi_a(\cdot)\colon a\in A)$ constitute the (full) Bayesian prior on rewards.% denoted $\prior$.

\xhdr{Beliefs.}
Instead of knowing the principals' algorithms $\alg[1],\,\alg[2]$, the Bayesian prior $\priorMu$, and the response function $\respF$, \asedit{agents could share a common ``point belief" on these objects. That is: if all agents believe that these objects are, resp., $\alg[1]',\, \alg[2]', \priorMu'$, and $\respF'$, then all our results carry over with respect to this belief, even if it is incorrect. This is because the agents in our model do not observe realized history, and their behavior is completely determined by the ``initial conditions"  (i.e.,  $\alg[1],\, \alg[2],\, \priorMu$, and $\respF$).}

\xhdr{Limited non-stationarity in $\respF$.}
Different agents can have different response functions. For \HardMaxRandom, our results carry over if each agent $t$ has a \HardMaxRandom response function $\respF$ with parameter $\eps_t\geq \eps_0$. For \SoftMaxRandom, different agents can have different response functions that satisfy Definition~\ref{def:SoftMax} with the same parameters.

\xhdr{MAB extensions.} Our results carry over, with little or no modification of the proofs, to much more general versions of MAB, as long as it satisfies the i.i.d. property. In each round, an algorithm can see a \emph{context} before choosing an action (as in \emph{contextual bandits}) and/or additional feedback other than the reward after the reward is chosen (as in, \eg \emph{semi-bandits}), as long as the contexts are drawn from a fixed distribution, and the (reward, feedback) pair is drawn from a fixed distribution that depends only on the context and the chosen action. The Bayesian prior $\prior$ needs to be a more complicated object, to make sure that \PMR and \BIR are well-defined. Mean rewards may also have a known structure, such as Lipschitzness, convexity, or linearity; such structure can be incorporated via $\prior$. All these extensions have been studied extensively in the literature on MAB, and account for a substantial segment thereof;
see \citep{slivkins-MABbook,LS19bandit-book} for background.

%Background can be found in any of the recent books on MAB \citep{Bubeck-survey12,slivkins-MABbook,LS19bandit-book}.

\xhdr{\BIR can depend on $T$.}
Many MAB algorithms are parameterized by the time horizon $T$, and their regret bounds include $\polylog(T)$. In particular,  a typical regret bound for \BIR is
\begin{align}
    \BIR(n\mid T)\leq \polylog(T)\cdot n^{-\gamma}
    \quad \text{for some $\gamma\in(0,\tfrac12]$}.
\end{align}
We write $\BIR(n\mid T)$ to emphasize the dependence on $T$. Accordingly, BIR-dominance can be redefined: there exists a number $T_0$ and a function $n_0(T)\in \polylog(T)$
such that
\begin{align}\label{eq:random-better-messy}
(\forall T\geq T_0,\; n\geq n_0(T)) \quad
\BIR_1(\eps_0 n /2\mid T)\,/\, \BIR_2(n\mid T) <\nicefrac12.
\end{align}
%If this holds, we say that \alg[1] \emph{BIR-dominates} \alg[2].
%The proof of this extension is very similar and is omitted.

\noindent Weak BIR-dominance extends similarly.
Theorem~\ref{thm:random-clean} and~\ref{thm:SoftMax-weak} easily carry over.

%\begin{theorem}\label{thm:random-messy}
%Assume \HardMaxRandom response function, and algorithms that satisfy \eqref{eq:random-better-messy} and~\eqref{eq:random-assn}. Then each agent
%    $t\geq n_0(T)$ chooses principal $1$ with probability $\respF(1)$.
%\end{theorem}

%Similarly, weak BIR-dominance can be redefined as follows: there exist some $T_0$, a function
%$n_0(T)\in \polylog(T)$, and constants $\beta_0,\alpha_0\in (0, 1/2)$, such that
%\begin{align}\label{eq:SoftMax-better-weaker}
%(\forall T\geq T_0,  n\geq n_0(T)) \quad
%\frac{\BIR_1((1-\beta_0)\, n\mid T)}{\BIR_2(n\mid T)} <1-\alpha_0.
%\end{align}

%Theorem~\ref{thm:random-clean} and Theorem~\ref{thm:SoftMax-weak} extend to these generalized versions; the easy modifications are omitted.

%% file: content/sim_details.tex
In this section we present our numerical simulations. As discussed in the Introduction, we focus on the \ExptsModel, whereby each agent chooses the firm with a maximal reputation score, modeled as a sliding window average of its rewards. While we experiment with various MAB instances and parameter settings, we only report on selected, representative experiments. Additional plots and tables are provided in Appendix~\ref{app:expts}. Unless noted otherwise, our findings are based on and consistent with all these experiments.

%In this section we move to the reputation choice variant. Recall that in this variant of the model, instead of making choices between firms based on the Bayesian expected reward, each agent chooses the firm with a maximal reputation score (breaking ties uniformly). The reputation score is simply a sliding window average: an average reward of the last $M$ agents that chose this firm. We focus on numerical investigation of this model instead of analytical characterizations of equilibrium strategies.

\subsection{Experiment setup}
\label{expts-prelims}

%The timing of the model is the same as before where each firm commits to a MAB algorithm before the game starts and uses this algorithm to choose its actions. We focus on i.i.d. Bernoulli rewards: the  reward of each arm $a$ is drawn from $\{0,1\}$ independently with expectation $\mu(a)$. The mean rewards $\mu(a)$ are the same for all rounds and both firms, but initially unknown. However, instead of starting from some prior, we suppose that each firm has a uniform, ``fake", prior and that the initial information set of the firm is given by a ``warm start". Each algorithm receives a ``warm start": additional $T_0$ agents that arrive before the game starts, and interact with the firm as described above. The warm start ensures that each firm has a meaningful reputation and initial prior when competition starts..

\xhdr{Challenges.} An ``atomic experiment" is a competition game between a given pair of bandit algorithms, in a given competition model, on a given
%instance of a
multi-armed bandit problem (and each such experiment is run many times to reduce variance). Accordingly, we have a three-dimensional space of atomic experiments one needs to run and interpret: \{pairs of algorithms\} x \{competition models\} x \{bandit problems\}, and we are looking for findings that are consistent across this entire space. It is essential to keep each of the three dimensions small yet representative. In particular, we need to capture a huge variety of bandit algorithms and bandit instances with only a few representative examples. Further, we need a succinct and informative summarization of results within one atomic experiment and across multiple experiments (\eg see Table~\ref{fig:market_share}).

\xhdr{Competition model.} All experiments use \HardMax response function (without mentioning it), except Section~\ref{sec:non_greedy} where we use \HardMaxRandom agents. In some of our experiments, one firm is the ``incumbent" who enters the market before the other (``late entrant"), and therefore enjoys a \emph{first-mover advantage}. Formally, the incumbent enjoys additional $X$ rounds of the ``warm start". We treat $X$ as an exogenous element of the model, and study the consequences for a fixed $X$.

\xhdr{MAB algorithms.} In abstract terms, we posit three types of technology, from ``low" to ``medium" to ``high". Concretely, we consider three essential classes of bandit algorithms: ones that never explicitly explore (\emph{greedy algorithms}), ones that explore without looking at the data (\emph{exploration-separating algorithms}), and ones where exploration gradually zooms in on the best arm (\emph{adaptive-exploration algorithms}). In the absence of competition, these classes are fairly well-understood: greedy algorithms are terrible for a wide variety of problem instances, exploration-separated algorithms learn at a reasonable but mediocre rate across all problem instances, and adaptive-exploration algorithms are optimal in the worst case, and exponentially improve for ``easy" problem instances (see Appendix~\ref{app:bg}).

We look for qualitative differences between these three classes under competition. We take a representative algorithm from each class. Our pilot experiments indicate that our findings do not change substantially if other representative algorithms are chosen. We use
\DynGreedy (\DG) algorithm  as in Section~\ref{sec:theory-prelims},
\DynamicEpsGreedy (\DEG) from the ``exploration-separating" algorithms,
and \Thompson (\TS) from the ``adaptive-exploration" algorithms.%
\footnote{In each round $t$, \Thompson computes a Bayesian posterior on $\mu_a$ for each arm $a$ and draws an independent sample $\tilde{\mu}_{a,t}$ from this posterior; it chooses an arm which maximizes $\tilde{\mu}_{a,t}$.
    \newline\indent
\DynamicEpsGreedy proceeds as follows. In each round, with probability $\eps$ it explores by choosing an arm from the full set of arms uniformly at random. With the remaining probability, it ``exploits" by choosing an arm with maximal posterior mean reward given the current data. We use $\eps=5\%$ throughout. Our pilot experiments show that choosing a different $\eps$ does not qualitatively change the results.}
For ease of comparison, all three algorithms are parameterized with the same ``fake" Bayesian prior: namely, the mean reward of each arm is drawn independently from a $\Beta(1,1)$ distribution. Recall that Beta priors with 0-1 rewards form a conjugate family, which allows for simple posterior updates.

%Self-contained background can be found in Appendix~\ref{sec:related-classes}.

\xhdr{MAB instances.}
We consider bandit problems with $K=10$ arms and Bernoulli rewards. The \emph{\MRV} $(\mu(a):\; a\in A)$ is initially drawn from some distribution, termed \emph{MAB instance}. We consider three MAB instances:
\begin{enumerate}
\item \emph{Needle-In-Haystack}: one arm (the ``needle") is chosen uniformly at random. This arm has mean reward $.7$, and the remaining ones have mean reward $.5$.

\item \emph{Uniform instance}: the mean reward of each arm is drawn independently and uniformly from $[\nicefrac{1}{4}, \nicefrac{3}{4}]$.
\item \emph{Heavy-Tail instance}: the mean reward of each arm is drawn independently from $\Beta(.6,.6)$ distribution (which is known to have substantial ``tail probabilities").
\end{enumerate}
We argue that these MAB instances are (somewhat) representative. Consider the ``gap" between the best and the second-best arm, an essential parameter in the literature on MAB. The ``gap" is fixed in Needle-in-Haystack, spread over a wide spectrum of values under the Uniform instance, and is spread but  focused on the large values under the Heavy-Tail instance. We also ran smaller experiments with versions of these instances, and achieved similar qualitative results.

\xhdr{Simulation details.}
For each MAB instance we draw $N = 1000$ \MRVs independently from the corresponding distribution. We use this same collection of \MRVs for all experiments with this MAB instance. For each \MRV we draw a table of realized rewards (\emph{realization table}), and use this same table for all experiments on this \MRV. This ensures that differences in algorithm performance are solely due to differences in the algorithms in the different experimental settings.

More specifically, the realization table is a 0-1 matrix $W$ with $K$ columns which correspond to arms, and $T+T_{\max}$ rows, which correspond to rounds. Here $T_{\max}$ is the maximal duration of the ``warm start" in our experiments, \ie the maximal value of $X+T_0$. For each arm $a$, each value $W(\cdot,a)$ is drawn independently from Bernoulli distribution with expectation $\mu(a)$. Then in each experiment, the reward of this arm in round $t$ of the warm start is taken to be $W(t,a)$, and its reward in round $t$ of the game is $W(T_{\max}+t,a)$.

For the reputation scores, we fix the sliding window size $M = 100$. We found that lower values induced too much random noise in the results, and increasing $M$ further did not make a qualitative difference. Unless otherwise noted, we used $T = 2000$.

\xhdr{Terminology.}
A particular instance of the competition game is specified by the MAB instance and the game parameters, as described above. Recall that firms are interested in maximizing their expected market share at the end of the game.
%Following a standard game-theoretic terminology,
Thus, for a given instance of the game and a given firm, algorithm $\alg[1]$ \emph{(weakly) dominates} algorithm $\alg[2]$ if $\alg[1]$ provides a larger (or equal) expected final market share than $\alg[2]$, no matter that the opponent does. An algorithm is a (weakly) dominant strategy for the firm if it (weakly) dominates the other two algorithms.
%This is for a particular MAB instance and a particular selection of the game parameters.

\OMIT{Even with a stylized model, numerical investigation is quite challenging. An ``atomic experiment" is a competition game between a given pair of bandit algorithms, in a given competition model, on a given instance of a multi-armed bandit problem.%
\footnote{Each such experiment is run many times to reduce variance.}
Accordingly, we have a three-dimensional space of atomic experiments one needs to run and interpret: \{pairs of algorithms\} x \{competition models\} x \{bandit instances\}, and we are looking for findings that are consistent across this entire space. It is essential to keep each of the three dimensions small yet representative. In particular, we need to capture a huge variety of bandit instances with only a few representative examples. Further, one needs succinct and informative summarization of results within one atomic experiment and across multiple experiments (\eg see Table~\ref{sim_table}).}

\OMIT{
\xhdr{Running time.}
The simulations are computationally intensive. An experiment on a particular MAB instance comprised multiple runs of the competition game: $N$ mean reward vectors times $9$ pairs of algorithms times three values for the warm start. We used a parallel implementation over a cluster of 12 2.2 GHz CPU cores, with 8 GB RAM per core. Each experiment took about $10$ hours.
}

%% file: ec19paper/content/perf_in_iso.tex
\subsection{Performance in Isolation}\label{sec:isolation}

\begin{figure}
\centering
\includegraphics[scale=0.35]{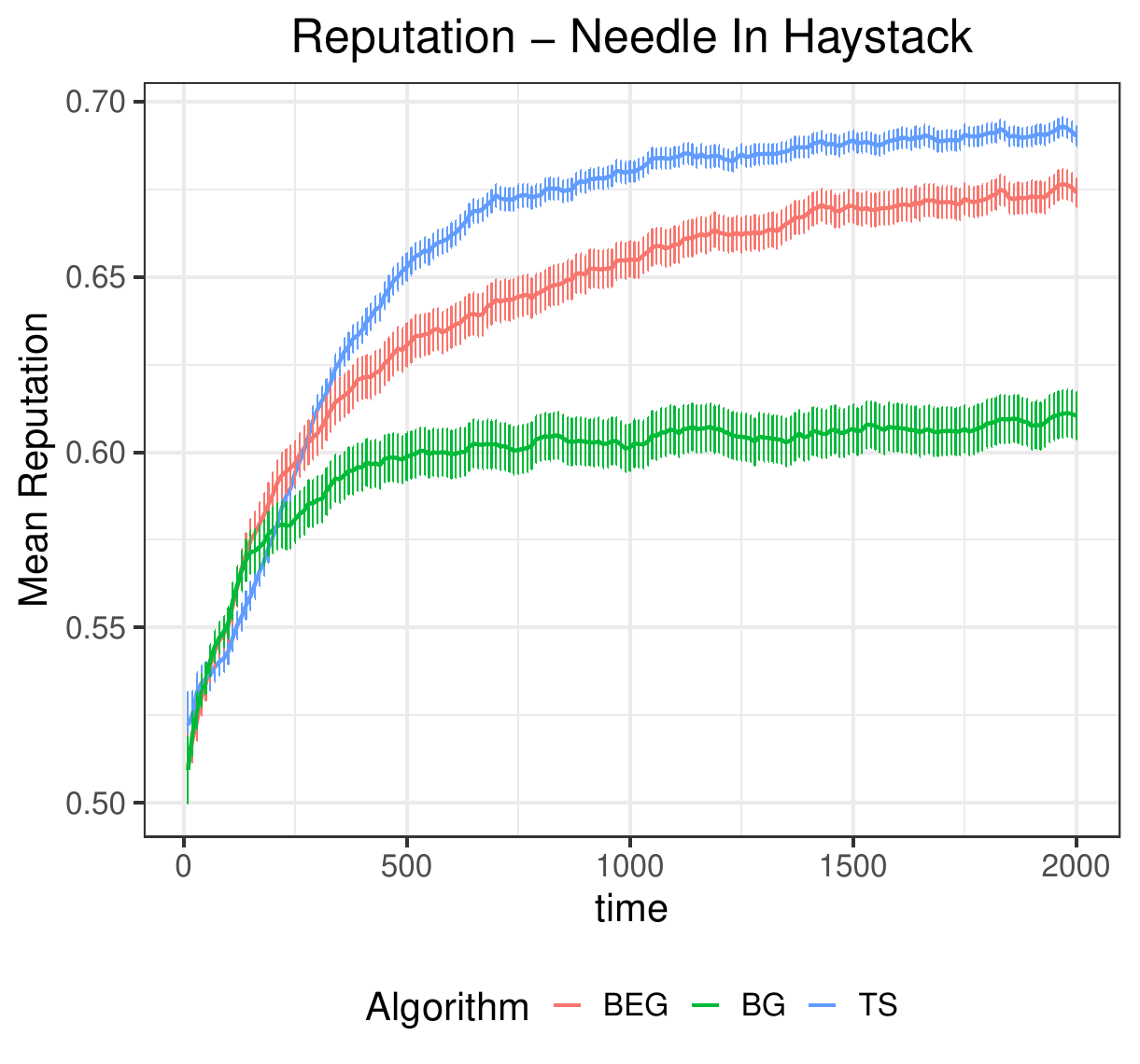}
\includegraphics[scale=0.35]{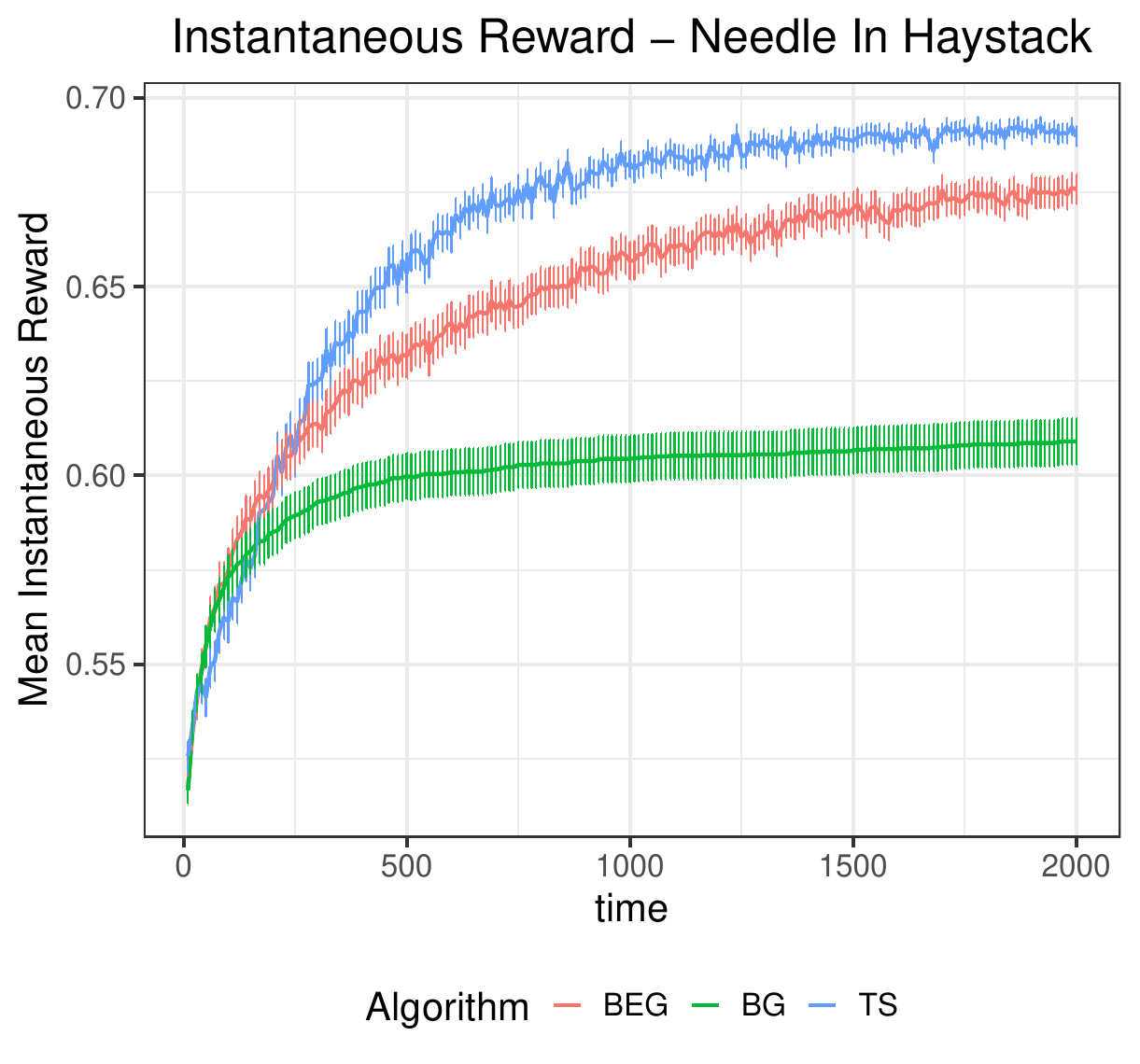}
%\caption*{\tiny{Mean reputation trajectory The plots contain the average reputation over $1000$ runs for a memory size of $100$ where, for a given $t$, we record the reputation of a given algorithm on a given instance and then average this value across all the runs. The shaded area display 95\% confidence intervals.}}
\caption{\footnotesize Mean reputation trajectory (left) and mean instantaneous reward trajectory (right) for Needle-in-Haystack. The shaded area shows 95\% confidence intervals. The shorthand for the algorithms is the same as in the main text: resp., $\DynamicEpsGreedy (\DEG)$, $\DynamicGreedy (\DG)$, and $\Thompson (\TS)$.}
\label{prelim_means}
\end{figure}

We start with a pilot experiment in which we investigate each algorithm's performance ``in isolation": in a stand-alone MAB problem without competition. We focus on reputation scores generated by each algorithm. We confirm that algorithms' performance is ordered as we'd expect:
    $\Thompson > \DynamicEpsGreedy > \DynamicGreedy$
for a sufficiently long time horizon. For each algorithm and each MAB instance, we compute the mean reputation score at each round, averaged over all \MRVs. We plot the \emph{mean reputation trajectory}: how this score evolves over time. We also plot the trajectory for instantaneous rewards (\emph{not} averaged over the previous time-periods), which provides a better view into algorithm's performance at a given time.%
\footnote{For ``instantaneous reward" at a given time $t$, we report the average (over all \MRVs) of the mean rewards at this time, instead of the average of the \emph{realized} rewards, so as to decrease the noise.}
Figure \ref{prelim_means} shows these trajectories for the Needle-in-Haystack instance; for other MAB instances the plots are similar.
We summarize this finding as follows:

\begin{finding}
\textit{The mean reputation trajectories and the instantaneous reward trajectories are arranged as predicted by prior work:
    $\Thompson > \DynamicEpsGreedy > \DynamicGreedy$ for a sufficiently long time horizon $T$.}
\end{finding}

We also use Figure~\ref{prelim_means} to choose a reasonable time-horizon for the subsequent experiments, as $T=2000$. The idea is, we want $T$ to be large enough so that algorithms performance starts to plateau, but small enough such that algorithms are still learning.

The mean reputation trajectory is probably the most natural way to represent an algorithm's performance on a given MAB instance. However, we found that the outcomes of the competition game are better explained with a different ``performance-in-isolation" statistic that is more directly connected to the game. Consider the  performance of two algorithms, $\alg[1]$ and $\alg[2]$, ``in isolation" on a particular MAB instance. The \emph{relative reputation} of $\alg[1]$ (vs. $\alg[2]$) at a given time $t$ is the fraction of \MRVs /realization tables for which $\alg[1]$ has a higher reputation score than $\alg[2]$. The intuition is that agent's selection in our model depends only on the comparison between the reputation scores.

\begin{figure}[ht]
\centering
\includegraphics[scale=0.35]{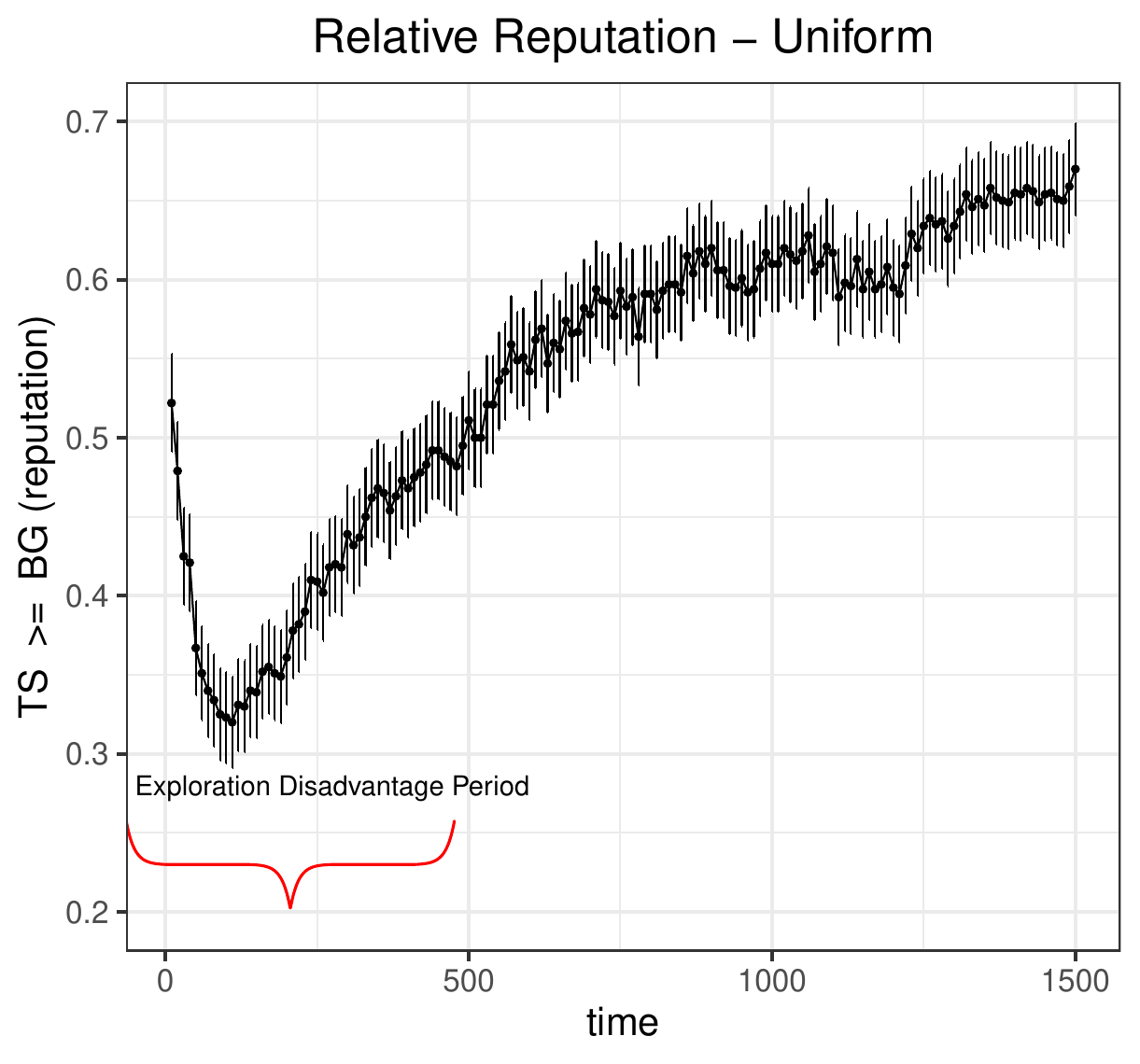}
\includegraphics[scale=0.35]{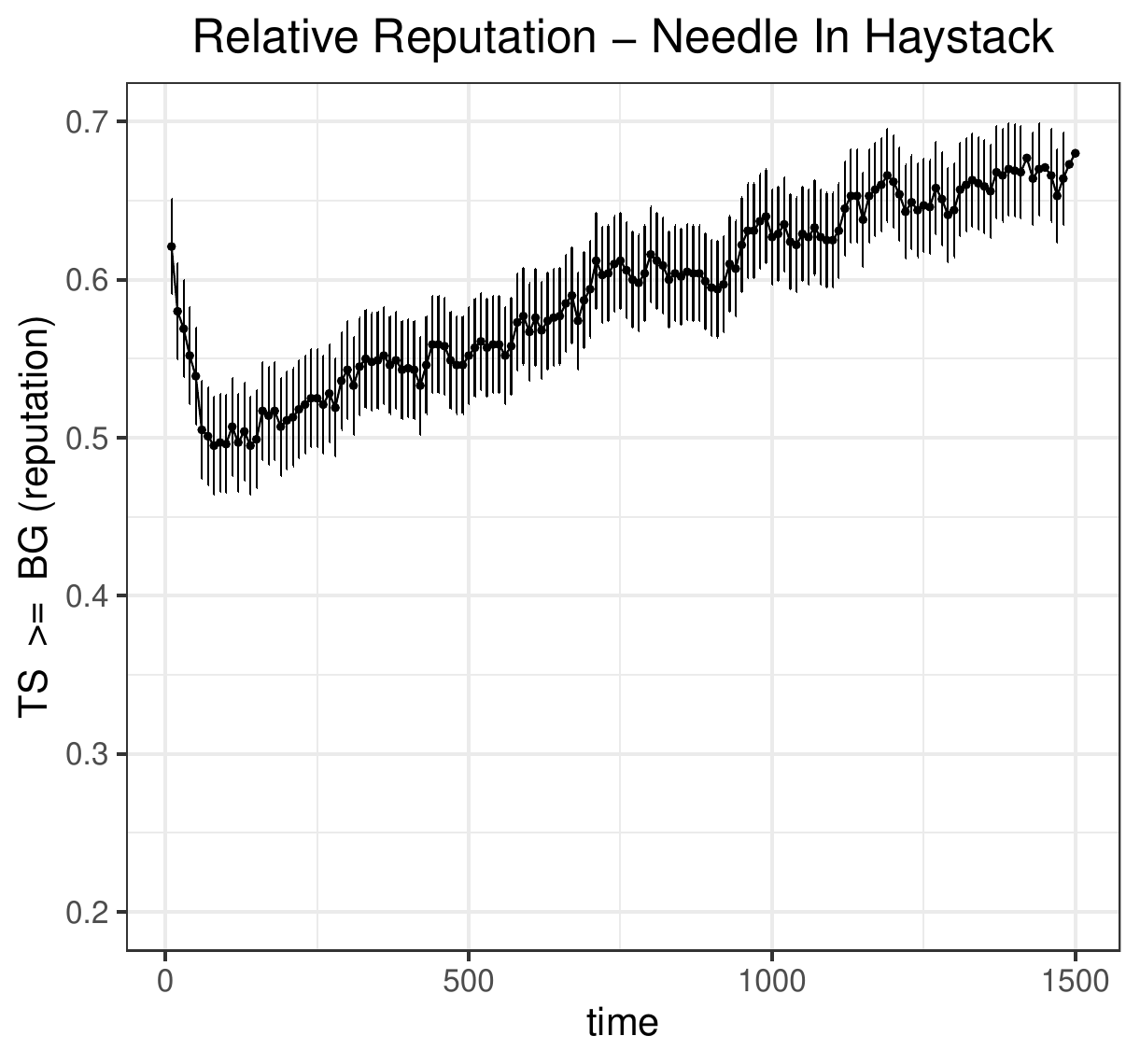}
%\caption*{\tiny{The plots contain the average reputation over $1000$ runs for a memory size of $100$ where, for a given $t$, we record the reputation of both of the algorithms on a given instance and then calculate the proportion of runs where $\Thompson \geq \DynamicGreedy$. The shaded area display 95\% confidence intervals.}}
\caption{\footnotesize Relative reputation trajectory for $\Thompson$ vs $\DynamicGreedy$, on Uniform instance (left) and Needle-in-Haystack instance (right). Shaded area display 95\% confidence intervals. The relative reputation at time $t$ is the fraction of \MRVs for which, at time $t$, $\Thompson$ has a higher reputation score than $\DynamicGreedy$.}
\label{relative_rep_plots}
\end{figure}

This angle allows a more nuanced analysis of reputation costs vs. benefits under competition. Figure \ref{relative_rep_plots} (left) shows the relative reputation trajectory for $\Thompson$ vs $\DynamicGreedy$ for the Uniform instance. The relative reputation is less than $\tfrac12$ in the early rounds, meaning that $\DynamicGreedy$ has a higher reputation score in a majority of the simulations, and more than $\tfrac12$ later on. The reason is the exploration in \Thompson leads to worse decisions initially, but allows for better decisions later. The time period when relative reputation vs. \DynamicGreedy dips below $\tfrac12$ can be seen as an explanation for the competitive disadvantage of exploration. Such period also exists for the Heavy-Tail instance. However, it does not exist for the Needle-in-Haystack instance, see Figure \ref{relative_rep_plots}.%
\footnote{We see two explanations for this: $\Thompson$ identifies the best arm faster for the Needle-in-Haystack instance, and there are no ``very bad" arms to make exploration expensive in the near term.}

\begin{finding}\label{find:period}
\textit{Exploration can lead to relative reputation vs. $\DynamicGreedy$ going below $\tfrac12$ for some initial time period. This happens for some MAB instances but not for some others.}
\end{finding}

\begin{definition}\label{def:exploration_disadvantage_period}
For a particular MAB algorithm, a time period when relative reputation vs. \DynamicGreedy goes below $\tfrac12$ is called {\em exploration disadvantage period}. An MAB instance is called \emph{exploration-disadvantaged} if such period exists.
\end{definition}

\noindent Note that Uniform and Heavy-tail instances are exploration-disadvantaged, but Needle-in-Haystack instance is not.

%% file: ec19paper/content/inverted_u.tex
\subsection{Competition vs. Better Algorithms}\label{sec:competition}

Our main experiments concern the duopoly game defined in Section~\ref{sec:model}. As the ``intensity of competition" varies from  monopoly to ``incumbent" to ``simultaneous entry" to ``late entrant", we find a stylized inverted-U relationship as in Section~\ref{sec:theory-welfare}. We look for equilibria in the duopoly game, where each firm's choices are limited to \DynamicGreedy, \DynamicEpsGreedy and \Thompson. We do this for each ``intensity level" and each MAB instance, and look for findings that are consistent across MAB instances.
%For cleaner results,
We break ties towards less advanced algorithms, as they tend to have lower adoption costs \citep{DS-arxiv}. \DynamicGreedy is then the dominant strategy under monopoly.

%(1) Permanent monopoly - indifferent between all algorithms but break indifference towards DG cause of deployment costs
%(2) Temporary monopoly - TS is the dominant strategy for the incumbent
%(3) Permanent duopoly -
%   (a) Under NIH, TS is dominant
%   (b) Under Heavy Tail, DG is weakly dominant
%   (c) Under Uniform, DG "beats" TS, but DEG "beats" TS by more than DG and DEG vs DG leads to ~50% market share (it's slightly in favor of DG but 50/50 is within the confidence band). If we break indifference towards DG cause of deployment costs, then we have that the unique PSNE is (DG, DG). If we don't add the indifference breaking, then we have four PSNE, (DG, DEG), (DEG, DG), (DG, DG), (DEG, DEG)

\xhdr{Simultaneous entry.}
The basic scenario is when both firms are competing from round $1$. A crucial distinction is whether an MAB instance is exploration-disadvantaged:

\begin{finding}\label{find:duopoly}
\textit{Under simultaneous entry:
\begin{itemize}
\item[(a)] (\DynamicGreedy,\DynamicGreedy) is the unique pure-strategy Nash equilibrium for exploration-disadvantaged MAB instances with a sufficiently small ``warm start".
\item[(b)] This is not necessarily the case for MAB instances that are not exploration-disadvantaged. In particular, \Thompson is a weakly dominant strategy for Needle-in-Haystack.
\end{itemize}
}
\end{finding}

\begin{table*}[t]
\centering
\begin{adjustbox}{width=\textwidth,center}
\begin{tabular}{|c|c|c|c||c|c|c||c|c|c|}
  \hline
  & \multicolumn{3}{c||}{Heavy-Tail}
  & \multicolumn{3}{c|}{Needle-in-Haystack}
  & \multicolumn{3}{c|}{Uniform}\\
  \hline
  & $T_0$ = 20 & $T_0$ = 250 & $T_0$ = 500
   & $T_0$ = 20 & $T_0$ = 250 & $T_0$ = 500
  & $T_0$ = 20 & $T_0$ = 250 & $T_0$ = 500 \\
  \hline
\TS vs \DG
  & \textbf{0.31} $\pm$0.03
  & \textbf{0.72} $\pm$0.02
  & \textbf{0.75} $\pm$0.02
  & \textbf{0.68} $\pm$0.03
  & \textbf{0.62} $\pm$0.03
  & \textbf{0.65} $\pm$0.03
  & \makecell{\textbf{0.44} $\pm$0.03}
 & \makecell{\textbf{0.52} $\pm$0.02}
 & \makecell{\textbf{0.58} $\pm$0.02} \\
\hline
  $\TS$ vs $\DEG$
  & \textbf{0.3} $\pm$0.03
  & \textbf{0.89} $\pm$0.01
  & \textbf{0.9} $\pm$0.01
  & \textbf{0.6} $\pm$0.03
  & \textbf{0.52} $\pm$0.03
  & \textbf{0.55} $\pm$0.02
 & \makecell{\textbf{0.41} $\pm$0.03}
 & \makecell{\textbf{0.47} $\pm$0.02}
 & \makecell{\textbf{0.55} $\pm$0.02} \\ \hline
  $\DG$ vs $\DEG$
  & \textbf{0.63} $\pm$0.03
  & \textbf{0.6} $\pm$0.02
  & \textbf{0.56} $\pm$0.03
  & \textbf{0.42} $\pm$0.03
  & \textbf{0.41} $\pm$0.03
  & \textbf{0.39} $\pm$0.02
   & \makecell{\textbf{0.5} $\pm$0.03}
 & \makecell{\textbf{0.46} $\pm$0.02}
 & \makecell{\textbf{0.45} $\pm$0.02} \\ \hline
\end{tabular}
\end{adjustbox}
\caption{\footnotesize {\bf Simultaneous Entry, Market Share}. Each cell describes a game between two algorithms, call them $\alg[1]$ vs. $\alg[2]$ for a particular value of the warm start $T_0$. Each cell contains the market share of $\alg[1]$: the average (in bold) and the 95\% confidence band.
%For example, the cell in the top left indicates that TS gets on average 64\% of the market when played against DG.
 The time horizon is $T=2000$.}
\label{fig:market_share}
\end{table*}

\normalsize

%\begin{finding}
%\textit{For sufficiently low warm start under the instances that are exploration-disadvantaged, the unique incentivized strategies in the competition game are:\footnote{The uniqueness comes from the fact that we break indifferences towards easier to deploy algorithms}
%\begin{center}
%\textbf{Permanent Monopoly} - $\DG$ \\
%\textbf{Temporary Monopoly} (incumbent)- $\TS$ \\
%\textbf{Permanent Duopoly} - $(\DG, \DG)$ \\
%\end{center}
%The conditions on $\TS$ being dominant under the temporary monopoly are that the incumbent is a temporary monopoly for sufficiently many periods.}
%\end{finding}

%\xhdr{Permanent Monopoly.} Since there is only a single firm in the
%market for the entire period, the firm can take the entire market
%regardless of what algorithm it deploys. \swedit{Since we assume that exploration algorithms ($\DEG$ and $\TS$) would incur deployment cost and firms break indifferences towards easier to deploy algorithms}, the firm would choose to deploy $\DG$.

We investigate the firms' market shares when they choose different algorithms (otherwise, by symmetry both firms get half of the agents). We report the market shares for each instance in Table~\ref{fig:market_share}. We find that $\DG$ is a weakly dominant strategy for the Heavy-Tail and Uniform instances, as long as $T_0$ is sufficiently small. However, \Thompson is a weakly dominant strategy for the Needle-in-Haystack instance. We find that for a sufficiently small $T_0$, \DynamicGreedy yields more than half the market against \Thompson,  but achieves similar market share vs. \DynamicGreedy and \DynamicEpsGreedy. By our tie-breaking rule, (\DynamicGreedy,\DynamicGreedy) is the only pure-strategy equilibrium.

\OMIT{This establishes our claim that $\DG$ is the incentivized algorithm.\footnote{We defer the table for uniform instances to the appendix. Summarizing the results, we see that, for low warm start, \DG yields more than half the market against \TS but is indifferent between \DG and \DEG. Since we break indifference towards easier to deploy algorithms, we also find that \DG is the incentivized algorithm in equilibrium}}

We attribute the prevalence of \DynamicGreedy on exploration-disadvantaged MAB instances to its prevalence on the initial ``exploration disadvantage period", as described in Section~\ref{sec:isolation}. Increasing the warm start length $T_0$ makes this period shorter: indeed, considering the relative reputation trajectory in Figure~\ref{relative_rep_plots} (left), increasing $T_0$ effectively shifts the starting time point to the right. This is why it helps \DynamicGreedy if $T_0$ is small.

\xhdr{First-Mover.}
We turn our attention to the first-mover scenario. Recall that the incumbent firm enters the market and serves as a monopolist until the entrant firm enters at round $X$. We make $X$ large enough, but still much smaller than the time horizon $T$. We find that the incumbent is incentivized to choose \Thompson, in a strong sense:

\begin{finding}\label{find:temp-monopoly}
\textit{Under first-mover, \Thompson is the dominant strategy for the incumbent. This holds across all MAB instances, if $X$ is large enough.
}
\end{finding}

The simulation results for the Heavy-Tail MAB instance are reported in Table~\ref{tab:ht-incum}, for a particular $X=200$. We see that \Thompson is a dominant strategy for the incumbent. Similar tables for the other MAB instances and other values of $X$ are reported in the supplement, with the same conclusion.

\begin{table}[H]
\centering
\begin{tabular}{|c|c|c|c|}
\hline
   & $\TS$  & $\DEG$  & $\DG$ \\ \hline
$\TS$
    & \makecell{\textbf{0.003}$\pm$0.003}
    & \makecell{\textbf{0.083}$\pm$0.02}
    & \makecell{\textbf{0.17}$\pm$0.02} \\\hline
$\DEG$
    & \makecell{\textbf{0.045}$\pm$0.01}
    & \makecell{\textbf{0.25}$\pm$0.02}
    & \makecell{\textbf{0.23}$\pm$0.02} \\\hline
$\DG$
    & \makecell{\textbf{0.12}$\pm$0.02}
    & \makecell{\textbf{0.36}$\pm$0.03}
    & \makecell{\textbf{0.3}$\pm$0.02} \\\hline
\end{tabular}
\caption{\footnotesize Market share of row player (entrant), 200 round head-start, Heavy-Tail Instance}
%\caption{{\bf Temporary monopoly}, with $X=200$ (and $T_0=20$), for the Heavy-Tail MAB instance. Each cell describes the duopoly game between the entrant's algorithm (the row) and the incumbent's algorithm (the column). The cell specifies the entrant's market share (fraction of rounds in which it was chosen) for the rounds in which he was present. We give the average (in bold) and the 95\% confidence interval. NB: smaller average is better for the incumbent.}
\label{tab:ht-incum}
\end{table}

\DynamicGreedy is a weakly dominant strategy for the entrant, for Heavy-Tail instance in Table~\ref{tab:ht-incum} and the Uniform instance, but not for the Needle-in-Haystack instance. We attribute this finding to exploration-disadvantaged property of these two MAB instance, for the same reasons as discussed above.

\begin{finding}\label{find:temp-monopoly-entrant}
\textit{Under first-mover, \DynamicGreedy is a weakly dominant strategy for the entrant for exploration-disadvantaged MAB instances.
}
\end{finding}

\tikzstyle{level 1}=[level distance=3.5cm, sibling distance=4.0cm]
\tikzstyle{level 2}=[level distance=3.5cm, sibling distance=2cm]
\tikzstyle{below} = [align=center]

\begin{figure}[t]
\begin{center}
\begin{tikzpicture}
      \draw[->] (-.5,0) -- (7,0) node[right] {};
      \draw[->] (0,-.5) -- (0,3) node[above] {Better algorithms};
      \draw[scale=0.6,domain=0.7:9.8,smooth,variable=\x,blue, line width=0.3mm] plot ({\x},{4.5 - 0.18 * (\x - 5.25)^2});
     \node[below] at (0.7, -0.22) {\footnotesize monopoly};
     \node[below] at (2.8, -0.2) {\footnotesize incumbent};
     \node[below] at (4.5, -0.65) {\footnotesize simultaneous entry};
     \node[below] at (6.2, -0.2) {\footnotesize entrant};
 \end{tikzpicture}
 \caption{\footnotesize A stylized ``inverted-U relationship" between strength of competition and ``level of innovation".}
\label{fig:inverted-U-expts}
\end{center}
\end{figure}
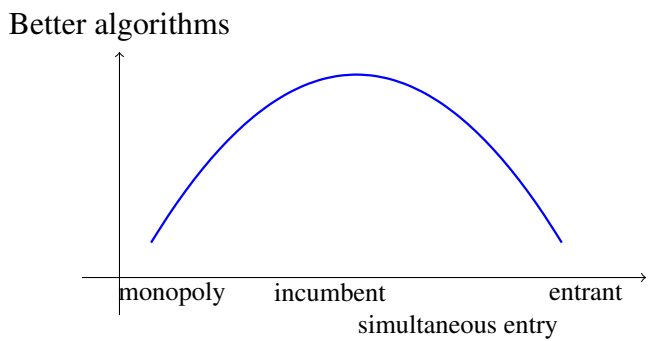

\xhdr{Inverted-U relationship.}
We interpret our findings through the lens of the inverted-U relationship between the ``intensity of competition" and the ``quality of technology". The lowest level of competition is monopoly, when \DynamicGreedy wins out for the trivial reason of tie-breaking. The highest levels are simultaneous entry and ``late entrant". We see that \DynamicGreedy is incentivized for exploration-disadvantaged MAB instances. In fact, incentives for \DynamicGreedy get stronger when the model transitions from simultaneous entry to ``late entrant".%
\footnote{For the Heavy-Tail instance, \DynamicGreedy goes from a weakly dominant strategy to a strictly dominant. For the Uniform instance, \DynamicGreedy goes from a Nash equilibrium strategy to a weakly dominant.}
Finally, the middle level of competition, ``incumbent" in the first-mover regime creates strong incentives for \Thompson. In stylized form, this relationship is captured in Figure~\ref{fig:inverted-U-expts}.%
\footnote{We consider the monopoly scenario for comparison only. We just assume that a monopolist chooses the greedy algorithm, because it is easier to deploy in practice. Implicitly, users have no ``outside option": the service provided is an improvement over not having it (and therefore the monopolist is not incentivized to deploy better learning algorithms). This is plausible with free ad-supported platforms such as Yelp or Google.}

% Transition Duopoly -> LateStart creates stronger incentives for DG.
% In Heavy Tail, DG is strictly dominant (before it was only weakly).
% In Uniform, DG is weakly dominant (before it was only PSNE).

Our intuition for why incumbency creates more incentives for exploration is as follows. During the period in which the incumbent is the only firm in the market, reputation consequences of exploration vanish. Instead, the firm wants to improve its performance as much as possible by the time competition starts. Essentially, the firm only faces a classical explore-exploit trade-off, and chooses algorithms that are best at optimizing this trade-off.

\xhdr{Death spiral effect.}
Further, we investigate the ``death spiral" effect mentioned in the Introduction. Restated in terms of our model, the effect is that one firm attracts new customers at a lower rate than the other, and falls behind in terms of performance because the other firm has more customers to learn from, and this gets worse over time until (almost) all new customers go to the other firm. With this intuition in mind, we define the following:
\begin{definition}\label{def:eeog}
The \emph{effective end of game} (\Eeog), for a particular \MRV and realization table, is \asedit{the earliest round such that all subsequent  agents (if any) choose the same firm.}
%the last round $t$ such that the agents at this and previous round choose different firms.
\end{definition}

\asedit{The terminology is self-explanatory: the game, effectively, ends this round. Note that our definition allows $\Eeog = T$; then, indeed, the game continues till the very last round.}

We interpret low \Eeog as a strong evidence of the ``death spiral" effect.
%\footnote{\gaedit{We provide no theoretical guarantees that this measure is always finite for all instances, but mainly use it as suggestive evidence for ``death spiral" effects.}}
Focusing on the simultaneous entry scenario, we specify the \Eeog values in Table~\ref{fig:eog}. We find that the \Eeog values are indeed small:

\begin{finding}
\textit{
Under simultaneous entry, \Eeog values tend to be much smaller than $T$.
}
\end{finding}

We also see that the \Eeog values tend to increase as the warm start $T_0$ increases. We conjecture this is because larger $T_0$ tends to be more beneficial for a better algorithm (as it tends to follow a better learning curve). Indeed, we know that the \Eeog in this scenario typically occurs when a better algorithm loses, and helping it merely delays the loss.

\begin{table*}[ht]
\centering
\begin{adjustbox}{width=\textwidth,center}
\begin{tabular}{|c|c|c|c||c|c|c||c|c|c|}
  \hline
  & \multicolumn{3}{c||}{Heavy-Tail}
  & \multicolumn{3}{c|}{Needle-in-Haystack}
  & \multicolumn{3}{c|}{Uniform}\\
  \hline
  & $T_0$ = 20 & $T_0$ = 250 & $T_0$ = 500
   & $T_0$ = 20 & $T_0$ = 250 & $T_0$ = 500
  & $T_0$ = 20 & $T_0$ = 250 & $T_0$ = 500 \\
  \hline
\TS vs \DG
  & 68 (0)  & 560 (8.5)  & 610 (86.5)
 & 180 (30)  & 380 (0)  & 550 (6.5)
  &  260 (0)
  &  780 (676.5)
  &  880 (897.5) \\
\hline
  $\TS$ vs $\DEG$
 & 37 (0)  & 430 (0)  & 540 (105)
 & 150 (10)  & 460 (25)  & 780 (705)
 & 230 (0)  & 830 (772)  & 980 (1038) \\ \hline
  $\DG$ vs $\DEG$
 & 340 (110)  & 640 (393)  & 670 (425)
  & 410 (8.5)  & 760 (666)  & 740 (646)
  & 530 (101)  & 990 (1058)  & 1000 (1059) \\ \hline
\end{tabular}
\end{adjustbox}
\caption{\footnotesize {\bf Simultaneous Entry, \Eeog}. Each cell describes a game between two algorithms, call them $\alg[1]$ vs. $\alg[2]$, for a particular value of the warm start $T_0$. Each cell specifies the ``effective end of game" (\Eeog): the average and the median (in brackets).
%For example, the cell in the top left indicates that TS gets on average 64\% of the market when played against DG.
The time horizon is $T=2000$.}
\label{fig:eog}
\end{table*}

\xhdr{Welfare implications.}
We study the effects of competition on consumer welfare: the total reward collected by the users over time. Rather than welfare directly, we find it more lucid to consider
\emph{market regret}:
$ \textstyle T\, \max_a \mu(a) - \sum_{t\in [T]} \mu(a_t), $
where $a_t$ is the arm chosen by agent $t$. This is a standard performance measure in the literature on multi-armed bandits. Note that smaller regret means higher welfare.

We assume that both firms play their respective equilibrium strategies.
% for the corresponding competition level.
As discussed previously, it is
\DynamicGreedy in the monopoly scenario, and
\DynamicGreedy for both firms for simultaneous entry (Finding \ref{find:duopoly}).
For the first-mover scenario, it is \Thompson for the incumbent (Finding \ref{find:temp-monopoly}) and \DynamicGreedy for the entrant (Finding \ref{find:temp-monopoly-entrant}).

\begin{figure}[bth]
\centering
\includegraphics[scale=0.35]{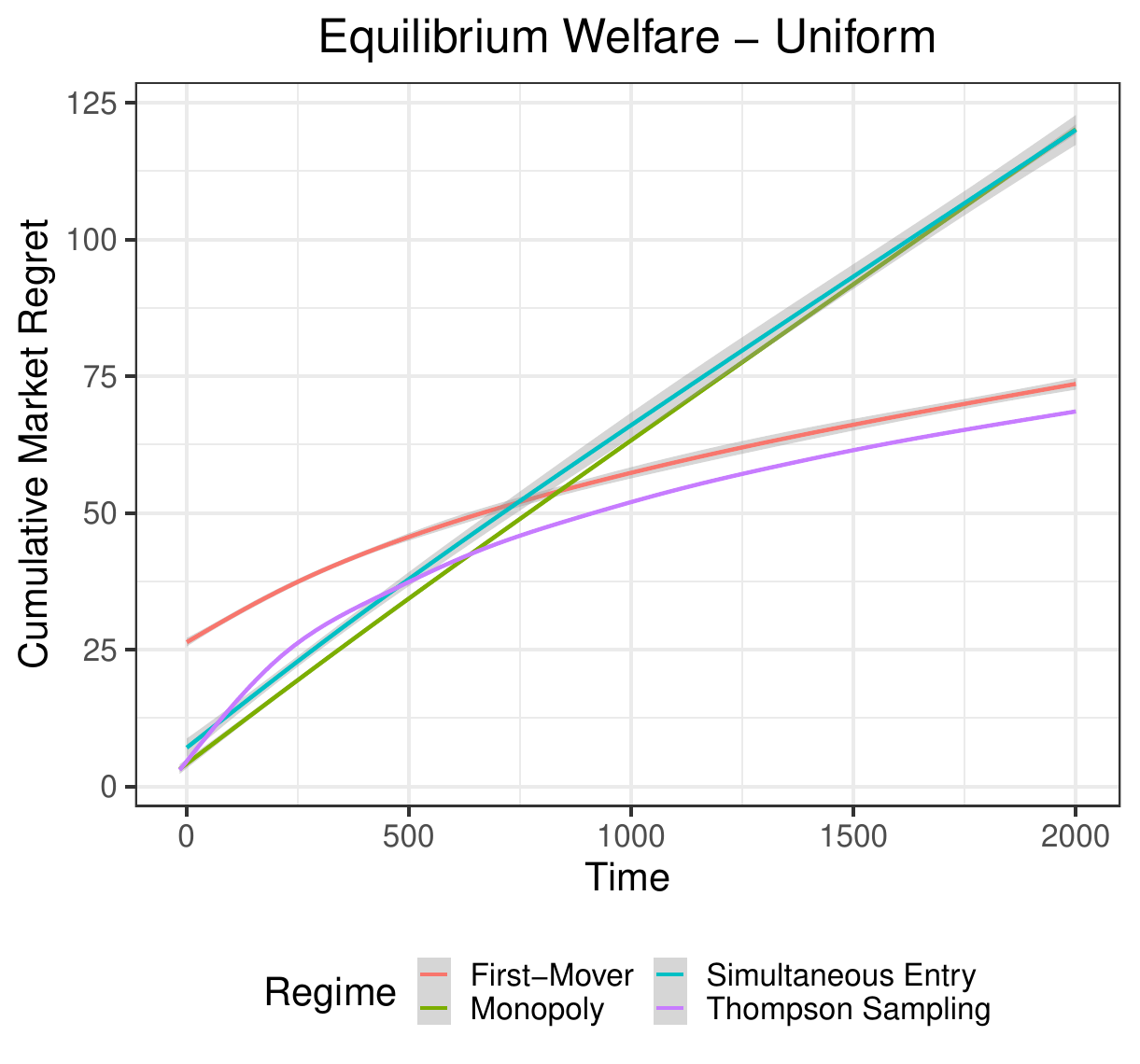}
\includegraphics[scale=0.35]{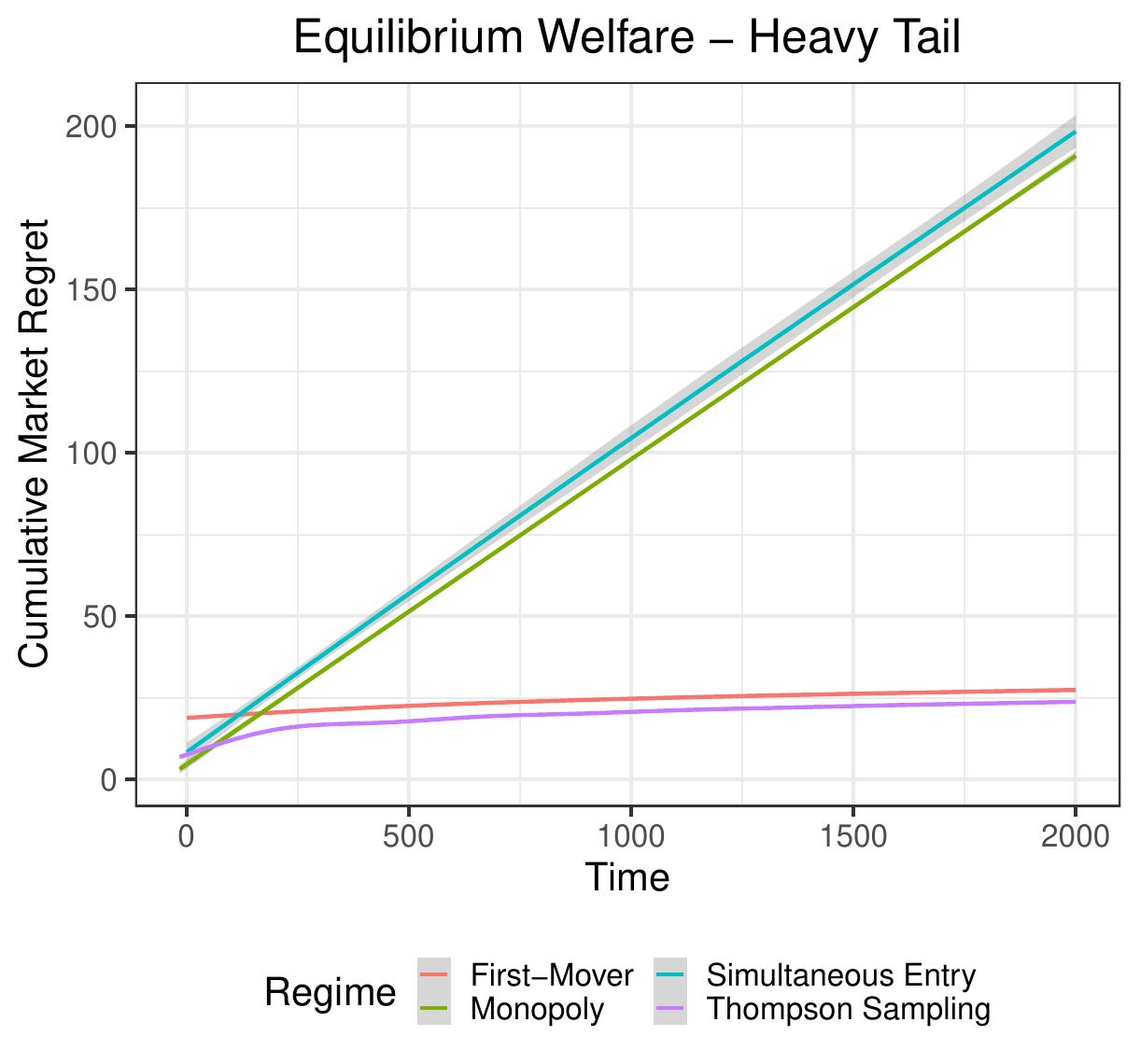}
\caption{\footnotesize Smoothed welfare plots resulting from equilibrium strategies in the different market structures. Note that welfare at $t = 0$ incorporates the regret incurred during the incumbent and warm start periods. The Thompson Sampling trajectory displays the regret incurred by running Thompson Sampling in isolation on the given instances.}
\label{eq_regret}
\end{figure}

Figure \ref{eq_regret} displays the market regret (averaged
  over multiple runs) under different levels of competition.
Consumers are \textit{better off} in the first-mover case than in
the simultaneous entry case. Recall that under first-mover, the incumbent is incentivized to play \Thompson. Moreover, we find that the welfare is close to that of having a single firm for all agents and running \Thompson. We also observe that monopoly and simultaneous entry achieve similar welfare.
%with monopoly being marginally better than duopoly.

\begin{finding}\label{find:welfare}
\textit{In equilibrium, consumer welfare is (a) highest under first-mover, (b) similar for monopoly and simultaneous entry.
%monopoly being marginally better leading to marginally better welfare than duopoly.
}
\end{finding}

Finding~\ref{find:welfare}(b) is interesting because, in equilibrium, both firms play \DynamicGreedy in both settings, and one might conjecture that the welfare should increase with the number of firms playing \DynamicGreedy. Indeed, one run of \DynamicGreedy may get stuck on a bad arm. However, two firms independently playing \DynamicGreedy are less likely to get stuck simultaneously. If one firm gets stuck and the other does not, then the latter should attract most agents, leading to improved welfare.

%\gadelete{
%\begin{figure}
%\centering
%\includegraphics[scale=0.3]{ec19paper/figures/eeog_vs_num_firms_ht}
%\includegraphics[scale=0.3]{ec19paper/figures/ht_many_firm_welfare}\\
%\caption{Average welfare and \Eeog as we increase the number of firms playing \DynamicGreedy}
%\label{many_firm_welfare}
%\end{figure}
%
%To study this phenomenon further, we go beyond the duopoly setting to more than two firms playing \DynamicGreedy (and starting at the same time). Figure~\ref{many_firm_welfare} reports the average welfare
%across these simulations. Welfare not only does not get better, \textit{but is weakly worse} as we increase the number of firms.
%
%\begin{finding}
%\textit{When all firms deploy \DynamicGreedy, and start at the same time, welfare is weakly decreasing as the number of firms increases.}
%\end{finding}
%
%
%We track the average \Eeog in each of the
%simulations and notice that it \textit{increases} with the number of firms.
%This observation also runs counter of the intuition that with more firms running \DynamicGreedy, one of them is more likely to ``get lucky" and take over the market (which would cause \Eeog to \emph{decrease} with the number of firms).
%}

%% file: ec19paper/content/barriers.tex
\subsection{Data as a Barrier to Entry}\label{sec:barriers}

%We explore what factors drive the large market share for the incumbent in the temporary monopoly model.

In the first-mover regime, the incumbent can explore without incurring immediate reputational costs, and build up a high reputation before the entrant appears. Thus, the early entry gives the incumbent both a \textit{data} advantage and a \textit{reputational} advantage. We explore which of the two factors is more significant.  Our findings provide a quantitative insight into the role of the classic ``first mover advantage" phenomenon in the digital economy.

%Second, our findings provide a viewpoint in terms of thinking about the role that data can play as a competitive advantage.

For a more succinct terminology, recall that the incumbent enjoys an extended warm start of $X+T_0$ rounds. Call the first $X$ of these rounds the \emph{monopoly period} (and the rest is the proper ``warm start"). The rounds when both firms are competing for customers are called \emph{competition period.}

We run two additional experiments to isolate the effects of the two
advantages mentioned above. The \emph{data-advantage experiment} focuses on the data advantage by, essentially, erasing the reputation advantage. Namely, the data from the monopoly period is not used in the computation of the incumbent's reputation score. Likewise, the \emph{reputation-advantage experiment} erases the data advantage and focuses on the reputation advantage: namely, the incumbent's algorithm `forgets' the data gathered during the monopoly period.

We find that either data or reputational advantage alone gives a substantial boost to the incumbent, compared to simultaneous entry duopoly. The results for the Heavy-Tail instance are presented in Table~\ref{barrier_exp}, in the same structure as Table~\ref{tab:ht-incum}. For the other two instances, the results are qualitatively similar.

\begin{table*}[h]
\centering
\begin{tabular}{|c|c|c|c||c|c|c|}
\hline
  & \multicolumn{3}{c||}{Reputation advantage (only)}
  & \multicolumn{3}{c|}{Data advantage (only)}\\
\hline
& $\TS$  & $\DEG$  & $\DG$
& $\TS$  & $\DEG$  & $\DG$
\\\hline
$\TS$
    & \makecell{\textbf{0.021}$\pm$0.009}
    & \makecell{\textbf{0.16}$\pm$0.02}
    & \makecell{\textbf{0.21} $\pm$0.02}
    & \makecell{\textbf{0.0096}$\pm$0.006}
    & \makecell{\textbf{0.11}$\pm$0.02}
    & \makecell{\textbf{0.18}$\pm$0.02}
    \\ \hline
$\DEG$
    & \makecell{\textbf{0.26}$\pm$0.03}
    & \makecell{\textbf{0.3}$\pm$0.02}
    & \makecell{\textbf{0.26}$\pm$0.02}
    & \makecell{\textbf{0.073}$\pm$0.01}
    & \makecell{\textbf{0.29}$\pm$0.02}
    & \makecell{\textbf{0.25}$\pm$0.02}
    \\ \hline
$\DG$
    & \makecell{\textbf{0.34}$\pm$0.03}
    & \makecell{\textbf{0.4}$\pm$0.03}
    & \makecell{\textbf{0.33}$\pm$0.02}
    & \makecell{\textbf{0.15}$\pm$0.02}
    & \makecell{\textbf{0.39}$\pm$0.03}
    & \makecell{\textbf{0.33}$\pm$0.02}
    \\\hline
\end{tabular}
\caption{\footnotesize Data advantage vs. reputation advantage experiment, on Heavy-Tail MAB instance. Each cell describes the duopoly game between the entrant's algorithm (the {\bf row}) and the incumbent's algorithm (the {\bf column}). The cell specifies the entrant's market share for the rounds in which hit was present: the average (in bold) and the 95\% confidence interval. NB: smaller average is better for the incumbent.}
\label{barrier_exp}
\end{table*}

We can quantitatively define the data (resp., reputation) advantage as the incumbent's market share in the competition period in the data-advantage (resp., reputation advantage) experiment, minus the said share under simultaneous entry duopoly, for the same pair of algorithms and the same problem instance. In this language, our findings are as follows.

\begin{finding}\label{barrier-find}
\textit{
(a) Data advantage and reputation advantage alone are large, across all algorithms and MAB instances. (b) The data advantage is larger than the reputation advantage when the incumbent chooses \Thompson. (c) The two advantages are similar in magnitude when the incumbent chooses \DynamicEpsGreedy or \DynamicGreedy.
}
\end{finding}

Our intuition for Finding~\ref{barrier-find}(b) is as follows. Suppose the incumbent switches from \DynamicGreedy to \Thompson. This switch allows the incumbent to explore actions more efficiently -- collect better data in the same number of rounds -- and therefore should benefit the data advantage. However, the same switch increases the reputation cost of exploration in the short run, which could weaken the reputation advantage.

%% file: ec19paper/content/non_greedy_choice.tex
\subsection{Non-deterministic choice model (\HardMaxRandom)}\label{sec:non_greedy}

Let us consider an extension in which the agents' response function \eqref{eq:model-f} is no longer deterministic. We focus on \HardMaxRandom model, where each agent selects between the firms uniformly with probability $\eps\in (0,1)$, and takes the firm with the higher reputation score with the remaining probability.

One can view \HardMaxRandom as a version of ``warm start", where a firm receives some customers without competition, but these customers are dispersed throughout the game. The expected duration of this ``dispersed warm start" is $\eps T$. If this quantity is large enough, we expect better algorithms to reach their long-term performance and prevail in competition. We confirm this intuition; we also find that this effect is negligible for smaller (but relevant) values of $\eps$ or $T$.

\OMIT{In Section \ref{sec:competition} we show that in the permanent duopoly
case exploration can lead to a death spiral which eventually starves
the firm of agents. However, giving one firm a small head start or
enough free agents via the warm start incentivizes it to play \TS
since it could recover the reputation costs it incurred from early
exploration. An interpretation of \HMR~ is that instead of
concentrating the free agents as arriving in the beginning of the game
instead they are dispersed throughout the game.}

\begin{finding}\label{find:non_greedy_choice}
\textit{\Thompson is weakly dominant under \HardMaxRandom, if and only if $\eps T$ is sufficiently large. Moreover, \HardMaxRandom leads to lower variance in market share, compared to \HardMax.}
\end{finding}

\footnotesize
\begin{table*}[t]
\centering
\begin{adjustbox}{width=\textwidth,center}
\begin{tabular}{|c|c|c|c||c|c|c|}
  \hline
  & \multicolumn{3}{c||}{Heavy-Tail (\HMR with $\eps=.1$)}
  & \multicolumn{3}{c|}{Heavy-Tail (\HM)}\\
  \hline
  & \TS vs \DG & \TS vs \DEG  & \DG vs \DEG
 & \TS vs \DG & \TS vs \DEG  & \DG vs \DEG  \\
  \hline
$T = 2000$
 & \makecell{ \textbf{0.43} $\pm$ 0.02 \\Var: 0.15 }
  & \makecell{ \textbf{0.44} $\pm$ 0.02 \\Var: 0.15 }
  & \makecell{ \textbf{0.6} $\pm$ 0.02 \\Var: 0.1 }
 &  \makecell{ \textbf{0.29} $\pm$ 0.03 \\Var: 0.2 }
  & \makecell{ \textbf{0.28} $\pm$ 0.03 \\Var: 0.19 }
  & \makecell{ \textbf{0.63} $\pm$ 0.03 \\Var: 0.18 }
    \\
\hline
  $T= 5000$
   & \makecell{ \textbf{0.66} $\pm$ 0.01 \\Var: 0.056 }
  & \makecell{ \textbf{0.59} $\pm$ 0.02 \\Var: 0.092 }
  & \makecell{ \textbf{0.56} $\pm$ 0.02 \\Var: 0.098 }
 & \makecell{ \textbf{0.29} $\pm$ 0.03 \\Var: 0.2 }
 & \makecell{ \textbf{0.29} $\pm$ 0.03 \\Var: 0.2 }
 & \makecell{ \textbf{0.62} $\pm$ 0.03 \\Var: 0.19 }
 \\
  \hline
  $T = 10000$
  & \makecell{ \textbf{0.76} $\pm$ 0.01 \\Var: 0.026 }
 & \makecell{ \textbf{0.67} $\pm$ 0.02 \\Var: 0.067 }
 & \makecell{ \textbf{0.52} $\pm$ 0.02 \\Var: 0.11 }
  & \makecell{ \textbf{0.3} $\pm$ 0.03 \\Var: 0.21 }
  & \makecell{ \textbf{0.3} $\pm$ 0.03 \\Var: 0.2 }
  & \makecell{ \textbf{0.6} $\pm$ 0.03 \\Var: 0.2 }
  \\
   \hline
\end{tabular}
\end{adjustbox}
\caption{\footnotesize \HardMax (\HM) and \HardMaxRandom (\HMR) choice models on the Heavy-Tail MAB instance. Each cell describes the market shares in a game between two algorithms, call them $\alg[1]$ vs. $\alg[2]$, at a particular value of $t$. Line 1 in the cell is the market share of $\alg[1]$: the average (in bold) and the 95\% confidence band.
%For example, the cell in the top left indicates that TS gets on average 64\% of the market when played against DG.
Line 2 specifies the variance of the market shares across the simulations. The results reported here are with $T_0 = 20$.}
\label{tab:non_greedy_table}
\end{table*}

\normalsize

\footnotesize
\begin{table*}[t]
\centering
\begin{adjustbox}{width=\textwidth,center}
\begin{tabular}{|c|c|c|c||c|c|c|}
  \hline
  & \multicolumn{3}{c||}{Uniform (\HMR  with $\eps=.1$)}
  & \multicolumn{3}{c|}{Needle-In-Haystack (\HMR  with $\eps=.1$)}\\
  \hline
  & \TS vs \DG & \TS vs \DEG  & \DG vs \DEG
 & \TS vs \DG & \TS vs \DEG  & \DG vs \DEG  \\
 \hline
$T = 2000$
 & \makecell{ \textbf{0.42} $\pm$ 0.02 \\Var: 0.13 }
 & \makecell{ \textbf{0.45} $\pm$ 0.02 \\Var: 0.13 }
 & \makecell{ \textbf{0.49} $\pm$ 0.02 \\Var: 0.093 }
  & \makecell{  \textbf{0.55} $\pm$ 0.02 \\Var: 0.15 }
  & \makecell{  \textbf{0.61} $\pm$ 0.02 \\Var: 0.13 }
  & \makecell{  \textbf{0.46} $\pm$ 0.02 \\Var: 0.12 }
    \\
\hline
  $T= 5000$
 & \makecell{ \textbf{0.48} $\pm$ 0.02 \\Var: 0.089 }
 & \makecell{ \textbf{0.53} $\pm$ 0.02 \\Var: 0.098 }
 & \makecell{ \textbf{0.46} $\pm$ 0.02 \\Var: 0.072 }
 & \makecell{  \textbf{0.56} $\pm$ 0.02 \\Var: 0.13 }
 & \makecell{  \textbf{0.63} $\pm$ 0.02 \\Var: 0.12 }
 & \makecell{  \textbf{0.43} $\pm$ 0.02 \\Var: 0.11 }
 \\
  \hline
  $T = 10000$
& \makecell{ \textbf{0.54} $\pm$ 0.01 \\Var: 0.055 }
& \makecell{  \textbf{0.6} $\pm$ 0.02 \\Var: 0.073 }
& \makecell{  \textbf{0.44} $\pm$ 0.02 \\Var: 0.064 }
  & \makecell{ \textbf{0.58} $\pm$ 0.02 \\Var: 0.083 }
  & \makecell{ \textbf{0.65} $\pm$ 0.02 \\Var: 0.096 }
  & \makecell{ \textbf{0.4} $\pm$ 0.02 \\Var: 0.1 }
  \\
   \hline
\end{tabular}
\end{adjustbox}
\caption{\footnotesize \HardMaxRandom (\HMR) choice model for Uniform and Needle-In-Haystack MAB instances.}
%Same semantics as in Table \ref{tab:non_greedy_table}.
\label{tab:additional_results}
\end{table*}
\normalsize

Table \ref{tab:non_greedy_table} shows the average market shares under
\HardMax vs \HardMaxRandom. In contrast to what happens under \HardMax,
  \TS becomes weakly dominant under \HardMaxRandom, as $T$ gets
  sufficiently large. These findings hold across all problem
instances, see Table \ref{tab:additional_results} (with the same semantics as in Table \ref{tab:non_greedy_table}).

\OMIT{\footnote{The results here are pulled using different \MRV
  and realizations than the results pulled previously (due to the
  larger $T$). However, they are drawn from the same prior instances
  and so qualitatively are the same but the quantitative results are
  not directly comparable to those from the previous
  section.\swcomment{not sure we need this}}}

\normalsize
\OMIT{The intuition for why \TS
should become the dominant strategy eventually is simple. The
consistent stream of random agents ensures that each principal is
chosen at least $\Omega(\eps t)$ times at every time step $t$. As
a result, each algorithm should eventually converge to its asymptotic
performance in isolation.}

\OMIT{Finding \ref{find:non_greedy_choice} implies that we can re-interpret the inverted-U findings from before in terms of the number of agents that a firm receives without having to worry about incentives. In the extreme when the firm gets all agents for free as in the monopoly case then it is incentivized to play \DG. When it only gets some of the agents for free, either via a large warm start, a temporary monopoly, or non-deterministic choice, then \TS is incentivized. However, if the number of free agents gets small enough then \DG is incentivized as in the permanent duopoly analysis from before.\swcomment{I don't understand this paragraph; I think it's too vague. Perhaps remove?}}

However, it takes a significant amount of randomness and a relatively large time horizon for this effect to take place. Even with $T = 10000$ and $\eps = 0.1$ we see that \DEG still outperforms \DG on the Heavy-Tail MAB instance as well as that \TS only starts to become weakly dominant at $T = 10000$ for the Uniform MAB instance.

\OMIT{Table \ref{tab:non_greedy_table} also shows that another difference between the two choice rules is that \HMR leads to lower variance in market shares across simulations compared to \HM.}

%% file: ec19paper/content/revisited.tex
\subsection{Performance in Isolation, Revisited}\label{sec:revisited}

We saw in Section~\ref{sec:competition} that mean reputation trajectories do not suffice to explain the outcomes under competition. Let us provide more evidence and intuition for this.

Mean reputation trajectories are so natural that one is tempted to conjecture that they determine the outcomes under competition. More specifically:
\begin{conjecture}\label{conj:mean-trajectories}
If one algorithm's mean reputation trajectory lies above another, perhaps after some initial time interval (\eg as in Figure~\ref{prelim_means}), then the first algorithm prevails under competition, for a sufficiently large warm start $T_0$.
\end{conjecture}

However, we find a more nuanced picture. For example, in Figure \ref{fig:market_share} we see that $\DynamicGreedy$ attains a larger market share than $\DynamicEpsGreedy$ even for large warm starts. We find that this also holds for $K = 3$ arms and longer time horizons, see the supplement for more details. We conclude that Conjecture~\ref{conj:mean-trajectories} is false:

\begin{figure}[ht]
\centering
\includegraphics[scale=0.35]{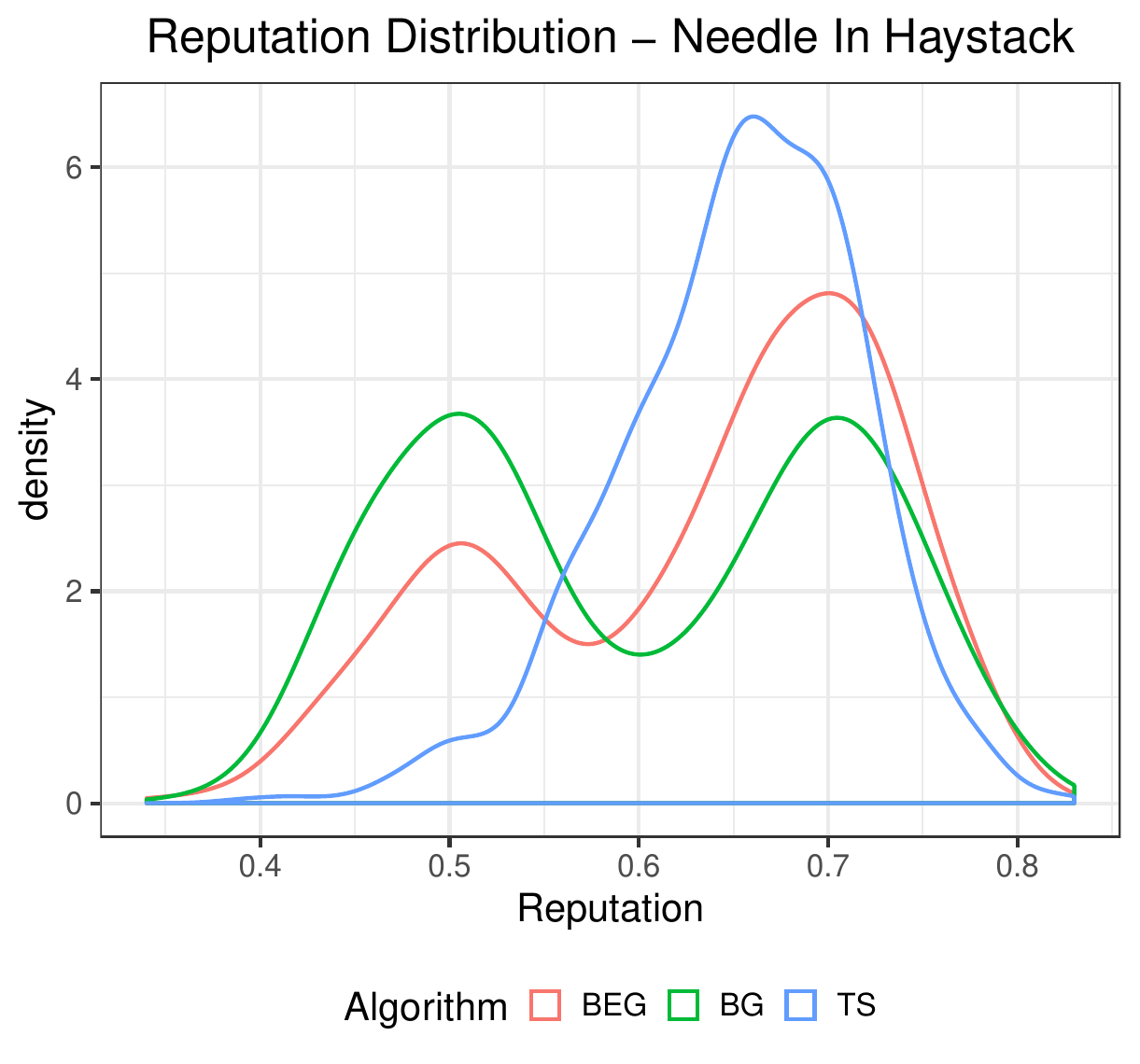}
\includegraphics[scale=0.35]{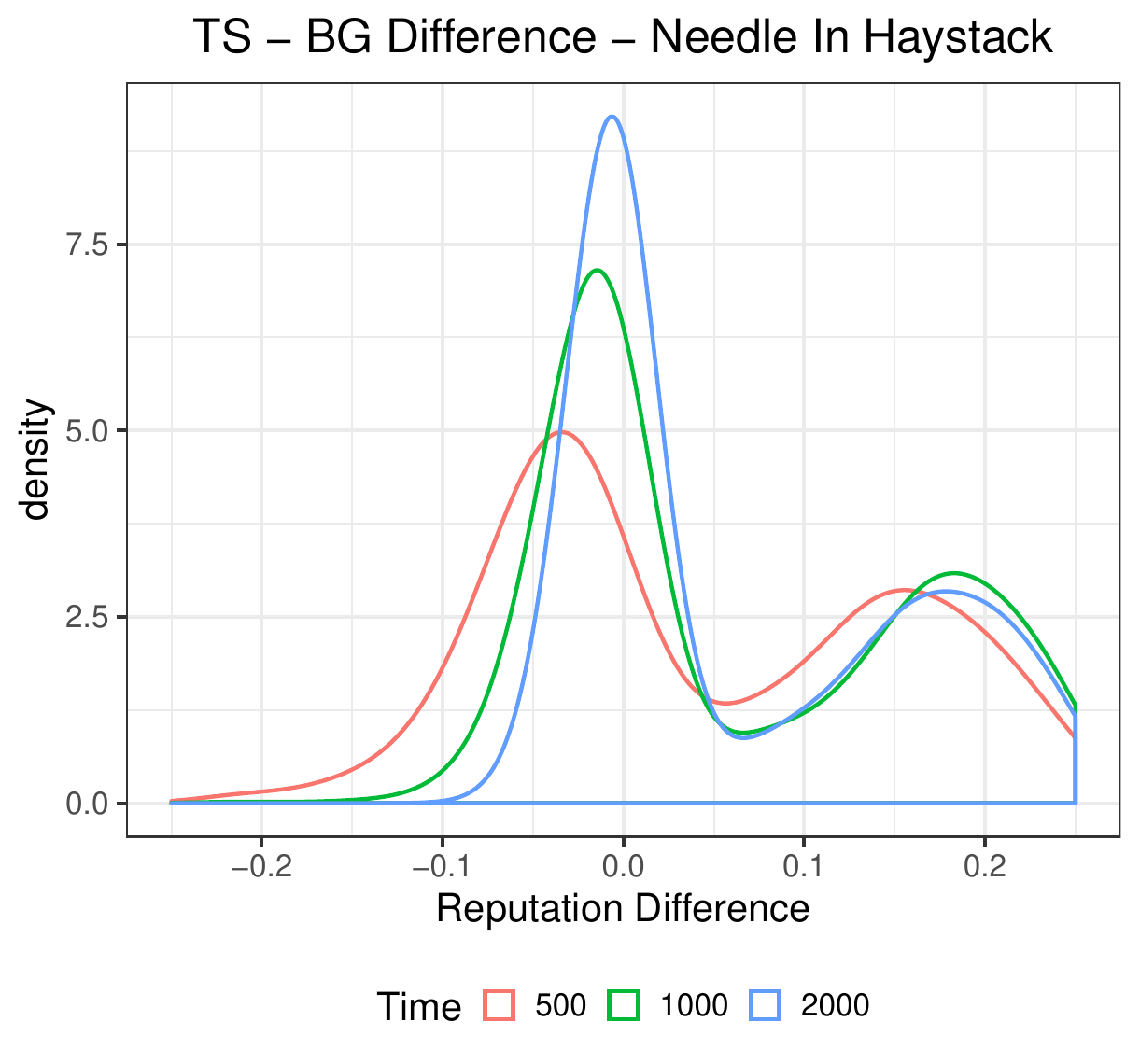}
\caption{\footnotesize Needle-in-Haystack: reputation scores at $t=500$ (left), reputation difference $\Thompson - \DynamicGreedy$ (right). Both are smoothed using a kernel density estimate.}
\label{fig:rep_dist_nih}
%\caption*{\tiny{The plots contain a kernel density estimate of the reputation distribution at $t = 500$}}
\end{figure}

\begin{finding}
\textit{
Mean reputation trajectories do not explain the outcomes under competition.}
\end{finding}

To see what could go wrong with Conjecture~\ref{conj:mean-trajectories}, consider how an algorithm's reputation score is distributed at a particular time. That is, consider the empirical distribution of this score over different \MRVs.%
\footnote{Recall that each \MRV in our experimental setup comes with a separate realization table.} For concreteness, consider the Needle-in-Haystack instance at time $t=500$, plotted in Figure~\ref{fig:rep_dist_nih} (left). (The other MAB instances lead to a similar intuition.)

We see that the ``naive" algorithms $\DynamicGreedy$ and $\DynamicEpsGreedy$ have a bi-modal reputation distribution, whereas $\Thompson$ does not. The reason is that for this MAB instance, $\DynamicGreedy$ either finds the best arm and sticks to it, or gets stuck on the bad arms. In the former case \DynamicGreedy does slightly better than $\Thompson$, and in the latter case it does substantially worse. However, the mean reputation trajectory fails to capture this complexity since it takes average over different \MRVs. This is inadequate for explaining the outcome of the duopoly game, given that the latter is determined by a comparison between the firm's reputation scores.

To further this intuition, consider the difference in reputation scores (\emph{reputation difference}) between \Thompson and \DynamicGreedy on a particular \MRV. Let's plot the empirical distribution of the reputation difference (over the \MRVs) at a particular time point. Figure~\ref{fig:rep_dist_nih} (right) shows such plots for several time points. We observe that the distribution is skewed to the right, precisely due to the fact that $\DynamicGreedy$ either does slightly better than $\Thompson$ or does substantially worse. So, the mean is not a good measure of the central tendency of this distribution.

%% file: content/conclusion.tex
\section{Conclusions}\label{sec:conclusion}

We study the tension between exploration and competition. We consider a stylized duopoly model in which two firms face the same multi-armed bandit problem and compete for a stream of users. A firm makes progress on its learning problem only if it attracts users.
%\asdelete{We investigate two variants: an analytical variant where users do not observe any performance signals and choose between firms according to their Bayesian-expected rewards, and numerical simulations where a reputation score is observed for each firm, based on the average reward of its recent users.  In both variants,}
We find that firms are incentivized to adopt a ``greedy algorithm" which does no purposeful exploration and leads to welfare losses for users. We then consider two relaxations of competition: we soften users' decision rule and give one firm a first-mover advantage. Both relaxations induce firms to adopt ``better" bandit algorithms, which benefits user welfare.

Our results have two economic interpretations. The first is that they can be framed in terms of the classic inverted-U relationship between innovation and competition, where \innovation refers to the adoption of better bandit algorithms. Unlike other models in the literature, what prevents innovation is not its direct costs, but the short-term reputation consequences of exploration. The second interpretation concerns the role of data in the digital economy. We find that even a small initial disparity in data or reputation gets amplified under competition to a very substantial difference in the eventual market share. Thus, we endogenously obtain ``network effects" without explicitly baking them into the model, and elucidate the role of data as a barrier to entry.

With this paper as a departure point, there are several exciting directions to explore.
First, consider more than two competing firms. One could potentially interpret all but one firms as a "joint" bandit algorithm and thus reduce the situation to the two-firm case studied in this paper, but it is unclear how to characterize the behavior of this "joint algorithm" so that it nicely plugs back into our analyses.
Second, the firms could acquire initial traffic by offering payments / discounts to early users, and more generally choose prices as a part of their overall strategy in the competition game. While strategic price setting is not included in our model, we do study the \emph{effects} of acquiring initial traffic (by varying the warm-start length $T_0$ in the  \ExptsModel).
%
%First, when the firms can set prices, they may be able to compensate early users for exploration, and potentially prevent the ``death spiral" effects. \gaedit{Furthermore, allowing firms to set prices may result in interesting price dynamics (e.g., low prices during exploration and higher prices during exploitation) that would lead to some ambiguity regarding the optimal market structure for consumer welfare.} (Our paper zeroes in on competition between free, ad-supported platforms that primarily compete on quality.)
%
%\gaedit{Second, our model assumes perfect alignment between the utility of the principals and the agents. This can arise even if the principal has no explicit incentive to deviate from maximizing agent's utility as agents may have time-inconsistent preferences \citep{kleinberg2024challenge}. An interesting direction is to consider both long-lived agents and how the platform's equilibrium algorithm choice depends on the magnitude of agents' time-inconsistency.}
%
Third, horizontally differentiated user preferences may help explain how competition may encourage specialization, \ie how the firms may \emph{learn to specialize} under competition.
Fourth, while we focus on a stationary world, another well-motivated scenario is that
%is ``continuous learning", when 
exploration continuously counteracts change. The economic story here would be about competition between relatively mature firms.
%
%\footnote{One difficulty is that the ``bandit model" becomes considerably more complicated: there are many reasonable ways to deal with a continuously changing world, starting from \citet{DynamicMAB-colt08}, and the distinctions between better and worse algorithms are not as clear and well-established.}
%
Finally, in some applications agents' immediate rewards are not quite aligned with their long-term satisfaction, and the firms' algorithms may strive to optimize the latter. However, such algorithms are much more difficult to design ``in isolation", and appear difficult to analyze in competition.

%\item \emph{Real-life data.}
%Numerical experiments based on real-life datasets would arguably be more realistic. To deal with real-life datasets, bandit algorithms would probably need to accommodate change over time and \emph{contexts} (auxiliary signals available before each round).

%\end{itemize}

%\noindent One difficulty with the last two directions is that the bandit ``side" of the model becomes considerably more complicated, both in terms of the rewards and in terms of the algorithms. Indeed, there are many reasonable ways to model a changing world (resp., dependence on  contexts), and several substantially different algorithmic approaches to deal with these models. Consequently, the distinctions between better and worse algorithms are not as clear and well-established.

%% file: content/app-tables.tex
\begin{longtable}{|p{7cm}|p{7cm}|}
\caption{Definition Table} \label{tab:definition-table} \\
\hline
\textbf{Term} & \textbf{Definition Link} \\
\hline
Deviates from another algorithm & Definition \ref{def:deviates} \\
\hline
$alg_1$ BIR-dominates $alg_2$ & Definition \ref{def:bir_dominate} \\
\hline
SoftMax Response & Definition \ref{def:SoftMax} \\
\hline
$alg_1$ weakly BIR-dominates $alg_2$ & Definition \ref{def:weak_bir_dominate} \\
\hline
Exploration-Disadvantaged Period & Definition \ref{def:exploration_disadvantage_period} \\
\hline
Effective End of Game (\Eeog) & Definition \ref{def:eeog} \\
\hline
\end{longtable}

\begin{longtable}{|c|p{12cm}|}
\caption{Notation Table} \label{tab:notation-table} \\
\hline
\textbf{Notation} & \textbf{Description} \\
\hline
\( n \) & Local step or round number in a multi-armed bandit problem. \\
\hline
\( t \) & Global round number, indicating the order in which users arrive. \\
\hline
\( T \) & Time horizon, i.e., the total number of global rounds. \\
\hline
\( \text{EST}_i(t) \) & Reward estimate for principal \( i \) at global round \( t \). \\
\hline
\( \Delta_t \) & Difference between reward estimates of principal 1 and principal 2, \( \Delta_t = \text{EST}_1(t) - \text{EST}_2(t) \). \\
\hline
\( \text{PMR}_i(t) \) & Posterior Mean Reward for principal \( i \) at global round \( t \). \\
\hline
\( \text{BIR}_i(n) \) & Bayesian Instantaneous Regret for principal \( i \) at local step \( n \). \\
\hline
\( f_{resp}(x) \) & Response function that maps the difference in reward estimates to the probability of choosing a principal. \\
\hline
\( p_t \) & Probability of choosing principal 1 at global round \( t \), defined as \( p_t = f_{resp}(\Delta_t) \). \\
\hline
\( \epsilon_0 \) & Baseline selection probability for random agents in the HardMax\&Random response function model. \\
\hline
\( \delta_0, c_0, c'_0 \) & Constants used to define the smoothness of the SoftMax response function around 0. \\
\hline
\( \text{BReg}(T) \) & Bayesian regret over time horizon \( T \), defined as the expected cumulative regret over \( T \) rounds. \\
\hline
\( \text{reg}_i(n) \) & Frequentist instantaneous regret of algorithm \( i \) at step \( n \). \\
\hline
\( \mu_a \) & Mean reward of arm \( a \). \\
\hline
\( \text{rew}_i(n) \) & Reward observed by principal \( i \) at local step \( n \). \\
\hline
\( T_0 \) & Warm start period, during which the principal operates without competition before the entrant arrives. \\
\hline
\( M \) & Size of the sliding window for calculating reputation scores. \\
\hline
\end{longtable}

%% file: content/sec-bg.tex
We present self-contained background on multi-armed bandits (MAB), to make the paper accessible to researchers who are not experts on MAB. More details can be found in books
\citep{Bubeck-survey12,slivkins-MABbook,LS19bandit-book}.

We focus on three algorithm classes, as in Section~\ref{sec:sim}:

\begin{itemize}
\item \emph{Greedy algorithms} that strive to maximize the reward for the next round given the available information. Thus, they always ``exploit" and never explicitly ``explore".
    %\asmargincomment{copied the list from the intro.}

\item \emph{Exploration-separating algorithms}
 that separate exploration and exploitation: essentially, each round is dedicated to one and completely ignores the other.

\item \emph{Adaptive-exploration} algorithms that combine exploration and exploitation, and gradually sway the exploration choices towards more promising alternatives.
\end{itemize}

Below we discuss which algorithms are better than others, and what does it \emph{mean} for one bandit algorithm to be better than another. This is a rather subtle issue, because some algorithms may be better for some problem instances and/or time intervals, and worse for some others. In particular, ``better" algorithms are better in the long run, but could be worse initially.

While we list precise upper and lower bounds on the regret rates, the main goal is to illustrate how the three algorithm classes are separated from one another; the exact results are not essential for this paper. For ease of presentation, we use standard asymptotic notation from computer science: $O(f(t))$ and $\Omega(f(t))$ means at most (resp., at least) $f(n)$, up to constant factors, starting from large enough $t$. Likewise, $\tildeO(f(t))$ notation suppresses the $\polylog(t)$ factors.

\xhdr{Fundamentals.}
We are concerned with the following problem. There are $T$ rounds and $K$ \emph{arms} to choose from. In each round $t\in [T]$, the algorithm chooses an arm and receives a reward $r_t\in[0,1]$ for this arm, drawn from a fixed but unknown distribution.%
\footnote{All ``negative" results (\ie lower bounds on regret) assume reward distributions with constant variance.}
The algorithm's goal is to maximize the total reward.

A standard performance measure is \emph{regret}, defined as the difference in the total expected reward between the algorithm and the best arm. In a formula, regret is
    $T\cdot \max_{\text{arms $a$}}\mu_a
    -  \E\sbr{ \sum_{t\in[T]} r_t }$,
where $\mu_a$ is the mean reward of arm $a$.
Normalized by the best arm, regret allows to compare algorithms across different problem instances.
The primary concern is the asymptotic growth rate of regret as a function of $T$.

The three classes of algorithms perform very differently in terms of regret: adaptive-exploration algorithms are by far the best, greedy algorithms are by far the worst, and exploration-separating ones are in the middle. Adaptive-exploration algorithms achieve optimal regret rates:
    $\tildeO(\sqrt{KT})$
for all problem instances, and simultaneously a vastly improved regret rate of
    $O(\tfrac{K}{\Delta}\log T)$
for all problem instances with $\text{gap}\geq \Delta$ (``easy" instances), without knowing the $\Delta$ in advance
\citep{Lai-Robbins-85,bandits-ucb1,bandits-exp3}.%
\footnote{The \emph{gap} is the difference in mean reward between the best arm and the second-best arm.}
Exploration-separating algorithms can only achieve regret $\tildeO(T^{2/3})$ across all problem instances. They can achieve the ``gap-dependent" regret rate stated above, but \emph{only} if they know the $\Delta$ in advance, and with terrible regret $\Omega(\Delta T)$ for some other problem instances \citep{MechMAB-ec09}. Finally, the greedy algorithm is terrible on a wide variety of problem instances, in the sense that with constant probability it fails to try the best arm even once, and therefore suffers regret $\Omega(T)$
\citep[see Chapter 11.2 in][]{slivkins-MABbook}.

The optimal regret rates are achieved by several adaptive-exploration  algorithms, of which the most known are
Thompson Sampling \citep{Thompson-1933,TS-survey-FTML18},%
\footnote{While Thompson Sampling dates back to 1933 and is probably the best-known bandit algorithm, its regret has not been understood until recently \citep{Shipra-colt12,Kaufmann-alt12,Shipra-aistats13}.}
UCB1 \citep{bandits-ucb1},
and Successive Elimination \citep{EvenDar-icml06}.%
\footnote{A substantial follow-up work on more ``refined" regret rates is not as relevant to this paper.}
These algorithms are very simple to describe. Focus on one round and consider the posterior distribution and/or the confidence interval on each arm's mean reward. Thompson Sampling draws a sample (``score") from each arm's posterior distribution, and picks an arm with the largest score. UCB1 picks an arm with the largest upper confidence bound. Successive Elimination eliminates an arm once it is worse than some other arm with high confidence, and chooses uniformly among the remaining arms.

Exploration-separating algorithms completely separate exploration and exploitation. Ahead of time, each round is either selected for exploration, in which case  the distribution over arms does not depend on the observed data, or it is assigned to exploitation, in which case the data from this round is discarded. The simplest approach, called \emph{Explore-First}, explores uniformly for a predetermined number of rounds, then chooses one arm for ``exploitation" and uses it from then on. A more refined approach, called \emph{Epsilon-Greedy}, explores uniformly in each round with a predetermined probability, and ``exploits" with the remaining probability. Both algorithms, and the associated $\tildeO(T^{2/3})$ regret bounds, have been ``folklore knowledge" for decades. The general definition and lower bounds trace back to \citet{MechMAB-ec09}.%
\footnote{\citet{MechMAB-ec09} consider a closely related, but technically different setting, which can be easily ``translated" into ours (either as a corollary or as another application of the same proof technique).}

% add: this is all more general

\xhdr{Advanced aspects.}
Switching from ``greedy" to ``exploration-separating" to ``adaptive-exploration" algorithms involves substantial adoption costs in infrastructure and personnel training \citep{DS-arxiv}. Inserting exploration into a complex decision-making pipeline necessitates a substantial awareness of the technology and a certain change in mindset, as well as an infrastructure to collect and analyze the data. Adaptive exploration requires the said infrastructure to propagate the data analysis back to the ``front-end" where the decisions are made, and do it on a sufficiently fast and regular cadence. Framing the problem (\eg choosing modeling assumptions and action features) and debugging the machine learning algorithms tend to be quite subtle, too.

The lower bounds mentioned above are fairly typical: while they are usually (and most cleanly) presented as worst-case, the actually hold for a wide variety of problem instances. The $\Omega(\sqrt{T})$ lower bound from \citet{bandits-exp3} can be extended to hold for most problem instances, in the following sense: for each instance $\mI$ there exists a ``decoy instance" $\mI'$ such that any algorithm incurs regret $\Omega(\sqrt{T})$ on at least one of them. The ``gap-dependent" lower bound of
 $\Omega(\tfrac{K}{\Delta}\log T)$
in fact holds for all problem instances and all algorithms that are not \emph{terrible} on the large-gap instances \citep{Lai-Robbins-85}. The $\Omega(T^{2/3})$ lower bound for exploration-separating algorithms in fact applies to all problem instances, as long as the algorithm achieves $\tildeO(T^{2/3})$ regret rate in the worst case \citep{MechMAB-ec09}.%
\footnote{Moreover, there is a tradeoff between the worst-case upper bound on the regret rate and a lower bound that applies for all problem instances \citep[Theorem 4.3 in][]{MechMAB-ec09}.}

Some MAB algorithms, \eg Thompson Sampling, are Bayesian: they input a prior on mean rewards, and attain strong Bayesian guarantees (in expectation over the prior) when the prior is correct. Such algorithms can also be initialized with some simple `fake' priors; in fact, this is how Thompson Sampling can be made to satisfy the optimal regret bounds.

The intuition on (the separation between) the three algorithm classes applies more generally, far beyond the basic MAB model discussed above. In particular, all algorithms that we explicitly mentioned are in fact general algorithmic techniques that are known to extend to a variety of more general MAB scenarios, typically with a similarly stark separation in regret bounds.

The greedy algorithm can perform well \emph{sometimes} in a more general model of \emph{contextual bandits}, where auxiliary payoff-relevant signals, a.k.a. contexts, are observed before each round. This phenomenon has been observed in practice
\citep{practicalCB-arxiv18}, and in theory \citep{kannan2018smoothed,bastani2017exploiting,externalities-colt18} under (very) substantial assumptions. The prevalent intuition is that the diversity of contexts can --- under some conditions and to a limited extent --- substitute for explicit exploration.

\xhdr{Instantaneous regret.}
Cumulative performance measures such as regret are not quite appropriate for our setting, as we need to characterize interactions in particular rounds. Instead, our theoretical results focus on \emph{Bayesian instantaneous regret} (\BIR), as defined in Section~\ref{sec:theory-prelims}. Recall that we posit a Bayesian prior on the \MRVs. In the notation of this appendix, the \BIR is simply:
\begin{align*}%\label{eq:bg-BIR-defn}
\BIR(t) := \E_{\text{prior}}\sbr{ \max_{\text{arms $a$}} \mu_a - r_t}.
\end{align*}
\noindent Note that Bayesian regret (\ie regret in expectation over the prior) is precisely
\begin{align}\label{eq:bg-BReg}
  \BReg(T):=
    \E_{\text{prior}}\sbr{ T\cdot \max_{\text{arms $a$}}\mu_a - \sum_{t=1}^T r_t }
    = \sum_{t=1}^T \BIR(t).
\end{align}
We are primarily interested in how fast \BIR decreases with $t$, treating $K$ as a constant.

The three classes are well-separated in terms of \BIR, much like they are in terms of regret.
%(The precise regret rates are listed below only to illustrate this point.)
\begin{itemize}
\item \DynGreedy has at least a constant \BIR for many reasonable priors (where the constant can depend on $K$ and the prior, but not on $t$). The reason / proof is the same as for regret.

\item Exploration-separating algorithms can achieve
    $\BIR(t) = \tildeO\rbr{t^{-1/3}}$
for all priors, \eg by using Epsilon-Greedy algorithm with exploration probability $\eps_t = t^{-1/3}$ in each round $t$.
In the typical scenario when
 $\BReg(t)\geq \Omega(t^{2/3})$,
the \BIR rate of $t^{-1/3}$  cannot be improved by \eqref{eq:bg-BReg}, in the following sense:
if
    $\BIR(t) = \tildeO\rbr{t^{-\gamma}}$
for all $t$, then $\gamma\geq \nicefrac{1}{3}$.

\item Adaptive-exploration algorithms \emph{can} have an even better regret rate: $\BIR(t) = \tildeO\rbr{ t^{-1/2} } $. This holds for Successive Elimination \citep{EvenDar-icml06} and for Thompson Sampling \citep{Selke-PoIE-ec21}.%
    \footnote{However, such result is not known for UCB1 algorithm, to the best of our knowledge.}
    Any optimal MAB algorithm enjoys this regret rate ``on average" by \eqref{eq:bg-BReg}, since
        $\BReg(T)\leq \tildeO(\sqrt{T})$.
    In particular, if such algorithm satisfies
        $\BIR(t) = \tildeO\rbr{t^{-\gamma}}$
    for all rounds $t$ and some constant $\gamma$, then $\gamma \leq \nicefrac{1}{2}$.

\end{itemize}

\noindent This theoretical intuition is supported by our numerical simulations: see
Figure~\ref{relative_rep_plots} and Appendix~\ref{app:isol}.

%\xhdr{}

%%% Local Variables:
%%% mode: latex
%%% TeX-master: "../journal_main"
%%% End:

%% file: itcs18paper/app-examples.tex
\newcommand{\ExplorExploit}{\term{ExplorExploit}}
\newcommand{\PhasedExplorExploit}{\term{PhasedExplorExploit}}
\newcommand{\SuccesiveEliminationReset}{\term{SuccesiveEliminationReset}}

\newcommand{\IReg}{R^{\term{inst}}} % instantaneous regret

%This appendix provides some pertinent background on multi-armed
%bandits (\emph{MAB}). We discuss \BIR and monotonicity of several MAB algorithms, touching upon: \DynGreedy and \StaticGreedy (Section~\ref{sec:MAB-greedy}), ``naive" MAB algorithms that separate exploration and exploitation (Section~\ref{sec:MAB-naive}), and ``smart" MAB algorithms that combine exploration and exploitation (Section~\ref{sec:MAB-smart}). As we do throughout the paper, we focus on MAB with i.i.d. rewards and a Bayesian prior; we call it \emph{Bayesian MAB} for brevity.

% For a given mean reward vector $\mu$, the $n$-th step instantaneous regret is
%\begin{align}
%\regret(n\mid\mu) &:= \max_{a\in A} \mu_a - \rew(n\mid\mu),\\
%\regretWC(n)    &:=  \sup_{\text{mean reward vectors $\mu$}} \; \BIR(n\mid \mu).
%\end{align}

This appendix proves \bmonotonicity of \DynGreedy and \DynamicEpsGreedy.
(The former is needed in Section~\ref{sec:theory}, the latter merely adds motivation for our theoretical story.) Recall that an algorithm is called \bmonotone if its Bayesian-expected reward is non-decreasing in time. Note that \Thompson is known to be \bmonotone if the prior $\priorMu$ is independent across arms \citep{Selke-PoIE-ec21}.

%The result on \Thompson also implies that $\BIR(t)\leq \tildeO(t^{-1/2})$. 

We consider Bayesian MAB with Bernoulli rewards. There are $T$ rounds and $K$ arms. In each round $t\in [T]$, the algorithm chooses an arm $a_t\in A$ and receives a reward $r_t\in\{0,1\}$ for this arm, drawn from a fixed but unknown distribution. The set of all arms is $A$; mean reward of arm $a$ is denoted $\mu_a$. The mean reward vector $\mu = (\mu_a:\; a\in A)$ is drawn from a common Bayesian prior $\priorMu$.
We let
    $\rew(t) = \mu_{a_t}$
denote the instantaneous mean reward of the algorithm.

%Instantaneous regret is denoted
%    $\IReg(t) := \max_{a\in A} \mu_a - \mu_{a_t}$,
%so that
%    $\BIR(t) = \E[\IReg(t)]$.

%\subsection{Monotone algorithms}
%\label{app:MAB-mono}

%This subsection argues that ``monotonicity" is a mild assumption in our context. To this end, we present versions of some standard MAB algorithms that are monotone and satisfy ``standard" regret bounds.

\xhdr{Monotonicity for the greedy algorithm.}
We state the monotonicity-in-information result for the ``Bayesian-greedy step": informally, exploitation can only get better with more data.  We invoke this result directly in Section~\ref{sec:theory}, and use it to derive monotonicity of \DynGreedy and \DynamicEpsGreedy.

A formal statement needs some scaffolding. The $n$-step history is the random sequence
    $H_n = \rbr{ (a_t,r_t):\, t\in[n]}$.
Realizations of $H_n$ are called \emph{realized histories}.
Let $\mH_n$ be the set of all possible values of $H_n$. The \emph{Bayesian-greedy step} given an $n$-step history $h\in \mH_n$ is defined as
\[ \DG(h) := \argmax_{a\in A} \E\sbr{ \mu_a\mid H_n = h },
\quad\text{ties broken arbitrarily}.\]
(However, recall that such ties are ruled out by Assumption~\ref{eq:assn-distinct}.) Now, the result is as follows:

\begin{lemma}[\citet{ICexplorationGames-ec16}]\label{lm:MII}
Let $h,h'$ be two realized histories such that $h$ is a prefix of $h'$. Then
\[ \E\sbr{ \mu_{\DG(h)} } \leq \E\sbr{ \mu_{\DG(h')} }. \]
\end{lemma}

\begin{corollary}\label{dgmono}
\DynGreedy is \bmonotone. Moreover,
$\E[\rew(n)]$ strictly increases in each time step $n$ with $\Pr[a_n\neq a_{n+1}]>0$.
\end{corollary}

\begin{proof}
Bayesian-monotonicity follows directly. The ``strictly increases" statement holds because the arm chosen in a given round has a strictly largest Bayesian-expected reward for that round.
\end{proof}

\xhdr{Monotonicity for Epsilon-Greedy.} Lemma~\ref{lm:MII} immediately implies monotonicity of \DynamicEpsGreedy, for a generic choice of exploration probabilities. Recall that in each round $t$, \DynamicEpsGreedy algorithm explores uniformly with a predetermined probability $\eps_t$, and ``exploits" with the remaining probability using the Bayesian-greedy step:
    $a_t = \DG(\text{current data})$.

\begin{corollary}\label{cor:mono-eps-greedy}
\DynamicEpsGreedy is \bmonotone whenever probabilities $\eps_t$ are non-increasing.
\end{corollary}

%% file: itcs18paper/app-perturb.tex
We provide two examples when Assumption~\eqref{eq:assn-distinct} holds almost surely under a small random perturbation of the prior. We posit Bernoulli rewards, and assume that the prior $\priorMu$ is independent across arms.

% and has a finite support.%
%\footnote{The assumption of 0-1 rewards is for clarity. Our results hold under a more general assumption that for each arm $a$, rewards can only take finitely many values, and each of these values is possible (with positive probability) for every possible value of the mean reward $\mu_a$.}

\xhdr{Beta priors.} Suppose the mean reward $\mu_a$ for each arm $a$ is drawn from some Beta distribution $\text{Beta}(\alpha_a, \beta_a)$. Given any history $H$ that contains $h_a$ number of heads and $t_a$ number of tails from arm $a$, the posterior mean reward is
    $\E[\mu_a \mid H] =  \frac{\alpha_a + h_a}{\alpha_a + h_a + \beta_a + t_a}$.
Therefore, perturbing the
    parameters $\alpha_a$ and $\beta_a$ independently with any
    continuous noise will induce a prior with property
    \eqref{eq:assn-distinct} with probability 1.

\xhdr{A prior with a finite support.} Consider the probability vector in the prior for arm $a$:
\[ \vec{p}_a = \left(\; \Pr[\mu_a=\nu]:\; \nu\in \support(\mu_a)\; \right).\]
We apply a small random perturbation independently to each such vector:
\begin{align}\label{eq:app:perturb:noise}
\vec{p}_a \leftarrow \vec{p}_a + \vec{q}_a,
    \quad \text{where}\quad \vec{q}_a\sim  \mN_a.
\end{align}
Here $\mN_a$ is the noise distribution for arm $a$: a distribution over real-valued, zero-sum vectors of dimension $d_a = |\support(\mu_a)|$. We need the noise distribution to satisfy the following property:
\begin{align}\label{eq:app:perturb:noise-prop}
\forall x\in [-1,1]^{d_a}\setminus \{0\}\qquad
\Pr_{q\sim \mN_a}\left[ x\cdot(\vec{p}_a+ q) \neq 0 \right] =1.
\end{align}

\begin{theorem}\label{thm:perturb}
Consider an instance of MAB with 0-1 rewards. Assume that the prior $\priorMu$ is independent across arms, and each mean reward $\mu_a$ has a finite support that does not include $0$ or $1$. Assume that noise distributions $\mN_a$ satisfy property \eqref{eq:app:perturb:noise-prop}. If random perturbation~\eqref{eq:app:perturb:noise} is applied independently to each arm $a$, then \refeq{eq:assn-distinct} holds almost surely for each history $h$.
\end{theorem}

\begin{remark}
As a generic example of a noise distribution which satisfies
  Property \eqref{eq:app:perturb:noise-prop}, consider the uniform
  distribution $\mN$ over the bounded convex set
  $ Q = \left\{q \in \R^{d_a} \mid q \cdot \vec{1} = 0 \mbox{ and }
      \|q\|_2 \leq \eps\right\}, $
  where $\vec{1}$ denotes the all-1
  vector. If $x = a \vec{1}$ for some non-zero value of $a$, then
  \eqref{eq:app:perturb:noise-prop} holds because
  $x \cdot (p+q) = x \cdot p = a\neq 0.$
Otherwise, denote
  $p=\vec{p}_a$ and observe that $x\cdot({p}+ {q}) = 0$ only if
  $x \cdot q = c \triangleq x \cdot (-p)$. Since $x\neq \vec{1}$, the
  intersection $Q\cap\{ x\cdot q = c \}$ either is empty or has
  measure 0 in $Q$, which implies
  $\Pr_{{q}}\left[ x\cdot({p}+ {q}) \neq 0 \right] =1$.
\end{remark}

To prove Theorem~\ref{thm:perturb}, it suffices to focus on two arms, and perturb one. Since realized rewards have finite support, there are only finitely many possible histories. So, it suffices to focus on a fixed history $h$.

\begin{lemma}\label{lm:perturb}
Consider an instance of MAB with Bernoulli rewards. Assume that the prior $\priorMu$ is independent across arms, and that $\support(\mu_1)$ is finite and does not include $0$ or $1$. Suppose random perturbation~\eqref{eq:app:perturb:noise} is applied to arm $1$, with noise distribution $\mN_1$ that satisfies \eqref{eq:app:perturb:noise-prop}. Then
    $\E[\mu_1\mid h] \neq \E[\mu_2\mid h] $
almost surely for any fixed history $h$. 
\end{lemma}

\begin{proof}
Note that $\E[\mu_a\mid h]$ does not depend on the algorithm which produced this history. Therefore, for the sake of the analysis, we can assume w.l.o.g. that this history has been generated by a particular algorithm, as long as this algorithm can can produce this history with non-zero probability. Let us consider  the algorithm that deterministically chooses same actions as $h$. Let $S = \support(\mu_1)$. Then:
\begin{align*}
\E[\mu_1\mid h]
    &=\textstyle \sum_{\nu\in S}\; \nu \cdot \Pr[\mu_1 =\nu \mid h] \\
    &= \textstyle \sum_{\nu\in S}\; \nu \cdot \Pr[h \mid \mu_1 =\nu] \cdot \Pr[\mu_1=\nu]\;/\;\Pr[h], \\
\Pr[h]
    &=\textstyle \sum_{\nu\in S}\; \Pr[h \mid \mu_1 =\nu] \cdot \Pr[\mu_1=\nu].
\end{align*}
Therefore,
    $\E[\mu_1\mid h] = \E[\mu_2\mid h] $
if and only if
\begin{align*}
\textstyle
\sum_{\nu\in S} (\nu-C) \cdot \Pr[h \mid \mu_1 =\nu] \cdot \Pr[\mu_1=\nu] = 0,
\quad\text{ where }\quad C=\E[\mu_2\mid h].
\end{align*}
Since $\E[\mu_2\mid h]$ and $\Pr[h \mid \mu_1 =\nu]$ do not depend on the probability vector $\vec{p}_1$, we conclude that
\begin{align*}
  \E[\mu_1\mid h] = \E[\mu_2\mid h]
\quad\Leftrightarrow\quad x\cdot \vec{p}_1 =0,
\end{align*}
where vector
    \[ x := \left(\; (\nu-C) \cdot \Pr[h \mid \mu_1 =\nu]:\; \nu\in S\; \right) \in [-1,1]^{d_1}\]
does not depend on $\vec{p}_1$.

Thus, it suffices to prove that $x\cdot \vec{p}_1 \neq 0$ almost surely under the perturbation. In a formula:
\begin{align}\label{eq:lm:perturb-pf}
    \Pr_{q\sim \mN_1}\left[ x\cdot( \vec{p}_1+q) \neq 0 \right] =1
\end{align}

Note that $\Pr[h \mid \mu_1 =\nu]>0$ for all $\nu\in S$, because $0,1\not\in S$. It follows that at most one coordinate of $x$ can be zero. So  \eqref{eq:lm:perturb-pf} follows from property \eqref{eq:app:perturb:noise-prop}.
\end{proof}
%%% Local Variables:
%%% mode: latex
%%% TeX-master: "main.tex"
%%% End:

%% file: content/sec-theory-proofs.tex
\xhdr{Some notation.} Without loss of generality, we label actions as $A=[K]$ and sort them according to their prior mean rewards, so that
    $ \E[\mu_1] > \E[\mu_2] > \ldots > \E[\mu_K]$.

Fix principal $i\in \{1,2\}$ and (local) step $n$. The arm chosen by algorithm \alg[i] at this step is denoted $a_{i,n}$, and the corresponding \BIR is denoted $\BIR_i(n)$. History of \alg[i] up to this step is denoted $H_{i,n}$.

Fix agent $t$. Recall that $n_i(t)$ denotes the number of global rounds before $t$ in which principal $i$ is chosen. Let $\posteriorN{i}{t}$ denote the distribution of $n_i(t)$.

Write
    $\PMR(a\mid E) = \E[\mu_a \mid E]$
for posterior mean reward of action $a$ given event $E$.

\xhdr{Chernoff Bounds.}
We use an elementary concentration inequality known as {\em Chernoff Bounds}, in a formulation from~\cite{MitzUpfal-book05}.
\begin{theorem}[Chernoff Bounds]
\label{thm:chernoff}
Consider $n$ i.i.d. random variables $X_1 \ldots X_n$ with values in $[0,1]$. Let
    $X = \tfrac{1}{n} \sum_{i=1}^n X_i$ be their average, and let $\nu = \E[X]$. Then:
\[ \min\left(\; \Pr[ X-\nu > \delta \nu ],\quad
                \Pr[ \nu-X > \delta \nu ]
    \; \right)
    < e^{-\nu n \delta^2/3}
    \quad \text{for any $\delta\in (0,1)$.}
\]
\end{theorem}

\subsection{Main result on \HardMax: Proof of Theorem~\ref{thm:DG-dominance}}
\label{sec:proofs-HM-main}

\begin{proof}[Proof of Lemma~\ref{lm:DG-rew}]
Since the two algorithms coincide on the first $n_0-1$ steps, it follows by symmetry that histories $H_{1,n_0}$ and $H_{2,n_0}$ have the same distribution. We use a \emph{coupling argument}: w.l.o.g., we assume the two histories coincide,
    $H_{1,n_0} = H_{2,n_0} = H$.

At local step $n_0$, \DynGreedy chooses an action $a_{1,n_0} = a_{1,n_0}(H)$
which maximizes the posterior mean reward given history $H$: for any realized history $h\in \support(H)$ and any action $a\in A$
\begin{align}\label{eq:lm:DG-rew-1}
 \PMR(a_{1,n_0} \mid H = h) \geq \PMR(a \mid H=h).
\end{align}

By assumption \eqref{eq:assn-distinct}, it follows that
\begin{align}\label{eq:lm:DG-rew-2}
 \PMR(a_{1,n_0} \mid H = h) > \PMR(a \mid H=h)
 \quad \text{for any $h\in \support(H)$ and $a\neq a_{1,n_0}(h)$.}
\end{align}

Since the two algorithms deviate at step $n_0$, there is a set $S\subset \support(H)$ of step-$n_0$ histories such that $\Pr[S]>0$ and any history $h\in S$ satisfies
    $\Pr[a_{2,n_0}\neq a_{1,n_0} \mid H=h]>0$.
Combining this with \eqref{eq:lm:DG-rew-2},
\begin{align}\label{eq:lm:DG-rew-3}
 \PMR(a_{1,n_0} \mid H = h) > \E\left[ \mu_{a_{2,n_0}}\mid H=h \right]
 \quad\text{for each history $h\in S$}.
\end{align}
Using \eqref{eq:lm:DG-rew-1} and \eqref{eq:lm:DG-rew-3} and integrating over realized histories $h$, we obtain
    $\E[\rew_1(n_0)] > \E[\rew_2(n_0)]$.
\end{proof}

\begin{proof}[Proof of Lemma~\ref{lm:DG-sudden}]
Let us use induction on round $t\geq t_0$, with the base case $t=t_0$. Let $\mN=\mN_{1,t_0}$ be the agents' posterior distribution for $n_{1,t_0}$, the number of global rounds before $t_0$ in which principal $1$ is chosen. By induction, all agents from $t_0$ to $t-1$ chose principal $1$, so $\PMR_2(t_0)= \PMR_2(t)$. Therefore,
\[ \PMR_1(t)
    = \Ex{n\sim \mN}{\rew_1(n+1+t-t_0)}
    \geq \Ex{n\sim \mN}{\rew_1(n+1)}
    =\PMR_1(t_0) > \PMR_2(t_0)= \PMR_2(t), \]
where the first inequality holds because \alg[1] is \bmonotone, and the second one is the base case.
\end{proof}

\begin{proof}[Proof of Theorem~\ref{thm:DG-dominance}]
Since the two algorithms coincide on the first $n_0-1$ steps, it follows by symmetry that 
    $\E[\rew_1(n)] = \E[\rew_2(n)]$ for any $n< n_0$.
By Lemma~\ref{lm:DG-rew}, it holds that 
    $\E[\rew_1(n_0)] > \E[\rew_2(n_0)]$.

Recall that $n_i(t)$ is the number of global rounds $s<t$ in which principal $i$ is chosen, and $\posteriorN{i}{t}$ is the agents' posterior distribution for this quantity. By symmetry, each agent $t<n_0$ chooses a principal uniformly at random. It follows that
    $\posteriorN{1}{n_0} = \posteriorN{2}{n_0}$
(denote both distributions by $\mN$ for brevity), and $\mN(n_0-1)>0$.
Therefore:
\begin{align}
\PMR_1(n_0)
  &= \Ex{n\sim \mN} {\rew_1(n+ 1)} 
  = \sum_{n = 0}^{n_0-1} \mN(n) \cdot \E[\rew_1(n + 1)] \nonumber \\
  & > \mN(n_0-1)\cdot \E[\rew_2(n_0)] + \sum_{n = 0}^{n_0-2}  \mN(n)\cdot\E[\rew_2(n + 1)]
    \nonumber \\
  &= \Ex{n\sim \mN}{\rew_2(n + 1)} = \PMR_2(n_0) \label{eq:pf:thm:DG-dominance}
\end{align}
So, agent $n_0$ chooses principal $1$. By Lemma~\ref{lm:DG-sudden} {(noting that \DynGreedy is \bmonotone)}, all subsequent agents choose principal $1$, too.
\end{proof}

%%%%%%%%%%%%
\subsection{\HardMax with biased tie-breaking:
Proof of Theorem~\ref{thm:HardMax-biased}}
\label{sec:proofs-HM-main-biased}

The proof re-uses Lemmas~\ref{lm:DG-rew} and~\ref{lm:DG-sudden}, which do not rely on fair tie-breaking.

Recall that $i_t$ is the principal chosen in a given global round $t$.
Because of the biased tie-breaking,
\begin{align}\label{eq:thm:HardMax-biased-PMRtoPr}
\text{if $\PMR_1(t)\geq \PMR_2(t)$ then $\Pr[i_t=1]>\tfrac12$.}
\end{align}

Let $m_0$ be the first step when \alg[2] deviates from \DynGreedy, or \DynGreedy deviates from \StaticGreedy, whichever comes sooner. {Then \alg[2], \DynGreedy and \StaticGreedy coincide on the first $m_0-1$ steps. Moreover, $m_0\leq n_0$ (since \DynGreedy deviates from \StaticGreedy at step $n_0$), so \alg[1] coincides with \DynGreedy on the first $m_0$ steps.

So, $\E[\rew_1(n)]=\E[\rew_2(n)]$ for each step $n<m_0$, because \alg[1] and \alg[2] coincide on the first $m_0-1$ steps. Moreover, if \alg[2] deviates from \DynGreedy at step $m_0$ then
    $\E[\rew_1(m_0)] > \E[\rew_2(m_0)]$
by Lemma~\ref{lm:DG-rew}; else, we trivially have
    $\rew_1(m_0) = \rew_2(m_0)$.} To summarize:
\begin{align}\label{eq:thm:HardMax-biased-rew}
    \E[\rew_1(n)]\geq\E[\rew_2(n)]\quad \text{for all steps $n\leq m_0$}.
\end{align}

We claim that $\Pr[i_t=1]>\tfrac12$ for all global rounds $t\leq m_0$. We prove this claim using induction on $t$. The base case $t=1$ holds by \eqref{eq:thm:HardMax-biased-PMRtoPr} and the fact that in step 1, \DynGreedy chooses the arm with the highest prior mean reward. For the induction step, we assume that $\Pr[i_t=1]>\tfrac12$ for all global rounds $t<t_0$, for some $t_0\leq  m_0$. It follows that distribution $\posteriorN{1}{t_0}$ stochastically dominates distribution $\posteriorN{2}{t_0}$.%
\footnote{For random variables $X,Y$ on \R, we say that $X$ \emph{stochastically dominates} $Y$ if $\Pr[X\geq x] \geq \Pr[Y\geq x]$ for any $x\in \R$.}
Observe that
\begin{align}\label{eq:thm:HardMax-biased-PMR-aux}
\PMR_1(t_0)
  = \Ex{n\sim \posteriorN{1}{t_0}} {\rew_1(n+1)}
  \geq \Ex{n\sim \posteriorN{2}{t_0}} {\rew_2(n+1)}
  = \PMR_2(t_0).
\end{align}
So the induction step follows by \eqref{eq:thm:HardMax-biased-PMRtoPr}. Claim proved.

Now let us focus on global round $m_0$, and denote $\mN_i = \posteriorN{i}{m_0}$.  By the above claim,
\begin{align}\label{eq:thm:HardMax-biased-mN}
\text{$\mN_1$ stochastically dominates $\mN_2$, and moreover
    $\mN_i(m_0-1)>\mN_i(m_0-1)$.}
\end{align}

By definition of $m_0$, either (i) \alg[2] deviates from \DynGreedy starting from local step $m_0$, which implies $\rew_1(m_0)> \rew_2(m_0)$ by Lemma~\ref{lm:DG-rew}, or (ii) \DynGreedy deviates from \StaticGreedy starting from local step $m_0$, which implies $\E[\rew_1(m_0)]>\E[\rew_1(m_0-1)]$ by Lemma~\ref{dgmono}. In both cases, using \eqref{eq:thm:HardMax-biased-rew} and \eqref{eq:thm:HardMax-biased-mN}, it follows that the inequality in \eqref{eq:thm:HardMax-biased-PMR-aux} is strict for $t_0=m_0$.

Therefore, agent $m_0$ chooses principal $1$, and by Lemma~\ref{lm:DG-sudden} so do all subsequent agents.

%%%%%%%%%%%%
\subsection{The main result for \HardMaxRandom:
Proof of Theorem~\ref{thm:random-clean}}
\label{sec:proofs-HMR-main}

Without loss of generality, assume $m_0 = n_0$.
Consider global round $t\geq n_0$. Recall that each agent chooses principal $1$ with probability at least
    $\respF(-1)>0$.

Then
    $\E[n_1(t+1)] \geq 2\eps_0\,t $.
By Chernoff Bounds (Theorem~\ref{thm:chernoff}), we have that
    $n_1(t+1)\geq \eps_0 t$
holds with probability at least $1-q$,
where $q = \exp(-\eps_0 t/12)$.

We need to prove that
    $\PMR_1(t) - \PMR_2(t)>0$.
For any $m_1$ and $m_2$, consider the quantity
\[ \Delta(m_1,m_2) := \BIR_2(m_2+1) - \BIR_1(m_1+1).\]
Whenever $m_1 \geq \eps_0 t/2 -1$ and $m_2<t$, it holds that
\begin{align*}
\Delta(m_1,m_2) \geq \Delta(\eps_0 t / 2, t)
    \geq \BIR_2(t)/2.
\end{align*}
The above inequalities follow, resp., from algorithms' \bmonotonicity and \eqref{eq:random-better-weaker}. Now,
\begin{align*}
\PMR_1(t) - \PMR_2(t)
    &= \Ex{m_1\sim \posteriorN{1}{t},\;m_2\sim \posteriorN{2}{t}}{\Delta(m_1,m_2)} \\
    &\geq -q
        + \Ex{m_1\sim \posteriorN{1}{t},\;m_2\sim \posteriorN{2}{t}}
            {\Delta(m_1,m_2)\mid m_1 \geq \eps_0 t/2-1} \\
    &\geq \BIR_2(t)/2-q \\
    &> \BIR_2(t)/4 > 0
    &\EqComment{by \refeq{eq:random-assn}}. \qedhere
\end{align*}

%%%%%%%%%%%%
\subsection{A little greedy goes a long way
(Proof of Theorem~\ref{thm:random-greedy})}
\label{sec:proofs-HMR-greedy}

%\begin{proof}[Proof of Theorem~\ref{thm:random-greedy}]
  Let $\rewgr(n)$ denote the Bayesian-expected reward of the ``greedy
  choice'' after after $n-1$ steps of \alg[1]. Note that
  $\rew_1(\cdot)$ and $\rewgr(\cdot)$ are non-decreasing: the former
  because \alg[1] is \bmonotone and the latter because the ``greedy
  choice'' is only improved with an increasing set of
  observations, see Lemma~\ref{lm:MII}.
Using \eqref{eq:BIR-modification}, we conclude that
the greedy modification \alg[2] is \bmonotone.

  By definition of the ``greedy choice,'' $\rew_1(n)\leq \rewgr(n)$
  for all steps $n$. Moreover, by Lemma~\ref{lm:DG-rew},
  \alg[1] has a strictly smaller $\rew(n_0)$ compared to \DynGreedy;
  so, $\rew_1(n_0)<\rew_2(n_0)$.

Let $\alg$ denote a copy of \alg[1] that is running ``inside" \alg[2]. Let $m_2(t)$ be the number of global rounds before $t$ in which the agent chooses principal $2$ \emph{and} \alg is invoked; \ie it is the number of agents seen by \alg before global round $t$. Let $\mM_{2,t}$ be the agents' posterior distribution for $m_2(t)$.

We claim that in each global round $t\geq n_0$, distribution $\mM_{2,t}$ stochastically dominates distribution $\posteriorN{1}{t}$, and $\PMR_1(t)<\PMR_2(t)$. We use induction on $t$. The base case $t=n_0$ holds because $\mM_{2,t} = \posteriorN{1}{t}$ (because the two algorithms coincide on the first $n_0-1$ steps), and $\PMR_1(n_0)<\PMR_2(n_0)$ is proved as in \eqref{eq:pf:thm:DG-dominance}, using the fact that $\rew_1(n_0)<\rew_2(n_0)$.

The induction step is proved as follows. The induction hypothesis for global round $t-1$ implies that agent $t-1$ is seen by \alg with probability $(1-\eps_0)(1-p)$, which is strictly larger than $\eps_0$, the probability with which this agent is seen by \alg[2]. Therefore, $\mM_{2,t}$ stochastically dominates $\posteriorN{1}{t}$.
\begin{align}
\PMR_1(t)
  &= \Ex{n\sim \posteriorN{1}{t}} {\rew_1(n+1)} \nonumber \\
  &\leq \Ex{m\sim \mM_{2,t}} {\rew_1(m+1)}
    \label{eq:pf:thm:random-greedy-1}\\
  &< \Ex{m\sim \mM_{2,t}} {(1-p)\cdot \rew_1(m+1) + p\cdot \rewgr(m+1)}
    \label{eq:pf:thm:random-greedy-2} \\
  &= \PMR_2(t). \nonumber
\end{align}
Here \eqref{eq:pf:thm:random-greedy-1} holds because $\E[\rew_1(\cdot)]$ is \bmonotone and $\mM_{2,t}$ stochastically dominates $\posteriorN{1}{t}$, and inequality \eqref{eq:pf:thm:random-greedy-2} holds because $\E[\rew_1(n_0)]<\E[\rew_2(n_0)]$ and $\mM_{2,t}(n_0)>0$.%
\footnote{If $\E[\rew_1(\cdot)]$ is strictly increasing, then  \eqref{eq:pf:thm:random-greedy-1} is strict, too; this is because $\mM_{2,t}(t-1)>\posteriorN{1}{t}(t-1)$.  }
%\end{proof}

%%%%%%%%%%%%
\subsection{\SoftMaxRandom: proof of Theorem~\ref{thm:SoftMax-strong}}
\label{sec:proofs-SoftMax}

%\begin{proof}[Proof of Theorem~\ref{thm:SoftMax-strong}]
  Let $\beta_1 = \min\{c_0'\delta_0, \beta_0/20\}$ with $\delta_0$
  defined in~\eqref{eq:SoftMax-smooth}.  Recall each agent chooses
  \alg[1] with probability at least $\respF(-1)= \eps_0$.  By
By condition \eqref{eq:SoftMax-assn-strong} and the fact that
$\BIR_1(n) \to 0$,
  there exists some sufficiently large $T_1$ such that for any
  $t\geq T_1$, $\BIR_1(\eps_0 T_1/2) \leq \beta_1/c_0'$ and
  $\BIR_2(t) > e^{-\eps_0 t/12}$. Moreover, for any $t\geq T_1$, we
  know $\E[n_1(t+1)] \geq \eps_0\,t $, and by the Chernoff Bounds
  (Theorem~\ref{thm:chernoff}), we have $n_1(t+1) \geq \eps_0 t/2$
  holds with probability at least $1 - q_1(t)$ with
  $q_1(t) = \exp(-\eps_0 t/12) < \BIR_2(t)$. It follows that for any $t\geq T_1$,
\begin{align*}
  \PMR_2(t) - \PMR_1(t) &= \Ex{m_1\sim \posteriorN{1}{t},\;m_2\sim \posteriorN{2}{t}}{\BIR_1(m_1+ 1) - \BIR_2(m_2+1)} \\
                        &\leq q_1(t)  + \Ex{m_1\sim \posteriorN{1}{t}}{\BIR_1(m_1+ 1)\mid m_1 \geq \eps_0 t/2 - 1 } - \BIR_2(t)\\
                        &\leq \BIR_1(\eps_0 T_1/2) \leq \beta_1/c_0'
\end{align*}
% where the second inequality follows from~\eqref{eq:random-assn}.
Since the response function $\respF$ is $c_0'$-Lipschitz in the
neighborhood of $[-\delta_0, \delta_0]$, each agent after round $T_1$
will choose \alg[1] with probability at least
\[
  p_t \geq \tfrac{1}{2} - c_0'\left(\PMR_2(t) - \PMR_1(t)\right) \geq
  \tfrac{1}{2} - \beta_1.
\]

Next, we will show that there exists a sufficiently large $T_2$ such
that for any $t\geq T_1 + T_2$, with high probability
$n_1(t) > \max\{n_0, (1 - \beta_0)n_2(t)\}$, where $n_0$ is defined
in~\eqref{eq:SoftMax-better}. %  Let us first lower bound the number of
% agents received by \alg[1] after some number of rounds $t = T_1 + T'$
% for any $T' \geq T_1$.
Fix any $t \geq T_1 + T_2$.
Since each agent chooses \alg[1] with probability at least
$1/2 - \beta_1$, by Chernoff Bounds (Theorem~\ref{thm:chernoff}) we
have with probability at least $1 - q_2(t)$ that the number of agents
that choose \alg[1] is at least
    $\beta_0(\nicefrac{1}{2} - \beta_1)t/5$,
where %the function
$$
q_2(x) = \exp\rbr{ -\nicefrac{1}{3}\;
    (\nicefrac{1}{2} - \beta_1)(1 - \beta_0/5)^2x} .
$$
The number of agents received by \alg[2] is at most
$T_1 + (\nicefrac{1}{2} + \beta_1)t + (\nicefrac{1}{2} - \beta_1)(1 - \beta_0/5)t$.

Then as long as
% $T_2 \geq \max\left\{\frac{3T_1}{(1 - \beta_0)}, 8 n_0\right\}$, we
% can guarantee that for any $t \geq T_1 + T_2$,
% $n_1(t) > n_2(t) (1 - \beta_0)$ and $n_1(t) > n_0$ with probability at
% least $1 - q_2(t)$.
$T_2 \geq \frac{5T_1}{\beta_0}$, we can guarantee that
$n_1(t) > n_2(t) (1 - \beta_0)$ and $n_1(t) > n_0$ with probability at
least $1 - q_2(t)$ for any $t \geq T_1 + T_2$.
% for any $t\geq T_1 + T_2$, as long as $T_2 \geq .$ Moreover, to
% guarantee that , we will just need $T_2 \geq$.  Finally, we will
% argue that in each round $t \geq T_1 + T_2$, we recover the
% guarantee in \eqref{eq:thm:SoftMax-strong}.
% % \[
% %   \Pr[i_t = 1] \geq \frac{1}{2} + \frac{c_0\alpha_0\BIR_2(t)}{4}.
% % \]
Note that the weak BIR-dominance condition
in~\eqref{eq:SoftMax-better} implies that for any $t\geq T_1 + T_2$
with probability at least $1 - q_2(t)$, we have
$ \BIR_1(n_1(t)) < (1- \alpha_0)\,\BIR_2(n_2(t))$.

It follows that for any $t\geq T_1 + T_2$,
\begin{align*}
  \PMR_1(t) - \PMR_2(t) &= \Ex{m_1\sim \posteriorN{1}{t},\;m_2\sim \posteriorN{2}{t}}{\BIR_2(m_2+ 1) - \BIR_1(m_1+1)} \\
                        &\geq (1 - q_2(t))\,\alpha_0\, \BIR_2(t) - q_2(t)
                        \geq \alpha_0\, \BIR_2(t)/4,
\end{align*}
where the last inequality holds as long as
$q_2(t) \leq \alpha_0\BIR_2(t)/4$, and is implied by the condition
in~\eqref{eq:SoftMax-assn-strong} as long as $T_2$ is sufficiently
large. Hence, by the definition of our \SoftMaxRandom response
function and assumption in~\eqref{eq:SoftMax-smooth}, we have
$  \Pr[i_t = 1] \geq \nicefrac{1}{2} +
    \nicefrac{1}{4}\; c_0\,\alpha_0\,\BIR_2(t)$.
%\end{proof}

%% file: ec19paper/content/appendix_for_one_version.tex
In this appendix we provide full results for the experiments described in Section~\ref{sec:sim}.

\subsection{``Performance In Isolation" (Section~\ref{sec:isolation})}
\label{app:isol}
%\subsection{Additional plots for ``Performance In Isolation"}

% First, we provide mean reputation trajectories for the Heavy-Tail MAB instance. Second, we provide trajectories for the average of the instantaneous mean rewards, for all three MAB instances. the mean reward trajectories provides a better view into the direct performance of the algorithm at a given time $t$ as, unlike reputation, it does not aggregate performance across many previous time periods.

We present the full plots for Section \ref{sec:isolation}: mean reputation trajectories and instantaneous reward trajectories for all three MAB instances.  For ``instantaneous reward" at a given time $t$, we report the average (over all \MRVs) of the mean rewards at this time, instead of the average of the \emph{realized} rewards, so as to decrease the noise. In all plots, the shaded area represents 95\% confidence interval.

\begin{figure}[H]
\begin{center}
\includegraphics[scale=0.35]{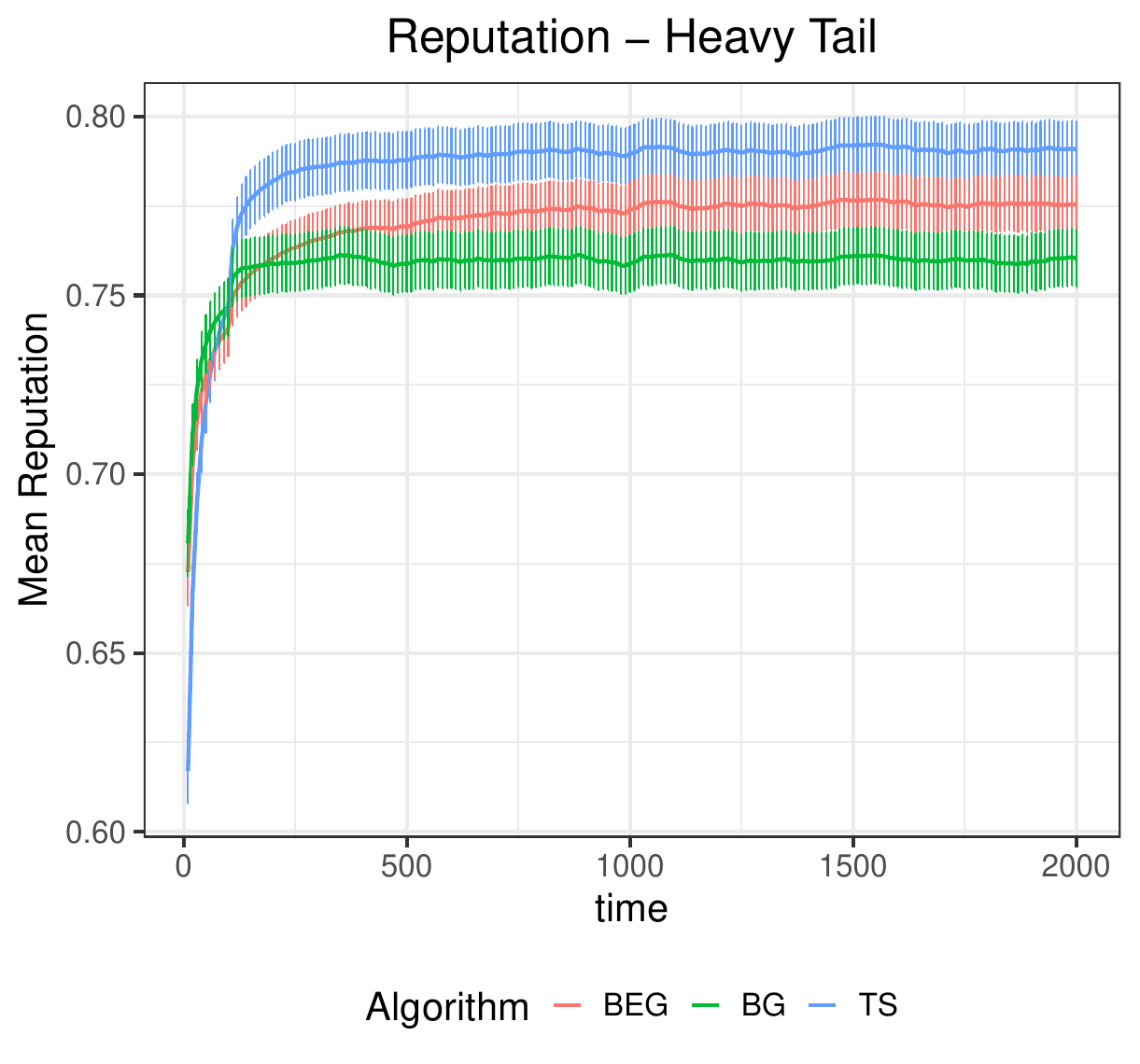}
\includegraphics[scale=0.35]{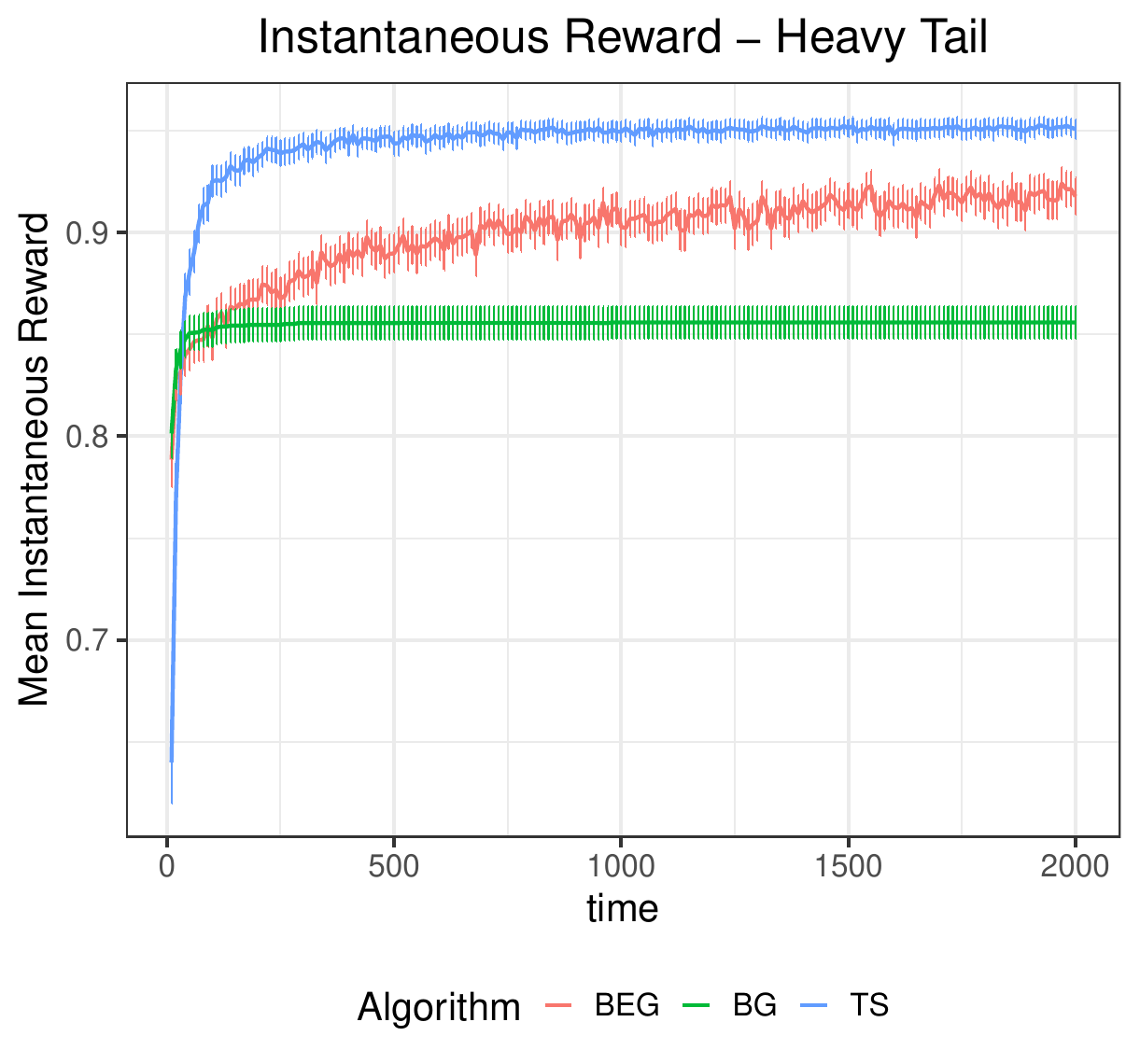}
\caption{Mean Reputation (left) and Mean Instantaneous Reward (right) for Heavy Tail Instance}
\end{center}
\end{figure}

\begin{figure}[H]
\begin{center}
\includegraphics[scale=0.35]{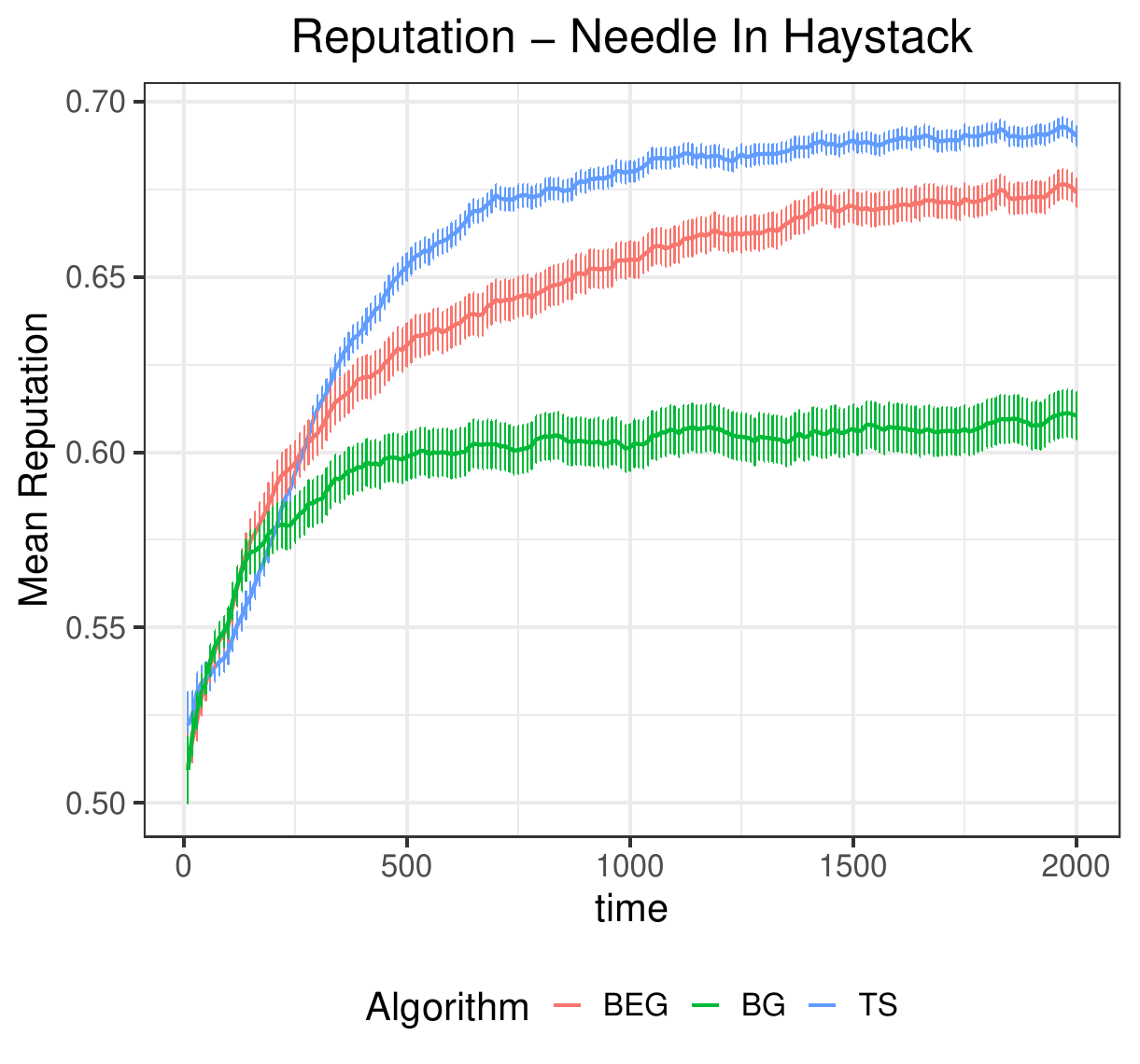}
\includegraphics[scale=0.35]{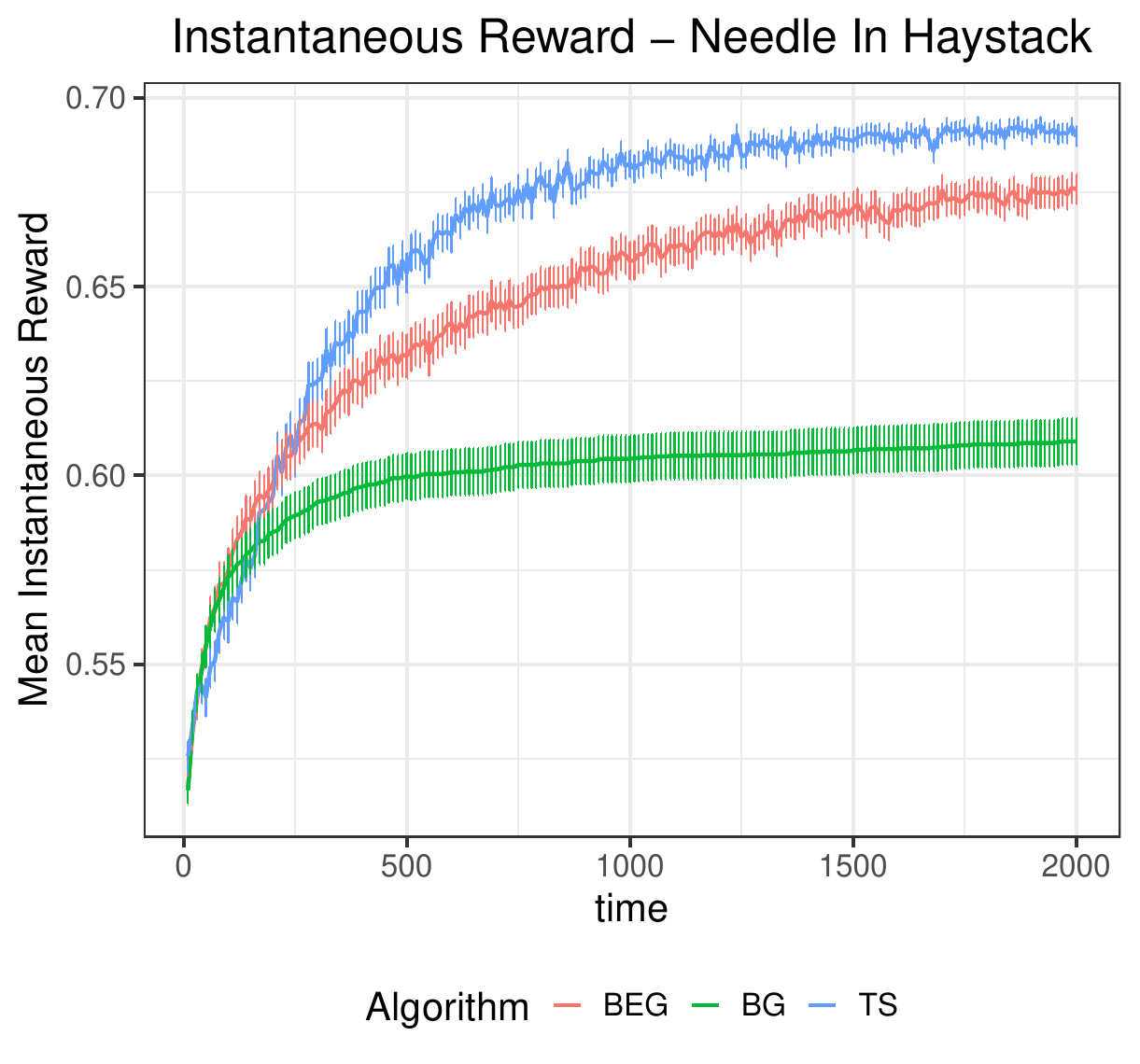}
\caption{Mean Reputation (left) and Mean Instantaneous Reward (right) for Needle In Haystack Instance}
\end{center}
\end{figure}

\begin{figure}[H]
\begin{center}
\includegraphics[scale=0.35]{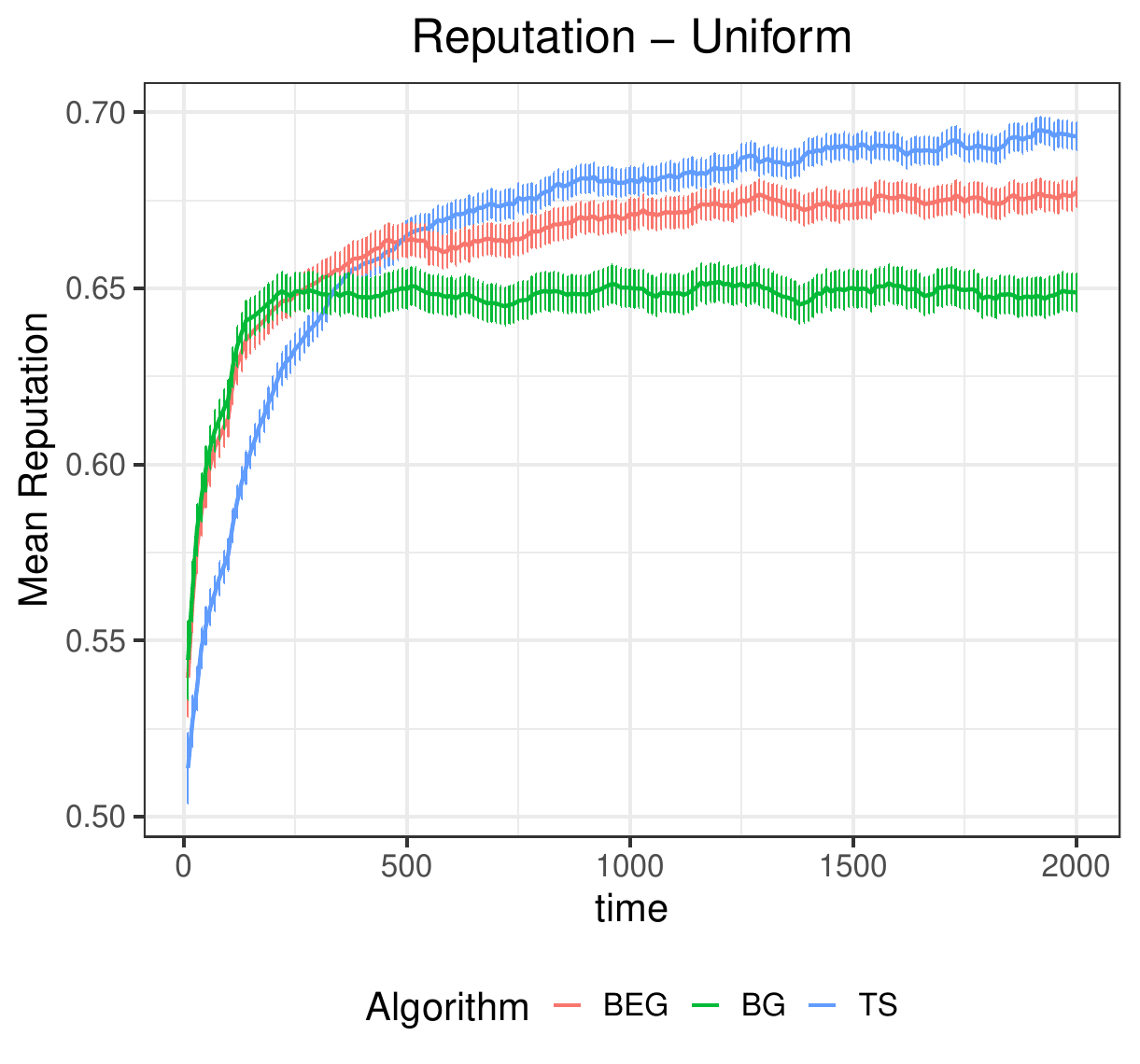}
\includegraphics[scale=0.35]{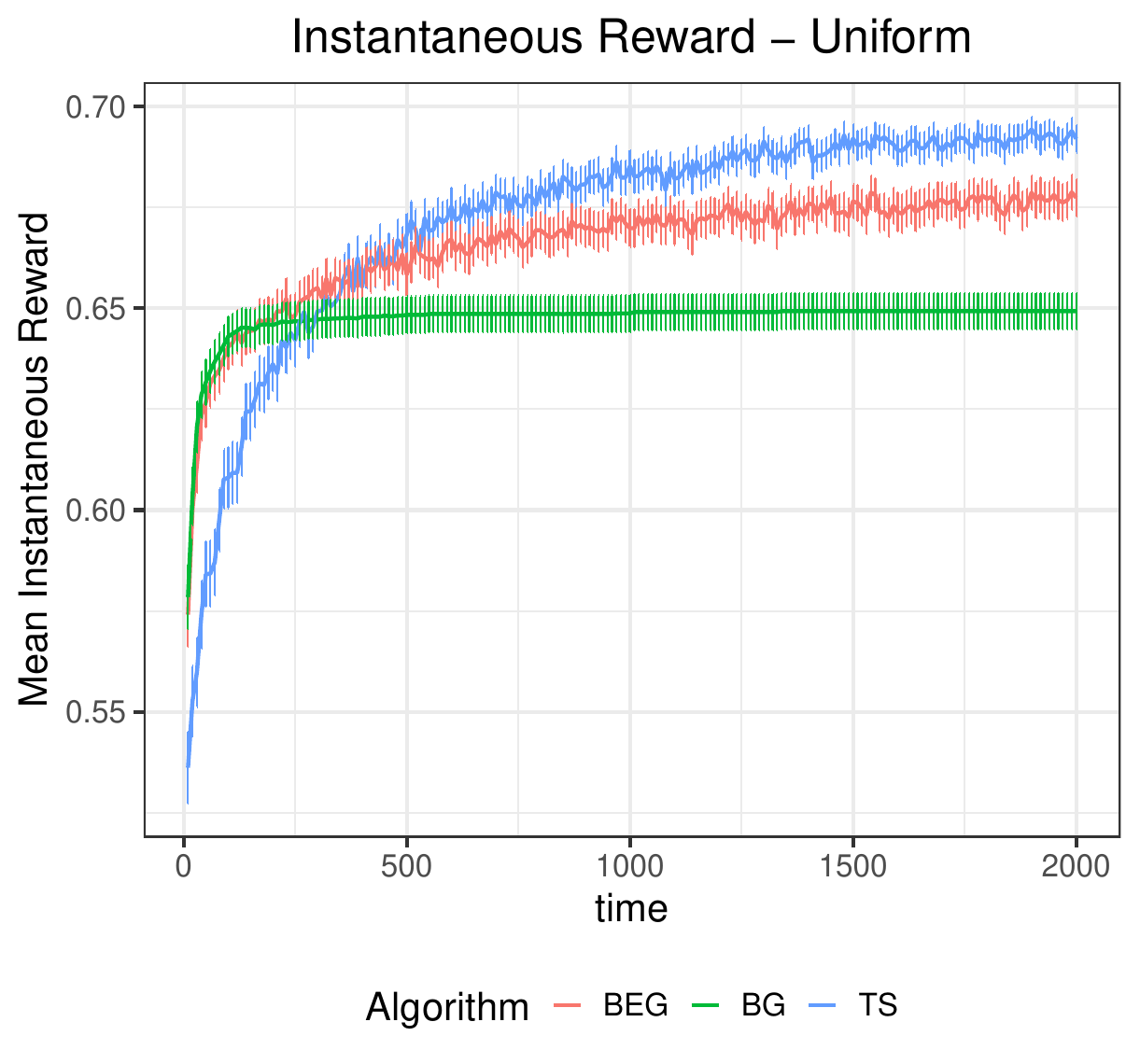}
\caption{Mean Reputation (left) and Mean Instantaneous Reward (right) for Uniform Instance}
\end{center}
\end{figure}

\subsection{First-mover regime (Section~\ref{sec:competition})}
%\subsection{Additional Experiments for First-Mover Regime}

We present additional experiments on the first-mover regime from Section~\ref{sec:competition}, across various MAB instances and various values of the incumbent advantage parameter $X$.

Each experiment is presented as a table with the same semantics as in the main text. Namely, each cell in the table describes the duopoly game between the entrant's algorithm (the row) and the incumbent's algorithm (the column). The cell specifies the entrant's market share (fraction of rounds in which it was chosen) for the rounds in which he was present. We give the average (in bold) and the 95\% confidence interval. NB: smaller average is better for the incumbent.

\OMIT{These results confirm the claim in the text that, for sufficiently large $X$, $\TS$ is preferred over all other algorithms for the incumbent. However, it also shows that, for smaller values of $X$ it is not necessarily the case that $\TS$ is the preferred algorithm. We provide many different parameterizations in order to check the robustness of the results. The interpretation of the tables is the same as those in the main text. The results presented here use the same instances and realization tables as those presented in the main text.}

\
\begin{table}[H]
\centering
\begin{adjustbox}{width=\textwidth,center}
\begin{tabular}{|c|c|c|c||c|c|c|}
  \hline
  & \multicolumn{3}{c||}{$X = 50$}
  & \multicolumn{3}{c|}{$X = 200$} \\
    \hline
  & $\TS$  & $\DEG$  & $\DG$
  & $\TS$  & $\DEG$  & $\DG$ \\
  \hline
  $\TS$
 & \makecell{\textbf{0.054} $\pm$0.01}
    & \makecell{\textbf{0.16} $\pm$0.02}
    & \makecell{\textbf{0.18} $\pm$0.02}
     & \makecell{\textbf{0.003} $\pm$0.003}
    & \makecell{\textbf{0.083} $\pm$0.02}
    & \makecell{\textbf{0.17} $\pm$0.02} \\\hline
    $\DEG$
    & \makecell{\textbf{0.33} $\pm$0.03}
    & \makecell{\textbf{0.31} $\pm$0.02}
    & \makecell{\textbf{0.26} $\pm$0.02}
    & \makecell{\textbf{0.045} $\pm$0.01}
    & \makecell{\textbf{0.25} $\pm$0.02}
    & \makecell{\textbf{0.23} $\pm$0.02} \\\hline
    $\DG$
    & \makecell{\textbf{0.39} $\pm$0.03}
    & \makecell{\textbf{0.41} $\pm$0.03}
    & \makecell{\textbf{0.33} $\pm$0.02}
    & \makecell{\textbf{0.12} $\pm$0.02}
    & \makecell{\textbf{0.36} $\pm$0.03}
    & \makecell{\textbf{0.3} $\pm$0.02} \\\hline
\end{tabular}
\end{adjustbox}
\caption{Heavy-Tail MAB Instance}
\end{table}

\begin{table}[H]
\centering
\begin{adjustbox}{width=\textwidth,center}
\begin{tabular}{|c|c|c|c||c|c|c|}
  \hline
  & \multicolumn{3}{c||}{$X = 300$}
  & \multicolumn{3}{c|}{$X = 500$} \\
    \hline
  & $\TS$  & $\DEG$  & $\DG$
  & $\TS$  & $\DEG$  & $\DG$ \\
  \hline
  $\TS$
 & \makecell{\textbf{0.0017} $\pm$0.002}
    & \makecell{\textbf{0.059} $\pm$0.01}
    & \makecell{\textbf{0.16} $\pm$0.02}
       & \makecell{\textbf{0.002} $\pm$0.003}
    & \makecell{\textbf{0.043} $\pm$0.01}
    & \makecell{\textbf{0.16} $\pm$0.02} \\\hline
    $\DEG$
        & \makecell{\textbf{0.029} $\pm$0.007}
    & \makecell{\textbf{0.23} $\pm$0.02}
    & \makecell{\textbf{0.23} $\pm$0.02}
     & \makecell{\textbf{0.03} $\pm$0.007}
    & \makecell{\textbf{0.21} $\pm$0.02}
    & \makecell{\textbf{0.24} $\pm$0.02} \\\hline
    $\DG$
    & \makecell{\textbf{0.097} $\pm$0.02}
    & \makecell{\textbf{0.34} $\pm$0.03}
    & \makecell{\textbf{0.29} $\pm$0.02}
    & \makecell{\textbf{0.091} $\pm$0.01}
    & \makecell{\textbf{0.32} $\pm$0.03}
    & \makecell{\textbf{0.3} $\pm$0.02}\\\hline
\end{tabular}
\end{adjustbox}
\caption{Heavy-Tail MAB Instance}
\end{table}

\begin{table}[H]
\centering
\begin{adjustbox}{width=\textwidth,center}
\begin{tabular}{|c|c|c|c||c|c|c|}
  \hline
  & \multicolumn{3}{c||}{$X = 50$}
  & \multicolumn{3}{c|}{$X = 200$} \\
    \hline
  & $\TS$  & $\DEG$  & $\DG$
  & $\TS$  & $\DEG$  & $\DG$ \\
  \hline
  $\TS$
 & \makecell{\textbf{0.34} $\pm$0.03}
    & \makecell{\textbf{0.4} $\pm$0.03}
    & \makecell{\textbf{0.48} $\pm$0.03}
      & \makecell{\textbf{0.17} $\pm$0.02}
    & \makecell{\textbf{0.31} $\pm$0.03}
    & \makecell{\textbf{0.41} $\pm$0.03} \\\hline
    $\DEG$
    & \makecell{\textbf{0.22} $\pm$0.02}
    & \makecell{\textbf{0.34} $\pm$0.03}
    & \makecell{\textbf{0.42} $\pm$0.03}
    & \makecell{\textbf{0.13} $\pm$0.02}
    & \makecell{\textbf{0.26} $\pm$0.02}
    & \makecell{\textbf{0.36} $\pm$0.03} \\\hline
    $\DG$
     & \makecell{\textbf{0.18} $\pm$0.02}
    & \makecell{\textbf{0.28} $\pm$0.02}
    & \makecell{\textbf{0.37} $\pm$0.03}
    & \makecell{\textbf{0.093} $\pm$0.02}
    & \makecell{\textbf{0.23} $\pm$0.02}
    & \makecell{\textbf{0.33} $\pm$0.03} \\\hline
\end{tabular}
\end{adjustbox}
\caption{Needle In Haystack MAB Instance}
\end{table}

\begin{table}[H]
\centering
\begin{adjustbox}{width=\textwidth,center}
\begin{tabular}{|c|c|c|c||c|c|c|}
  \hline
  & \multicolumn{3}{c||}{$X = 300$}
  & \multicolumn{3}{c|}{$X = 500$} \\
    \hline
  & $\TS$  & $\DEG$  & $\DG$
  & $\TS$  & $\DEG$  & $\DG$ \\
  \hline
  $\TS$
  & \makecell{\textbf{0.1} $\pm$0.02}
    & \makecell{\textbf{0.28} $\pm$0.03}
    & \makecell{\textbf{0.39} $\pm$0.03}
      & \makecell{\textbf{0.053} $\pm$0.01}
    & \makecell{\textbf{0.23} $\pm$0.02}
    & \makecell{\textbf{0.37} $\pm$0.03} \\\hline
    $\DEG$
         & \makecell{\textbf{0.089} $\pm$0.02}
    & \makecell{\textbf{0.23} $\pm$0.02}
    & \makecell{\textbf{0.36} $\pm$0.03}
      & \makecell{\textbf{0.051} $\pm$0.01}
    & \makecell{\textbf{0.2} $\pm$0.02}
    & \makecell{\textbf{0.33} $\pm$0.03} \\\hline
    $\DG$
     & \makecell{\textbf{0.05} $\pm$0.01}
    & \makecell{\textbf{0.21} $\pm$0.02}
    & \makecell{\textbf{0.33} $\pm$0.03}
    & \makecell{\textbf{0.031} $\pm$0.009}
    & \makecell{\textbf{0.18} $\pm$0.02}
    & \makecell{\textbf{0.31} $\pm$0.02} \\\hline
\end{tabular}
\end{adjustbox}
\caption{Needle In Haystack MAB Instance}
\end{table}

\begin{table}[H]
\centering
\begin{adjustbox}{width=\textwidth,center}
\begin{tabular}{|c|c|c|c||c|c|c|}
  \hline
  & \multicolumn{3}{c||}{$X = 50$}
  & \multicolumn{3}{c|}{$X = 200$} \\
    \hline
  & $\TS$  & $\DEG$  & $\DG$
  & $\TS$  & $\DEG$  & $\DG$ \\
  \hline
  $\TS$
  & \makecell{\textbf{0.27} $\pm$0.03}
    & \makecell{\textbf{0.21} $\pm$0.02}
    & \makecell{\textbf{0.26} $\pm$0.02}
   & \makecell{\textbf{0.12} $\pm$0.02}
    & \makecell{\textbf{0.16} $\pm$0.02}
    & \makecell{\textbf{0.2} $\pm$0.02} \\\hline
    $\DEG$
     & \makecell{\textbf{0.39} $\pm$0.03}
    & \makecell{\textbf{0.3} $\pm$0.03}
    & \makecell{\textbf{0.34} $\pm$0.03}
     & \makecell{\textbf{0.25} $\pm$0.02}
    & \makecell{\textbf{0.24} $\pm$0.02}
    & \makecell{\textbf{0.29} $\pm$0.02} \\\hline
    $\DG$
     & \makecell{\textbf{0.39} $\pm$0.03}
    & \makecell{\textbf{0.31} $\pm$0.02}
    & \makecell{\textbf{0.33} $\pm$0.02}
    & \makecell{\textbf{0.23} $\pm$0.02}
    & \makecell{\textbf{0.24} $\pm$0.02}
    & \makecell{\textbf{0.29} $\pm$0.02} \\\hline
\end{tabular}
\end{adjustbox}
\caption{Uniform MAB Instance}
\end{table}

\begin{table}[H]
\centering
\begin{adjustbox}{width=\textwidth,center}
\begin{tabular}{|c|c|c|c||c|c|c|}
  \hline
  & \multicolumn{3}{c||}{$X = 300$}
  & \multicolumn{3}{c|}{$X = 500$} \\
    \hline
  & $\TS$  & $\DEG$  & $\DG$
  & $\TS$  & $\DEG$  & $\DG$ \\
  \hline
  $\TS$
  & \makecell{\textbf{0.094} $\pm$0.02}
    & \makecell{\textbf{0.15} $\pm$0.02}
    & \makecell{\textbf{0.2} $\pm$0.02}
     & \makecell{\textbf{0.061} $\pm$0.01}
    & \makecell{\textbf{0.12} $\pm$0.02}
    & \makecell{\textbf{0.2} $\pm$0.02} \\\hline
    $\DEG$
      & \makecell{\textbf{0.2} $\pm$0.02}
    & \makecell{\textbf{0.23} $\pm$0.02}
    & \makecell{\textbf{0.29} $\pm$0.02}
     & \makecell{\textbf{0.17} $\pm$0.02}
    & \makecell{\textbf{0.21} $\pm$0.02}
    & \makecell{\textbf{0.29} $\pm$0.02} \\\hline
    $\DG$
  & \makecell{\textbf{0.21} $\pm$0.02}
    & \makecell{\textbf{0.23} $\pm$0.02}
    & \makecell{\textbf{0.29} $\pm$0.02}
  & \makecell{\textbf{0.18} $\pm$0.02}
    & \makecell{\textbf{0.22} $\pm$0.02}
    & \makecell{\textbf{0.29} $\pm$0.02} \\\hline
\end{tabular}
\end{adjustbox}
\caption{Uniform MAB Instance}
\end{table}

\subsection{Reputation Advantage vs. Data Advantage (Section \ref{sec:barriers})}
%\subsection{Reputation vs. Data Advantage

This section presents full experimental results on reputation advantage vs. data advantage.

Each experiment is presented as a table with the same semantics as in the main text. Namely, each cell in the table describes the duopoly game between the entrant's algorithm (the {\bf row}) and the incumbent's algorithm (the {\bf column}). The cell specifies the entrant's market share for the rounds in which hit was present: the average (in bold) and the 95\% confidence interval. NB: smaller average is better for the incumbent.

\begin{table}[H]
\centering
\begin{adjustbox}{width=\textwidth,center}
\begin{tabular}{|c|c|c|c||c|c|c|}
  \hline
  & \multicolumn{3}{c||}{Data Advantage}
  & \multicolumn{3}{c|}{Reputation Advantage} \\
    \hline
  & $\TS$  & $\DEG$  & $\DG$
  & $\TS$  & $\DEG$  & $\DG$ \\
  \hline
  $\TS$
   & \makecell{\textbf{ 0.0096 } $\pm$ 0.006}
    & \makecell{\textbf{ 0.11 } $\pm$ 0.02}
    & \makecell{\textbf{ 0.18 } $\pm$ 0.02}
       & \makecell{\textbf{ 0.021 } $\pm$ 0.009}
    & \makecell{\textbf{ 0.16 } $\pm$ 0.02}
    & \makecell{\textbf{ 0.21 } $\pm$ 0.02} \\\hline
    $\DEG$
     & \makecell{\textbf{ 0.073 } $\pm$ 0.01}
    & \makecell{\textbf{ 0.29 } $\pm$ 0.02}
    & \makecell{\textbf{ 0.25 } $\pm$ 0.02}
     & \makecell{\textbf{ 0.26 } $\pm$ 0.03}
    & \makecell{\textbf{ 0.3 } $\pm$ 0.02}
    & \makecell{\textbf{ 0.26 } $\pm$ 0.02} \\\hline
    $\DG$
   & \makecell{\textbf{ 0.15 } $\pm$ 0.02}
    & \makecell{\textbf{ 0.39 } $\pm$ 0.03}
    & \makecell{\textbf{ 0.33 } $\pm$ 0.02}
   & \makecell{\textbf{ 0.34 } $\pm$ 0.03}
    & \makecell{\textbf{ 0.4 } $\pm$ 0.03 }
    & \makecell{\textbf{ 0.33 } $\pm$ 0.02} \\\hline
\end{tabular}
\end{adjustbox}
\caption{Heavy Tail MAB Instance, $X = 200$}
\end{table}

\begin{table}[H]
\centering
\begin{adjustbox}{width=\textwidth,center}
\begin{tabular}{|c|c|c|c||c|c|c|}
  \hline
  & \multicolumn{3}{c||}{Data Advantage}
  & \multicolumn{3}{c|}{Reputation Advantage} \\
    \hline
  & $\TS$  & $\DEG$  & $\DG$
  & $\TS$  & $\DEG$  & $\DG$ \\
  \hline
  $\TS$
      & \makecell{\textbf{0.0017} $\pm$0.002}
    & \makecell{\textbf{0.06} $\pm$0.01}
    & \makecell{\textbf{0.18} $\pm$0.02}
    & \makecell{\textbf{0.022} $\pm$0.009}
    & \makecell{\textbf{0.13} $\pm$0.02}
    & \makecell{\textbf{0.21} $\pm$0.02} \\\hline
    $\DEG$
      & \makecell{\textbf{0.04} $\pm$0.009}
    & \makecell{\textbf{0.24} $\pm$0.02}
    & \makecell{\textbf{0.25} $\pm$0.02}
  & \makecell{\textbf{0.26} $\pm$0.03}
    & \makecell{\textbf{0.29} $\pm$0.02}
    & \makecell{\textbf{0.28} $\pm$0.02} \\\hline
    $\DG$
   & \makecell{\textbf{0.12} $\pm$0.02}
    & \makecell{\textbf{0.35} $\pm$0.03}
    & \makecell{\textbf{0.33} $\pm$0.02}
   & \makecell{\textbf{0.33} $\pm$0.03}
    & \makecell{\textbf{0.39} $\pm$0.03}
    & \makecell{\textbf{0.34} $\pm$0.02} \\\hline
\end{tabular}
\end{adjustbox}
\caption{Heavy Tail MAB Instance, $X = 500$}
\end{table}

\begin{table}[H]
\centering
\begin{adjustbox}{width=\textwidth,center}
\begin{tabular}{|c|c|c|c||c|c|c|}
  \hline
  & \multicolumn{3}{c||}{Data Advantage}
  & \multicolumn{3}{c|}{Reputation Advantage} \\
    \hline
  & $\TS$  & $\DEG$  & $\DG$
  & $\TS$  & $\DEG$  & $\DG$ \\
  \hline
  $\TS$
   & \makecell{\textbf{ 0.25 } $\pm$ 0.03}
    & \makecell{\textbf{ 0.36 } $\pm$ 0.03}
    & \makecell{\textbf{ 0.45 } $\pm$ 0.03}
     & \makecell{\textbf{ 0.35 } $\pm$ 0.03}
    & \makecell{\textbf{ 0.43 } $\pm$ 0.03}
    & \makecell{\textbf{ 0.52 } $\pm$ 0.03} \\\hline
    $\DEG$
    & \makecell{\textbf{ 0.21 } $\pm$ 0.02}
    & \makecell{\textbf{ 0.32 } $\pm$ 0.03}
    & \makecell{\textbf{ 0.41 } $\pm$ 0.03}
     & \makecell{\textbf{ 0.26 } $\pm$ 0.03 }
    & \makecell{\textbf{ 0.36 } $\pm$ 0.03}
    & \makecell{\textbf{ 0.43 } $\pm$ 0.03} \\\hline
    $\DG$
   & \makecell{\textbf{ 0.18 } $\pm$ 0.02}
    & \makecell{\textbf{ 0.29 } $\pm$ 0.03}
    & \makecell{\textbf{ 0.4 } $\pm$ 0.03}
    & \makecell{\textbf{ 0.19 } $\pm$ 0.02}
    & \makecell{\textbf{ 0.3 } $\pm$ 0.02}
    & \makecell{\textbf{ 0.36 } $\pm$ 0.02} \\\hline
\end{tabular}
\end{adjustbox}
\caption{Needle-in-Haystack MAB Instance, $X=200$}
\end{table}

\begin{table}[H]
\centering
\begin{adjustbox}{width=\textwidth,center}
\begin{tabular}{|c|c|c|c||c|c|c|}
  \hline
  & \multicolumn{3}{c||}{Data Advantage}
  & \multicolumn{3}{c|}{Reputation Advantage} \\
    \hline
  & $\TS$  & $\DEG$  & $\DG$
  & $\TS$  & $\DEG$  & $\DG$ \\
  \hline
  $\TS$
  & \makecell{\textbf{0.098} $\pm$0.02}
    & \makecell{\textbf{0.27} $\pm$0.03}
    & \makecell{\textbf{0.41} $\pm$0.03}
     & \makecell{\textbf{0.29} $\pm$0.03}
    & \makecell{\textbf{0.44} $\pm$0.03}
    & \makecell{\textbf{0.52} $\pm$0.03} \\\hline
    $\DEG$
      & \makecell{\textbf{0.093} $\pm$0.02}
    & \makecell{\textbf{0.24} $\pm$0.02}
    & \makecell{\textbf{0.38} $\pm$0.03}
    & \makecell{\textbf{0.19} $\pm$0.02}
    & \makecell{\textbf{0.35} $\pm$0.03}
    & \makecell{\textbf{0.42} $\pm$0.03} \\\hline
    $\DG$
    & \makecell{\textbf{0.064} $\pm$0.01}
    & \makecell{\textbf{0.22} $\pm$0.02}
    & \makecell{\textbf{0.37} $\pm$0.03}
    & \makecell{\textbf{0.15} $\pm$0.02}
    & \makecell{\textbf{0.27} $\pm$0.02}
    & \makecell{\textbf{0.35} $\pm$0.02} \\\hline
\end{tabular}
\end{adjustbox}
\caption{Needle-in-Haystack MAB Instance, $X=500$}
\end{table}

\begin{table}[H]
\centering
\begin{adjustbox}{width=\textwidth,center}
\begin{tabular}{|c|c|c|c||c|c|c|}
  \hline
  & \multicolumn{3}{c||}{Data Advantage}
  & \multicolumn{3}{c|}{Reputation Advantage} \\
    \hline
  & $\TS$  & $\DEG$  & $\DG$
  & $\TS$  & $\DEG$  & $\DG$ \\
  \hline
  $\TS$
  & \makecell{\textbf{ 0.2 } $\pm$ 0.02}
    & \makecell{\textbf{ 0.22 } $\pm$ 0.02}
    & \makecell{\textbf{ 0.27 } $\pm$ 0.03}
     & \makecell{\textbf{ 0.27 } $\pm$ 0.03}
    & \makecell{\textbf{ 0.23 } $\pm$ 0.02}
    & \makecell{\textbf{ 0.27 } $\pm$ 0.02}\\\hline
    $\DEG$
    & \makecell{\textbf{ 0.33 } $\pm$ 0.03}
    & \makecell{\textbf{ 0.32 } $\pm$ 0.03}
    & \makecell{\textbf{ 0.35 } $\pm$ 0.03}
     & \makecell{\textbf{ 0.4 } $\pm$ 0.03}
    & \makecell{\textbf{ 0.3 } $\pm$ 0.02 }
    & \makecell{\textbf{ 0.32 } $\pm$ 0.02} \\\hline
    $\DG$
    & \makecell{\textbf{ 0.32 } $\pm$ 0.03}
    & \makecell{\textbf{ 0.31 } $\pm$ 0.03}
    & \makecell{\textbf{ 0.35 } $\pm$ 0.03}
     & \makecell{\textbf{ 0.36 } $\pm$ 0.03}
    & \makecell{\textbf{ 0.29 } $\pm$ 0.02}
    & \makecell{\textbf{ 0.3 } $\pm$ 0.02} \\\hline
\end{tabular}
\end{adjustbox}
\caption{Uniform MAB Instance, $X = 200$}
\end{table}

\begin{table}[H]
\centering
\begin{adjustbox}{width=\textwidth,center}
\begin{tabular}{|c|c|c|c||c|c|c|}
  \hline
  & \multicolumn{3}{c||}{Data Advantage}
  & \multicolumn{3}{c|}{Reputation Advantage} \\
    \hline
  & $\TS$  & $\DEG$  & $\DG$
  & $\TS$  & $\DEG$  & $\DG$ \\
  \hline
  $\TS$
 & \makecell{\textbf{0.14} $\pm$0.02}
    & \makecell{\textbf{0.18} $\pm$0.02}
    & \makecell{\textbf{0.26} $\pm$0.03}
    & \makecell{\textbf{0.24} $\pm$0.02}
    & \makecell{\textbf{0.2} $\pm$0.02}
    & \makecell{\textbf{0.26} $\pm$0.02}\\\hline
    $\DEG$
  & \makecell{\textbf{0.26} $\pm$0.02}
    & \makecell{\textbf{0.26} $\pm$0.02}
    & \makecell{\textbf{0.34} $\pm$0.03}
   & \makecell{\textbf{0.37} $\pm$0.03}
    & \makecell{\textbf{0.29} $\pm$0.02}
    & \makecell{\textbf{0.31} $\pm$0.02} \\\hline
    $\DG$
    & \makecell{\textbf{0.25} $\pm$0.02}
    & \makecell{\textbf{0.27} $\pm$0.02}
    & \makecell{\textbf{0.34} $\pm$0.03}
     & \makecell{\textbf{0.35} $\pm$0.03}
    & \makecell{\textbf{0.27} $\pm$0.02}
    & \makecell{\textbf{0.3} $\pm$0.02}  \\\hline
\end{tabular}
\end{adjustbox}
\caption{Uniform MAB Instance, $X = 500$}
\end{table}

\newpage
\subsection{Mean Reputation vs. Relative Reputation}

We present the experiments omitted from Section~\ref{sec:revisited}. Namely, experiments on the Heavy-Tail MAB instance with $K=3$ arms, both for ``performance in isolation" and the permanent duopoly game. We find that $\DynamicEpsGreedy > \DynamicGreedy$ according to the mean reputation trajectory but that $\DynamicGreedy > \DynamicEpsGreedy$ according to the relative reputation trajectory \emph{and} in the competition game. As discussed in Section~\ref{sec:revisited}, the same results also hold for $K = 10$ for the warm starts that we consider.

The result of the permanent duopoly experiment for this instance  is shown in Table \ref{ht_k3}.

\begin{table}[h]
\centering
\begin{tabular}{|c|c|c|c|}
  \hline
  & \multicolumn{3}{c|}{Heavy-Tail} \\
\hline
   & $T_0$ = 20 & $T_0$ = 250 & $T_0$ = 500 \\ \hline
\TS vs. \DG
  & \makecell{\textbf{0.4} $\pm$0.02\\ \Eeog 770 (0)}
    & \makecell{\textbf{0.59} $\pm$0.01\\ \Eeog 2700 (2979.5)}
    & \makecell{\textbf{0.6} $\pm$0.01\\ \Eeog 2700 (3018)} \\ \hline
\TS vs. \DEG
    & \makecell{\textbf{0.46} $\pm$0.02 \\ \Eeog 830 (0)}
    & \makecell{\textbf{0.73} $\pm$0.01 \\ \Eeog 2500 (2576.5)}
    & \makecell{\textbf{0.72} $\pm$0.01 \\ \Eeog 2700 (2862)} \\ \hline
\DG vs. \DEG
    & \makecell{\textbf{0.61} $\pm$0.01 \\ \Eeog 1400 (556)}
    & \makecell{\textbf{0.61} $\pm$0.01 \\ \Eeog 2400 (2538.5)}
    & \makecell{\textbf{0.6} $\pm$0.01 \\ \Eeog 2400 (2587.5)} \\\hline
\end{tabular}
\caption{Duopoly Experiment: Heavy-Tail, $K=3$, $T=5000$.\\
Each cell describes a game between two algorithms, call them $\alg[1]$ vs. $\alg[2]$, for a particular value of the warm start $T_0$. Line 1 in the cell is the market share of $\alg[1]$: the average (in bold) and the 95\% confidence band.
%For example, the cell in the top left indicates that TS gets on average 64\% of the market when played against DG.
Line 2 specifies the ``effective end of game" (\Eeog): the average and the median (in brackets). }
\label{ht_k3}
\end{table}

The mean reputation trajectories for algorithms' performance in isolation and the relative reputation trajectory of \DynamicEpsGreedy vs. \DynamicGreedy:
\begin{figure}[H]
\centering
\includegraphics[scale=0.35]{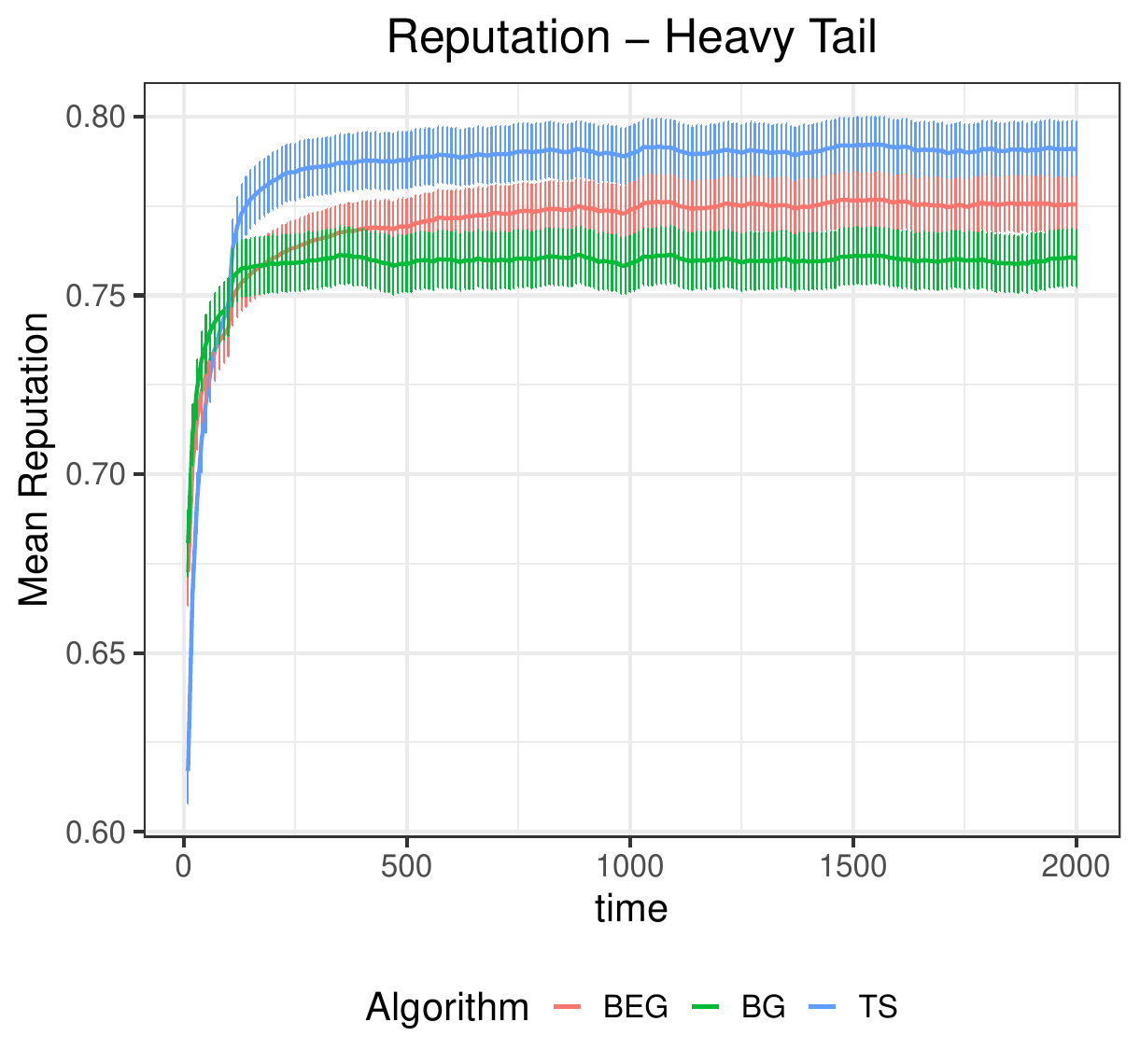}
\includegraphics[scale=0.35]{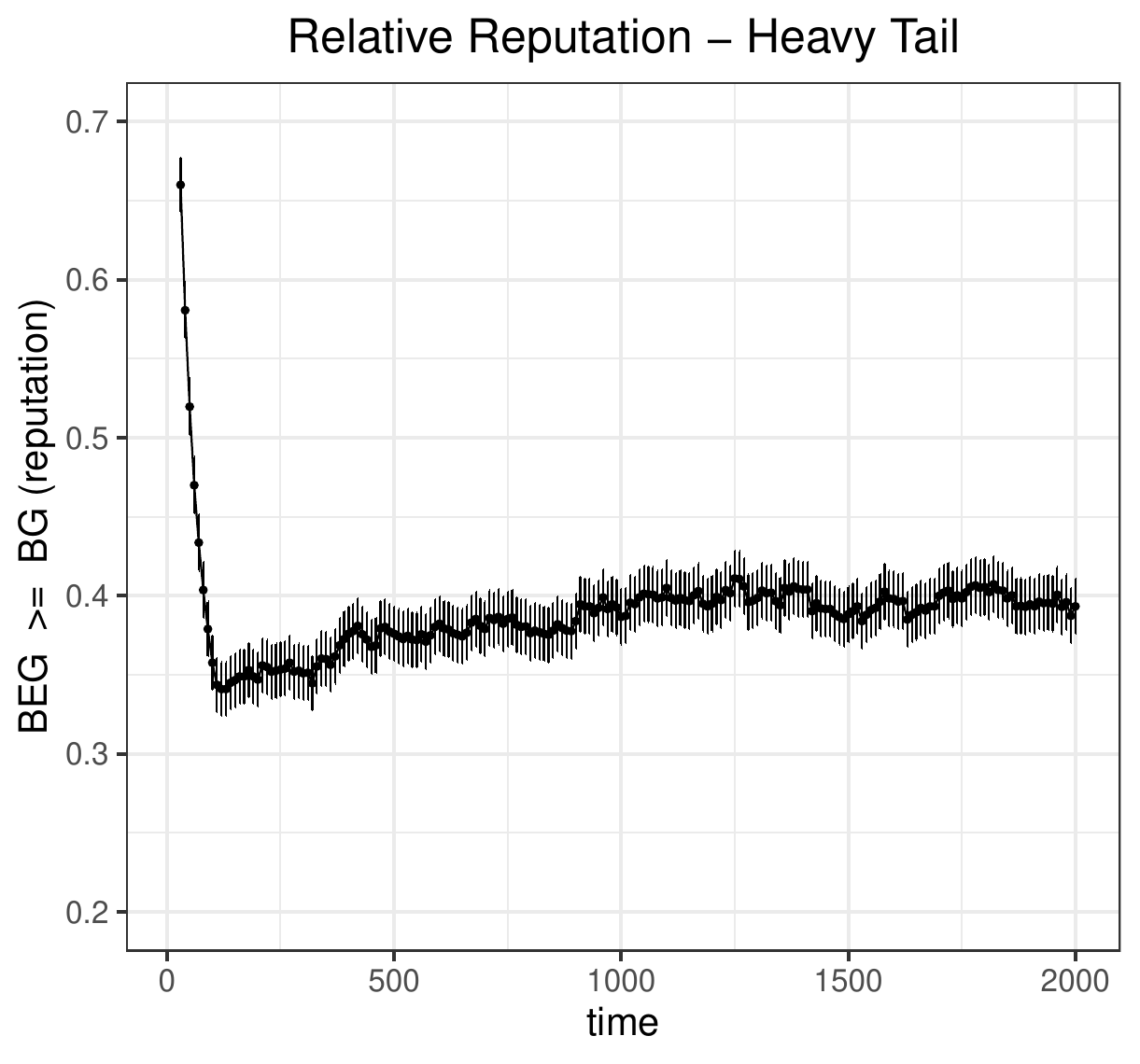}
\caption{Mean reputation (left) and relative reputation trajectory (right) for Heavy-Tail, $K = 3$}
\end{figure}